     \font\sm=cmbx5    
   \def\({\left(} \def\){\right)}  
   \def\lk{\,\left[ \,} \def\rk{\,\right] \,} 
   \def\rki{\,\right]}  \def\lb{\left\{} \def\rb{\right\}}  
   \def\lw{\left\langle} \def\rw{\right\rangle}

   \def\wu#1{\sqrt{{#1} \,}^{ \hbox to0.2pt{\hss$ 
     \vrule height 2.5pt width 0.6pt depth -.5pt $} \;\! }}
   \font\dick=cmbx12  \font\duenn=cmbx10 
   \def\dop{\duenn \raise0.6pt\hbox to 0.2pt{: \hss}}
   \def\dep{\duenn \raise0.6pt\hbox to 0.3pt{\hss :}}

   \def\cl#1{{\cal #1}}    \def\ct#1{\nz\centerline{#1}}
   \def\schl#1{\widetilde{#1}}  \def\ov#1{\overline{#1}}   
   \def\dis{\displaystyle}  \def\0{\over } \def\6{\partial }
   \def\ueb#1#2{{\buildrel{#1}\over{#2}}}
   \def\P{ {\mit\Pi} }  
   \def\gll{\; {\dop} =}   \def\glr{= {\dep} \;}
   \def\ln{{\rm ln}}   \def\det{{\rm det}}
   \def\Tr{\hbox{Tr}}  \def\Sp{\hbox{Sp}}
       
      \def\grad{{\rm grad}} 
   \def\div{{\rm div}} \def\rot{{\rm rot}}
   \let\a=\alpha  \let\b=\beta    \let\d=\delta  
   \let\g=\gamma  \let\l=\lambda  \let\o=\omega  
   \let\s=\sigma      \let\e=\varepsilon  
   \let\ph=\varphi   \let\ta=\vartheta
   \let\D=\Delta  \let\G=\Gamma  \let\L=\Lambda  
   \let\O=\Omega
   \def\mn{_{\mu\nu}}   \def\omn{^{\mu\nu}}
   \def\eq#1{(\ref{#1})}        \def\nonu{\nonumber}
   \def\be#1{\begin{equation} \label{#1}}
   \def\ben{\begin{equation}}   \def\ee{\end{equation}}
   \def\bea#1{\begin{eqnarray} \label{#1}}
   \def\bean{\begin{eqnarray}}  \def\eea{\end{eqnarray}}
   \let\thq=\theequation
   \def\pfeil{_\rightharpoonup}  \def\leer{\phantom{a}}
   \def\opf{\buildrel \pfeil \over \leer}
   \def\jvv{j \lower0.4pt\hbox to 2pt{\hss $\opf$}} 
   \def\jv{j \lower0.2pt\hbox to 1.4pt{\hss $\opf$}} 
   \def\ivv{i \lower0.4pt\hbox to 2pt{\hss $\opf$}} 
   \def\iv{i \lower0.2pt\hbox to 1.4pt{\hss $\opf$}} 
   \def\hq{h \raise0.2pt\hbox to 0.4pt{\hss $^-$}}   
   \def\vk#1{\hbox{$\buildrel           \pfeil \over #1$}}
   \def\vkk#1{\hbox{$\buildrel   \;     \pfeil \over #1$}}
   \def\vkkk#1{\hbox{$\buildrel  \, \;  \pfeil \over #1$}}
   \def\grpf{\displaystyle  _\rightharpoonup}
   \def\vg#1{\hbox{$\buildrel       \grpf \over #1$}}
   \def\vgg#1{\hbox{$\buildrel  \;  \grpf \over #1$}}
\def\fzz{f} \def\bzz{b} \def\dzz{d} \def\gzz{g} \def\hzz{h} 
\def\jzz{j} \def\kzz{k} \def\lzz{l} \def\mzz{m} \def\wzz{w} 
\def\tzz{t} \def\izz{i} 
\def\bezz{\beta} \def\dezz{\delta} \def\xizz{\xi} 
\def\pszz{\psi}  \def\vthzz{\vartheta}
\def\uph{ \! \mathop{\vphantom{a}} } \def\dph{ \vphantom{a} }
\def\vc#1{\def\tast{\noexpand#1} \def\test{#1}
\ifcat\tast\bzz 
\ifx\test\fzz \vkkk f \uph \else  \ifx\test\bzz \vkk b \uph \else
\ifx\test\dzz \vkkk d \uph \else  \ifx\test\gzz \vkk g \dph \else
\ifx\test\hzz \vkk h \uph \else   \ifx\test\izz \ivv \dph \else 
\ifx\test\jzz \jvv \dph \else     \ifx\test\kzz \vkk k \uph \else
\ifx\test\lzz \vkk l \uph \else   \ifx\test\tzz \vkk t \uph \else
\ifx\test\mzz \vg m \dph \else    \ifx\test\wzz \vg w \dph \else 
\ifnum \lq#1<91 \vgg #1 \uph \else \vk #1 \dph 
  \fi \fi \fi \fi \fi \fi \fi \fi \fi \fi \fi \fi \fi  \else
\ifx\test\bezz \vkk \beta \uph \else  
    \ifx\test\pszz \vkk \psi \dph \else 
\ifx\test\dezz \vkk \delta \uph \else 
    \ifx\test\xizz \vkk \xi \uph \else
\ifx\test\vthzz \vkk \vartheta \uph \else \vk #1 \dph
  \fi \fi \fi \fi \fi \fi }
\def\abst#1{\def\tast{\noexpand#1} \ifcat\tast\bzz 
    \ifnum \lq#1<91 \; \else \, \fi      \else \, \fi}
\def\absv#1{\def\tast{\noexpand#1} \def\test{#1}
\ifcat\tast\bzz \ifx\test\fzz \, \; \else  
    \ifx\test\dzz \, \; \else
\ifx\test\bzz \, \else \ifx\test\gzz \, \else 
       \ifx\test\hzz \, \else
\ifx\test\kzz \, \else \ifx\test\lzz \, \else 
       \ifx\test\tzz \, \else 
\ifnum \lq#1<91 \; \else   
  \fi \fi \fi \fi \fi \fi \fi \fi \fi \else
\ifx\test\bezz \, \else \ifx\test\pszz \, \else 
       \ifx\test\dezz \, \else
\ifx\test\xizz \, \else \ifx\test\vthzz \, \else 
   \fi \fi \fi \fi \fi \fi }
\def\vphan#1{\def\tast{\noexpand#1} \def\test{#1}
\ifcat\tast\bzz \ifx\test\fzz  \uph \else 
      \ifx\test\bzz  \uph \else
\ifx\test\dzz  \uph \else  \ifx\test\hzz  \uph \else
\ifx\test\kzz  \uph \else  \ifx\test\lzz  \uph \else
\ifx\test\tzz  \uph \else  \ifnum \lq#1<91 \uph \else \dph 
  \fi \fi \fi \fi \fi \fi \fi \fi \else
\ifx\test\bezz  \uph \else  \ifx\test\dezz  \uph \else
\ifx\test\xizz  \uph \else  \ifx\test\vthzz \uph \else \dph
  \fi \fi \fi \fi \fi }

  \def\pubox{\dick _{\raise 1pt\hbox{.}} }
  \def\ppubox{\dick _{\raise 1pt\hbox{..}} }
 \def\p#1{{\buildrel \abst #1 \pubox \over #1} \vphan #1 }

  \def\pbox{\dick _{\hbox{.}} } 
  \def\pvc#1{{\buildrel \absv #1 
      \pbox \over {\vc #1}} \vphan #1 }

  \def\vcsm#1{ \def\sm{\raise 1.6pt\hbox to 5pt{\hss $_#1$}} 
               {\buildrel \pfeil \over \sm} \>\!\! }

   \def\abpfeil{ \raise 1pt\hbox{ $_\vee$ \hskip -6.8pt
          \vrule depth 0.6pt height 4.6pt width 0.3pt} \;\; }
   \def\apf#1{ \buildrel \abpfeil \over {#1} }

   \font\log=logo10 scaled \magstep2 
   \def\MA{\hbox{\log A}}  \def\MiA{\hbox{\scriptsize $\MA$}} 
   \def\MMA{\hbox{\large $\MA$}}

\def\dod{\,D\hspace{-.42cm}I\hspace{.2cm}}
\def\sec#1{\let\dq=\thq 
   \renewcommand{\theequation}{\arabic{section}.\dq}    
   \setcounter{equation}{0} \vskip .1cm \section{#1}}         
\def\nz{$ $\\}  \def\ek#1{{\bf [$\,$#1$\,$]}}
\def\ft{\footnotesize}   \def\ou{^{[U]}}

\def\unt#1#2#3#4{\hbox{
  \vrule height .2cm width 0.4pt  depth -.5pt
  \vrule height .5pt width #1cm   depth 0pt
  \vrule height .2cm width 0.3pt  depth -.5pt
  \hskip -#1cm \hskip #3cm  \vrule width 0.3pt 
  height 0pt depth #2cm \lower #2cm\hbox{\lower .14cm\hbox{$
  \!\!\!\,\dis = #4$}}} \nonu }
  \def\I{\hbox{$\,\cl I \,$}}  \def\2{\hbox{${1\02}$}}
  \def\bp{\hbox{\boldmath$\partial$}} 
 \def\AA{\hbox{{\bf A}}}
  \def\fhsp{\hat{\bf T}{\bf r}}
 \def\gdw{\,\leftharpoonup \hspace*{-.3cm}\rightharpoondown\,}

 \def\ekk#1{{\bf \{$\,$#1$\,$\}}}
 \def\gef{\lower.5pt\hbox{$^\angle\hskip -.245cm
         $\raise .7pt\hbox{${}_{^\swarrow}$}}}
 \def\Sp{{\rm Tr}}
 \def\folg{\,\curvearrowright\,}

\documentclass[12pt,twoside]{article}   \usepackage{amssymb}
\begin{document}

\vspace*{-12mm}
\noindent 
Hermann Schulz\,\footnote{\ \parbox[t]{15cm}{
hschulz@itp.uni-hannover.de \\
Institut f\"ur Theoretische Physik, Universit\"at Hannover, 
Appelstr.~2, D--30167 Hannover, Germany}} \hfill 
LAPTH--808/00 \\ 
Annecy in summer 2000  \hfill ITP--UH 16/00\,\, \\
\rightline{hep-ph/0008239}

\vskip 8mm
\pagestyle{myheadings}
\markboth{\small YM\hspace*{14.76cm}}{\hspace*{14.76cm}\small YM}
\thispagestyle{empty}

\begin{center} 
{\Large\bf  The 3~D $\,$Yang--Mills system } \\[4mm]
Introductory Lecture Notes on the Schr\"odinger wave 
functional and Hamiltonian treatment by 
Karabali, Kim, Nair \cite{kkn}            \end{center}

\nz
Field theory, when viewed as a medium coupled to a thermal bath, 
gains a useful new parameter~: the temperature $T$. At the upper 
end of the $T$ axis ($g$ small) Yang--Mills theory (i.e. QCD without 
quarks, i.e. gluon black--body radiation) becomes susceptible
to perturbation theory. The corresponding diagrammatics, however,
runs against some ``barrier of calculability''. At order $g^6$ 
for the pressure and at $g^4$ for the self-energy infinitely many 
diagrams (the ``Linde sea'') contribute with the same order of
magnitude. Does QCD exist? \cite{linde}. The Linde sea represents
a physics by itself \cite{gpy}, namely that of a 3~D Yang--Mills 
system at zero temperature. Through learning about this system
(non-perturbatively, but not necessarily exact), the barrier 
is overcome.  

As usual, a lecture details the work done by others. In essence, 
there are only three others since, in the following, we shall focus 
on the paper \cite{kkn}
\\[8pt] 
\hspace*{1cm} {\small\bf 
   D. Karabali, C. Kim and V. P. Nair, Nucl. Phys. 
   {\bf B 524} (1998) 661 } \\[1pt]
\hspace*{1.8cm}{\small 
   ``Planar Yang--Mills theory$\,$: Hamiltonian, 
   regulators and mass gap''} \quad   
\\[8pt] 
referred to as KKN for brevity. Part I of these notes was originally 
put in german: hep-ph/9908527. While translating, we step back from 
any reorganization. Let the original step by step understanding of 
this ``very strange new matter'' be a suitable low level 
introduction automatically. KKN's \S~2 is merely an {\sl Outline 
of the main argument}. So, it has its preceeding papers \cite{vor}. 
The ``paper after'' \cite{nach} is examined in part II. Equations 
in \cite{kkn} are referred to in the form \ek{n.m}. But there might 
be no need for really looking into the original work. 

One aspect of KKN's treatment could be a bit fascinating, namely 
the ``unification'' of several areas of physics. The old idea 
of Feynman \cite{fey} in 1981 becomes explicit. The gauge orbit 
can be prepared. The (here Hermitean) Wess--Zumino--Witten action 
gets application, as does conformal field theory. Finally, the 
thermal field theory \ --- \  a bit sickly after its euphoric 
phase around 1990 (Braaten--Pisarski resummation) \ --- \  finds 
back to its original attitude of basically understanding reality. 
A circle gets closed.  
  
\newpage 

\noindent 
{\bf Deser, Jackiw, Templeton, 1982 \cite{deser} :} \\ {\sl
The study of vector and tensor gauge theories in 
three--dimensional space--time is motivated by their 
connection to high temperature be\-ha\-vior of 
four--di\-men\-sio\-nal models, and is justified by the 
special properties which they enjoy. }

\vskip 2mm
\noindent  
{\bf Karabali, Kim, Nair, 1998 \cite{kkn} :} \\ {\sl
... there is at least one interesting physical situation, viz., 
the high temperature phase of chromodynamics and associated
magnetic screening effects, to which the (2+1)--dimen\-sio\-nal 
theory can be directly applied. } 

\vskip 20mm

\tableofcontents

\newpage


\sec{ Two--dimensional classical electrodynamics}

\vspace*{-.6cm} \hspace*{2cm} \parbox{6.5cm}{ 
\bea{1.1} \div \vc E &=& \rho  \\  \label{1.2}
        \rot \vc E &=& - \pvc B \eea } \parbox{7cm}{ 
\bea{1.3} \div \vc B &=& 0  \\ \label{1.4}
        \rot \vc B &=& \vc j + \pvc E  \eea } \\
2D physics is a special case of 3D. Consider homogeneously 
charged straight lines (parallel to z--axis) to be the 
``point'' charges. They form the densities 
$\,\rho (x,y,t)\,$ and $\;\vc j =$ 
$(\, j_1(x,y,t)\, ,\, $ $ j_2(x,y,t)\, ,\, 0\, )$. 
Assuming that even $\rot \vc B$ attains the structure of 
$\vc j$, we have that
\vspace*{-.6cm}
\bean \hspace{1.2cm} 
      \hbox{\eq{1.1}, \eq{1.4} lead to } 
      \hspace{1cm} \vc E  &=& \big(\; E_1(x,y,t)\; ,
      \; E_2(x,y,t)\; ,\; 0\;\big) \;\; . \;\;  \nonu \\
      \hspace{1.2cm}
      \hbox{Now \eq{1.2}, \eq{1.3} show that } 
      \hspace{1cm} \vc B  &=& \big(\; 0\; ,
       \; 0\; ,\; B(x,y,t)\;\big) \nonu
\eea \\[-.8cm] 
has a third component only. This verifies the assumption. 
Maxwell has reduced to \eq{1.1}, \eq{1.2}, \eq{1.4}. 
No theorem is required to write $\, B = \big( \rot \vc A \big)_3 
= \6_x A_2(x,y,t) - \6_y A_1(x,y,t)\,$, hence introducing
a 2--component vector potential. But we need \eq{1.2} to allow 
for $\,\vc E =  - \pvc A - \grad \,\phi\;$. The three fields 
$E_1$, $E_2$, $B$ remain unchanged under regauging according to
$\vc A \to \vc A + \nabla \chi(x,y,t)\,$ and $\;\phi 
\to \phi - \6_t \chi(x,y,t)\,$. 

The strict temporal gauge $\phi=0$ (or Weyl gauge, or radiation
gauge, [Muta, S.51]) does not fix completely. Without changing
$\vc E =  - \pvc A$ or $\,\vc B=\nabla \times \vc A$, we may still
regauge by $\vc A \to \vc A + \nabla \chi(x,y)\,$. Note that
$\chi$ must not depend on time.

Four--notation turns into ``three--notation'', of course, with
metrics $+ - -\,$, $\mu=0,1,2\,$, $\big(\, \phi\, , \, 
\vc A \,\big) \glr A^\mu\;$ and $\6^\mu = \big(\,\6_0\, , 
\, -\nabla\,\big)$. Hence the connection between fields and
potentials reads $E^j=-\6^0A^j+\6^j A^0\,$, 
$B=-\6^1A^2+\6^2A^1 = - \e_{jk}\6^jA^k $ ($\e_{12} \gll 1$). 
As usual we define the field tensor $\, \6^\mu A^\nu 
-\6^\nu A^\mu \glr F\omn\,$ and enjoy the resulting
matrix version$\,$:
\be{1.5}  F\omn = \( 
     \matrix{  0   & - E_1 & - E_2 \cr
               E_1 &   0   & - B   \cr 
               E_2 &   B   &   0   \cr } \) \qquad , \qquad
    F\mn = \( 
     \matrix{   0   &  E_1  &  E_2  \cr
              - E_1 &   0   & - B   \cr 
              - E_2 &   B   &   0   \cr } \) \quad .
\ee 
Using this, we arrive at the Lagrangian 
\bea{1.6} 
  \cl L &=&  - {1\04}\, F\omn F\mn \; = \; 
  {1\04}\, \Sp \Bigg(\;\;\; 
\lower 7pt\vbox{\hbox{\scriptsize \ (1.5)}  \vskip -.3cm
                 \hbox{\scriptsize \ left}  \vskip -.3cm
                 \hbox{\scriptsize matrix }} \;\,\cdot \;\;
\lower 7pt\vbox{\hbox{\scriptsize \ (1.5)}  \vskip -.3cm
                 \hbox{\scriptsize right }  \vskip -.3cm
                 \hbox{\scriptsize matrix }} \;\; \Bigg)
  \;  = \; {1\02} \( \vc E^2 - B^2 \) \nonu \\
     &=& {1\02} \Big( \, [ - \pvc A - \grad \phi ]^2
     - {1\02} [ ( \rot \vc A )_3 ]^2 \,\Big) \quad . \quad 
\eea 
Now, when turning to the Hamiltonian density, the advantage 
of strict Weyl gauge $\phi=0$ becomes obvious$\,$:
\bea{1.7}  
  \cl L &=& {1\02} \pvc A^2 - {1\02} B^2 \qquad\; , 
  \;\;\qquad  \vc \P \; = \;\pvc A \; = \; - \vc E  \\
       \label{1.8} 
  \cl H &=& \Big[\; \pvc A \vc \P  - \cl L \;\Big]_{\rm 
  eliminate\; ...} \;  = \; {1\02} \big( \vc \P^2 + B^2 \big) 
  \quad . 
\eea 
Only the two real fields $A_1$ and $A_2$ and their generalized
momenta are left to work with.


\sec{ Yang--Mills fields in 2+1~D }

The specialities of non--abelian theory have nearly nothing to do 
with dimension. Just the Lorentz index $\mu$ now runs from 0 to 2. 
As if there were particles in the x-y plane ($\psi$ with $N$ 
colour components) too, we require invariance of any physics under 
$\vc r$--$t$ dependent changes of the $\psi$ phase. Changes of
notation are psychological warfare. We therefore first remember 
our familar Hannover notation. But let the coupling immediately
be denoted by $e$ (in place of $g\,$)$\,$:
\be{2.1} 
  \! \left. \parbox{14.6cm}{ \vspace*{-.6cm}
  \bean
  U = e^{-ie \L^a (x) T^a } 
  &,&  D_\mu = \6_\mu - i e A_\mu^a T^a   \nonu \\
  \MA_\mu \gll T^a A_\mu^a 
  &,&  \MA_\mu \to \MA_\mu\ou =  U\MA_\mu U^{-1} 
    - {i\0e} \, U_{\prime \mu} U^{-1} \nonu \\ 
   F\mn^{\;\, a} = \6_\mu A_\nu^{\; a}  - \6_\nu 
    A_\mu^{\; a} + e f^{abc} A_\mu^{\; b} A_\mu^{\; c} 
  &,& F\mn = \6_\mu \MA_\nu  - \6_\nu \MA_\mu 
      - i e \lk \MA_\mu , \MA_\nu \rk   \nonu \\
  & & \hspace*{-7cm}  \hbox{\eq{1.5} :} \qquad
  B^a = - F^{12\, a} = - \( \6^1 A^{2\,a}  - 
  \6^2 A^{1\,a} + e f^{abc} A^{1\, b} A^{2\, c}\) \nonu \\
  & & \hspace*{-7cm}  E^{j\, a} = - F^{0j\, a} 
      = - \( \6^0 A^{j\,a} - \6^j A^{0\,a} + e f^{abc} 
      A^{0\, b} A^{j\, c} \) \;\;\; , \;\;\;
      A^{0\, a} \equiv 0 : \; E^{j\, a} 
      = - \p A{}^{j\, a}   \nonu \\  
  & & \hspace*{-6.6cm} \cl L_{\rm strict Weyl} 
    = - {1\04} F^{\mu\nu \, a} F\mn^{\;\, a} = {1\02}
   E^{j\, a} E^{j\, a} - {1\02} F^{12\, a} F^{12\, a}
   = {1\02} \p A{}^{ja} \p A{}^{ja} - {1\02} B^a B^a 
   \quad \nonu 
   \eea \vspace*{-.6cm} } \!\!\right]
\ee 
Among certain people around Chern and Simons (but nevertheless
the article of Jackiw \cite{jack} in the Les Houches lectures of 
1983 is very nice) it is common, however, to absorb the coupling 
$e$ in the fields and to work with \ a n t i hermitean field matrices.
Then, the fields in \eq{2.1} are processed as follows$\,$:
\bea{2.2}
  \L^a &\gll& e \, \L^{a\; {\bf old}} \qquad , \qquad 
  A_j^a \gll e \, A_j^{a\; {\bf old}}  \qquad ,\qquad
  F\mn^a \gll e \, F\mn^{a\; {\bf old}} \qquad ,
       \;\; \nonu \\
  B^a &\gll& - e \, B^{a\; {\bf old}} \,\quad ,\qquad
  A_j \gll - i \, e \, \MA_j^{\; {\bf old}} 
       \quad , \qquad
  F\mn \gll - i \, e \, F\mn^{\; {\bf old}} 
  \quad . \quad
\eea 
In parallel with these translations, we now (and forever)
adopt the strict temporal gauge. Only the $2*n$ fields 
$A_j^a (\vc r )$ are left. They live in the plane
$\vc r = (x,y)$. Let changes of gauge lie in a finite region
$\,$: $\L^a \( \vc r \to \infty \) \to 0\,$, and with \eq{2.2}
it is
\be{2.3}
   U (\vc r ) = \exp{\( -i\L^a(\vc r ) T^a \)} \quad , 
   \quad a = 1, \ldots , N^2-1 \glr n \quad . 
\ee 
KKN are not among those, who even prefer antihermitean generators.
Hence, the following line, which was ``forgotten'' in \eq{2.1},
holds true old as new$\,$:
\be{2.4}
   \Sp \( T^a T^b \) = {1\02} \d^{ab} \quad , \quad
   \lk T^a , T^b \rk = i f^{abc} T^c  \quad . \quad
\ee 
By means of \eq{2.2} the covariant derivative becomes nice.
$\6_j$ is antihermitean, and this now harmonizes with the 
antihermiticity (and tracelessness) of the matrix fields$\,$:
\be{2.5}
  D_j = \6_j + A_j \quad , \quad 
  A_j = - i T^a A_j^a \quad . \quad 
\ee 
Their gauge transformation reads
\be{2.6}
   A_j \to A_j\ou = U A_j U^{-1} - U_{\prime j} 
   U^{-1}  \qquad , \qquad j=1,2 \quad . \quad
\ee 
Even field tensor and magnetic field loose ballast$\,$:
\be{2.7}
  F_{jk}^{\; a} = \6_j A_k^{\, a} - \6_k A_j^{\, a} 
         + f^{abc} A_j^{\, b} A_k^{\, c} 
   \quad ,\quad   F_{jk} = \6_j A_k - \6_j A_k
       + \lk A_j , A_k \rk \quad , \quad
\ee 
\be{2.8}
  B^a  = \6_1 A_2^{\, a} - \6_2 A_1^{\, a} 
   + f^{abc} A_1^{\, b} A_2^{\, c} \quad . \quad
\ee 
Finally, the Lagrangian becomes
\be{2.9}
   \cl L = {1\0 2 e^2} \p A{}_j^a \p A{}_j^a  - 
  {1\0 2 e^2} B^a B^a \;\; \glr \;\; \cl T - \cl V \quad .
\ee 
May be, we are the first step in. \eq{2.8} is found at KKN 
in the text below \ek{2.4}. \eq{2.6} is \ek{2.1}. But 
\eq{2.9} is not \ek{2.4}. Well, perhaps something was wrong
with the key board, setting $e^2$ in the numerator. 
\eq{2.9} is fine, because it leads by
\be{2.10}
  \P_j^a = \6_{\p A{}_j^a} \cl L = {1\0 e^2}
   \p A{}_j^a \quad , \quad 
   \cl H = \lk \P_j^a \p A{}_j^a - \cl L 
   \rki_{\hbox{\scriptsize eliminate $\p A$}} 
   \; = {e^2\0 2} \P_j^a\P_j^a \, + \,\cl V
\ee 
to KKN's  Hamiltonian density. Of course, the Lagrangian \eq{2.9}
can be shown to be invariant under the restricted 
$U$--transformations \eq{2.3}, \eq{2.6}, as usual.


\sec{ Matrix parametrisation }

The first important step into the KKN buisiness needs only some
rough philosophy. There are only the 2$*n$ fields $A_j^a$. 
They regauge according to \eq{2.6}. But quantization must be
restricted to physical fields (not connected through regauging). 
Hence, any better view into the space of fields $A_j^a$, 
any gain in harmony, would be fine.

The matrix parametrization could have been discovered as follows. 
We look at the gauge transformation \eq{2.6} and play around with. 
For instance, we may place a unit matrix anywhere in this line. 
Let us put $1=M M^{-1}$ in front of each $U^{-1}\,$,
\bea{3.1}
   A_j\ou &=& U A_j \, M \, M^{-1} \,U^{-1}
              - U_{\prime j}\, M \, M^{-1} \, U^{-1} 
          \nonu \\ 
          &=& U A_j \, M \, (UM)^{-1} 
              -  U_{\prime j}\, M \, (UM)^{-1} \quad , \quad 
\eea 
with the second line coming into mind for harmony. But
something is not yet good with the last term. $U$ likes $M$,
and $(UM)_{\prime j}$ might appear there. Well, we may write
$ U_{\prime j} \, M = (UM)_{\prime j} - U\, M_{\prime j}$ and 
insert: 
\be{3.2}
   A_j\ou = - (UM)_{\prime j} \, (UM)^{-1} \, + \,
   U \lk M_{\prime j} + A_j \,M \rk \, (UM)^{-1} \quad.
\ee 
The first term is ``harmonic'', and the second term we should
like to get rid of. But before reaching this there is one
more step towards harmony.

As the fields $A_1^a$ and $A_2^a$ are real, we may combine them
to a general complex field $A^a \,\gll\, {1\02}\,(A^a_1 + 
i A^a_2)\,$. Correspondingly, $A_1= -iA^a_1 T^a$ and 
$A_2= -iA^a_2 T^a$, which are antihermitean and traceless matrix
fields, combine to
\be{3.3}
    A \; \gll \; {1\02} \,\Big( \; A_1 \, + \, i \, A_2 
    \;\Big) \quad . \quad
\ee 
This is \ o n e \ field. It is traceless but otherwise a general 
complex matrix. Given $A$, by preparing its hermitean and 
antihermiean part, one is led back uniquely to $A_1$ and 
$A_2$ . With \eq{3.3}, and with the definition
\be{3.4}
  \6 \; \gll \; {1\02} \,\Big( \; \6_1 \, + \, i \, \6_2 
  \;\Big) \;\;\qquad \hbox{( and $\dis  \ov{\6} \; \gll \; 
   {1\02}\, \Big( \; \6_1 \, - \, i \, \6_2\,\Big)\,$ 
   for later use ) } \quad , \quad 
\ee 
we now turn back to \eq{3.2} and combine these two equations
(first one $+ i\,$the second) to a single one 
\be{3.5}
   A\ou \; = \; - (\6\; UM) \, (UM)^{-1} \, + \,
     U \lk \6\, M + A \,M \rk \, (UM)^{-1} \qquad 
\ee 
with again the desire to get rid of the second term.

So far, nothing was assumed on $M$, except that it is a $N 
\times N$ matrix with an existing inverse. The square bracket in 
\eq{3.5} vanishes, \ i f \ for any given traceless field $A$ 
there is a matrix field $M$ such that
\be{3.6}
   A \; = \; - \, \(\6\, M \)  \, M^{-1} \quad . \quad
\ee 
If so, \eq{3.5} becomes $A\ou =  - (\6\; UM) \, (UM)^{-1}$
and tells us that regauging amounts to
\be{3.7}
    M\;\; \to \;\;\,  M\ou \; = \; U\, M  \quad . \quad
\ee 
Moreover, (if so) we have that
\be{3.8}
  \( M^\dagger \, M \)\ou = (UM)^\dagger \, UM  = M^\dagger\, 
   U^\dagger\, U\, M  = M^\dagger M\,\glr\, H \qquad
\ee 
is an invariant under gauge transformations. 

To exhaust the above ~``~i~f~''~ we first realize intuitively 
that by running through the $M$ space any matrix $A$ is
reached. However, $A$ has to be traceless (but is 
unrestricted otherwise). Which way are the $M$'s to be 
restricted, correspondingly?  
\be{3.9} 
\hbox{\underline{\,The answer~:\,}} \quad 
  \det (M) \;\;\;\; \hbox{must be a function of only} \ \ \ 
     x+iy\,\glr\,\ov{z}  \quad , \quad \hspace*{15mm} 
\ee 
which is the only restriction on $M\,$. 
\be{3.10} \!\!
\hbox{\underline{\,Proof~:\,}} \quad 
  0 = \6 \,\ln \Big(\,\det(M)\,\Big)   = \6\,\Sp \Big(\,\ln (M)
  \,\Big)  = \Sp\Big(\, (\6 M )\,M^{-1} \,\Big) 
  \quad , \quad \hbox{q.e.d.} \quad 
\ee 
One may also avoid using the ln--det formula.
To reach a unique mapping from $A$ space to $M$ space, the
function just mentioned in \eq{3.9} can be even fixed:
\be{3.11}
   \det \( M \) = 1 \qquad , \;\; {\rm i.e.}
   \quad M\; \in \; \hbox{SL(N,C)} \quad .
\ee 
In passing, \eq{3.6} is \ek {2.6}, and \eq{3.7} is \ek{2.9}.

Anything depending on $x$, $y$ may be also understood to be a 
function of $\; z \gll x-iy\;$ and $\;\ov{z} \gll x+iy\;$, of 
course. For the (above) case that only one of these variables 
does occur, we have that
\bea{3.12}
   \hbox{\ft $z \gll x - iy \;\; , \;\; 
         \6 \,\gll {1\02} \(\6_1 + i \6_2 \)$} 
   \quad &,& \quad   
   \6 f(z) = f'(z) \quad , \quad  
           \ov{\6} f(z) = 0   \nonu \\ 
   \hbox{\ft $\ov{z} \gll x + i y \;\; , \;\; 
         \ov{\6} \,\gll {1\02} \(\6_1 - i \6_2 \)$} 
   \quad &,& \quad
   \ov{\6} f(\ov{z}) = f' (\ov{z}) \quad , \quad  
           \6 f(\ov{z}) = 0  \quad , \quad 
\eea 
i.e. differentiation with respect to the ``wrong'' variable 
gives zero.


\sec{ Solving \boldmath$A=-(\6M)M^{-1}$ for \boldmath$M$ }

Remember the way Born's approximation is derived$\,$: book down 
$(\D + k^2 )\,\psi = V \psi\,$, consider the r.h.s. as a known 
inhomogenity, solve for $\psi$ by means of the Greens function
of the Helmholtz operator $\D +k^2$ and then iterate by starting
with a physical $\psi$. In the present case, the equation is
$\6 \, M = - AM\,$, and the operator is $\6\,$.

To solve $\,\6\, G(\vc r) = \d (\vc r )\,$ in 2D we write 
$\,G(\vc r) = 2 (x-iy)\, f(r)\,$ to get 
\be{4.1}
  \6\, G\; = \; (\6_x+i\6_y)\, (x-iy)\,f(r) = (2+r\6_r)f(r)
         = {1\0r}\,\6_r\, r^2 f(r) \;\; 
        \ueb{!}{=}\;\; \d(\vc r) \quad . \quad
\ee 
The zero apart from the origin obviously needs that $f(r) \sim 1/r^2$. 
This function now must be embedded from the physical side, and the 
arising delta--function representation needs normalization$\,$:
\bea{4.2}
  f(r) &=& {\a \0 r^2 + \e^2} \quad , \quad  
  1 \;\ueb{!}{=}  2\pi \int_0^\infty \! dr \; r \; \( 
       {1\0r}\, \6_r \, r^2\) {\a \0 r^2 + \e^2} \quad 
   \folg \quad \a = {1\02\pi} \nonu \\ 
  G(\vc r ) \;& = &\; {1\0 \pi}\, {x-iy \0 r^2 + \e^2 }
  \;\; = \;\; {1\0 \pi}\, {z \0 z \ov{z} + \e^2 } \quad . \quad
\eea 
Of course, $\e \to +0$ is meant. The limit may be only performed, 
$G \to 1/(\pi \ov{z})$, if the pitfall ``$\6 G=0$'' is anyhow
excluded. Due to the translational invariance of $\6$ we may write
\be{4.3}
   \6 \, G (\vc r-\vc r') = \d (\vc r-\vc r') \;\folg \;  
  \6 \int \! d^2r' \; G(\vc r-\vc r') \( -  A (\vc r')  
  M(\vc r') \)  = - A(\vc r ) M(\vc r ) \;\; , \;\; 
\ee 
hence having obtained a special solution of the inhomogeous 
equation $\6\, M = - AM\,$:  
\be{4.4}
   M = M_{\rm hom} - \int' G A M \quad , \quad
  \6 M_{\rm hom} = 0 \quad , \quad 
   M = 1 - \int' G A M \quad . \quad
\ee 
The fact, that the homogeneous equation is solved by any matrix 
$M_{\rm hom}(\ov{z})\,$ is taken up again in \S~11.1$\,$. 
But here, to the right in \eq{4.4}, we specify to $M_{\rm hom} 
=1\,$, as being one allowed choice to get a unique mapping.

To iterate \eq{4.4}, we now use a matrix language also with 
respect to space. Integrals are omitted (sum convention). $M$ 
is a vector with the continuous idex $\vc r\,$. Even the 1 in
\eq{4.4} is such a vector (having equal components). $G$ is
matrix, and $A(\vc r' )$ may be replaced in \eq{4.4} by the 
matrix $\AA (\vc r' , \vc r'') \gll A(\vc r' )\,\d (\vc r' - 
\vc r'')$. Let \hbox{\bf 1} stand for $\d (\vc r - \vc r' )\,$.
The letter $A$ to the very right in each term of the following 
equation is only a vector again. Herewith the iteration of \eq{4.4} 
reads$\,$:
\bea{4.5}
  M &=& 1\, -\, GA\, +\, G\AA\, GA\,  - \, G\AA\, 
     G\AA\, GA\, + G\AA\, G\AA\, G\AA\, GA\, 
     - \, \ldots \nonu \\[2pt]
  &=& 1 - { 1 \0 \hbox{\bf 1} + G \AA }\, G A
        \, = \, 1 - {1\0 1/G + \AA\,}\, A
        \, = \, 1 - {1\0 \,\bp + \AA\,} \, A \quad , 
\eea 
where we had read off $\,1/G=\bp\,$ from $\,\bp\, G = 
\hbox{\bf 1}$. $\,\bp$ is matrix (!), namely $\6_{\vcsm r , 
\vcsm r'} = {1\02} \d' (x-x') + {i\02} \d' (y-y')\,$. Given 
a gauge field $A$, \eq{4.5} tells us the corresponding member 
in the ``underworld'' of $M$'s. Eqs. \ek{2.7}, \ek{2.8} are 
understood. 


\sec{ \boldmath$M=V\rho$ \ --- \ gauge invariant degrees of freedom }

The unique mapping, we have reached in the preceding two sections,
can be thought of in some analogy to the Fourier transformation.
Any member of the space of gauge fields ``knows'' of its 
partner in underworld, i.e. in the space of SL(N,C) matrices $M\,$,
and vice versa~:
\be{5.1}  
    M \;\; \ueb{ A = -(\6 M)\, M}{ \vrule depth -2.6pt 
    height 3pt width 2cm\!\!\longrightarrow} 
   \;\; A  \qquad , \qquad 
   A \;\; \ueb{M = 1 - {1\0 \6+A} \, A}{
   \vrule depth -2.6pt height 3pt width 2cm\!\!
   \longrightarrow}  \;\; M \quad .
\ee 
To any physics among the $A$ fields there is something
corresponding going on among the $M$'s. This also applies
to any manipulation as e.g. the preparing of the gauge orbit.
Now, in the underworld, the gauge transformation was seen to be 
extraordinarily simple: $M \to M\ou = UM\,$. So, this preparation 
and splitting off might be done there. Moreover, it reduces to a 
bit of thinking, if the following is true. Any SL(N,C) matrix 
$M$ can be uniquely written as the following \ p r o d u c t \ 
\be{5.2}
  M \; = \; V \; \rho \quad\qquad \vtop{\hbox{with \quad
  $V V^\dagger =1 \; , \;\; \det(V)=1$}
  \hbox{and \qquad $ \rho^\dagger 
        = \rho \; , \;\; \det(\rho) = 1$ \quad . }}
\ee 
Let the ``bit of thinking'' precede the proof. Given $M$,
the term ``uniquely'' means, that the corresponding matrix 
$\rho$ may be constructed. Many $M$'s have the same $\rho$. 
They differ by $V$. But $V$ is a gauge transformation. $M$'s with 
the same $\rho$ lie on the gauge orbit through $\rho$. Thats it.
Splitting off the gauge orbit means reducing the space SL(N,C)
to the hermitean matrices $\rho$, or with the words of KKN below 
\ek{2.9}: {\sl $\rho$ represents the 
gauge--invariant degrees of freedom}. \ --- \  {WOW$\,$!}

As we shall see shortly, $\rho$ is positive definit. Hence,
in place of $\rho$, one can equivalently work with
\be{5.3}
  \rho^2 \; = \;\rho^\dagger \,\rho 
   = M^\dagger V \, V^\dagger M  = M^\dagger M 
   \; = \; H \quad , \quad \det(H)=1 \quad . \quad 
\ee 
$H$ is the gauge invariant already noticed in \eq{3.8}. It is 
not hard to speculate that the Schr\"odinger wave functionals 
$\psi$ of the 2+1D functional quantum mechanics must not depend 
on $A$ or $M$ but on only the physical degrees of freedom:
$\psi (H)\,$. Don't we have already some very rough strategy? 
All we want to do can be formulated in the upper world.
But for really doing it, the underworld is appropriate. With 
the mapping between the two worlds at hand, things will be
managable anyhow. This strategy is more detailed in the next 
section and seen to be followed up through all the further
headlines.

Having problems, it sometimes helps asking around. In the present
case a few e--mails with York Schr\"oder (DESY) led to the 
\\[.4cm]  \hspace*{2cm} 
{\bf Proof of \boldmath$\; M=V\rho$ :}

\vskip -11mm
\begin{enumerate}
\item
$M^\dagger M$ is hermitean, hence can be diagonalized$\,$: 
\be{5.4}
      U \; M^\dagger\; M\; U^\dagger\; = \,\; \hbox{diag}
      \,(\l_1, \ldots, \l_N ) \;\;\glr\; {\rm diag} \quad . 
\ee 
\item
As $M^\dagger M \ph = \l \ph \; \folg \; \int | M \ph |^2 = \l$ 
shows, the diagonal elements $\l_j$ are non--negative. They are even
positive, because through           
\be{5.5}
  1\; = \; \det (UM^\dagger M U^\dagger)
   \; = \; \det ({\rm diag})\quad 
\ee 
zero--eigenvalues are forbidden. 
\item  
We now \ d e f i n e \ the hermitean matrix 
\be{5.6}
   \rho \; \gll \; U^\dagger \,\wu {\rm diag}\, U 
   \qquad \folg \quad  \det(\rho) \, 
   = \, 1 \;\; , \;\;\;  {1\0 \rho} \, 
   = \, U^\dagger\, {1\0 \wu {\rm diag} }\, U \quad , 
\ee 
where $\wu {\phantom{aa}} \gll + \wu {\phantom{aa}}$. Given 
$M$, the matrix $\rho$ is fixed uniquely, because, on one hand,  
$\rho^2=U^\dagger \wu {{\rm diag}} U U^\dagger \wu {{\rm diag}} U 
= U^\dagger {\rm diag} U = M^\dagger M$ due \eq{5.4} and, 
on the other hand, the eigenvalues of $\rho$ are all positive, 
since they are the elements of $\wu {\rm diag}\,$. In passing, 
$U$ is not fixed by $M$. Other than with a real rotation matrix, 
a diagonal matrix $U_{\rm ph}$ made up of phase factors can be 
split off from $U$ to the left, and these factors recombine in
$U_{\rm ph}^\dagger\, {\rm diag}\, U_{\rm ph}$. $U$ is not fixed, 
but $M^\dagger M$ and $\rho$ are.
\item  
Once knowing that $\rho$ has an inverse, we may solve 
$M \glr V \rho$ for $V$ and ask for its determinant and
unitarity$\,$: 
\bea{5.7}
  V &=& M \; {1\0\rho} \;\;\qquad \folg \qquad 
  \det(V)\, = \, 1 \quad \hbox{and} \quad \cr
  V^\dagger \, V \; &=& \; {1\0\rho}\, M^\dagger\, M\, 
   {1\0 \rho} \; = \; U^\dagger\, {1\0 \wu {\rm diag} }\, 
   U\, M^\dagger\, M\, U^\dagger\, {1\0 \wu {\rm diag} }\, 
   U \;\; = \;\; 1 \quad .
\eea  
Somewhat, that can be written down, does exist, \, q.e.d. and thanks.
\end{enumerate}

Initially we had some trouble to understand \eq{5.2}.
Start with counting real parameters, Sergei Ketov said, 
to see whether $M=V\rho$ is possible at all. Both, Ketov
and O. Lechtenfeld, referred to the relation to 
Lorentz transformations. Such counting is amusing, indeed.
A complex $N\!\times\!N$ matrix has $2N^2$ elements, and the
requirement $\,\det (M)\,\ueb{!}{=}\,1$ reduces to $2N^2\!-\!2 
= 2n\,$. $V$ is element of SU(N) and has $n$ real parameters$\,$: 
$V=e^{i \vcsm \L \vcsm T}\,$. The more interesting factor is
$\rho$ with $\det(\rho)=1$. Since it is positive definit, one 
may write
\be{5.8}
  0 = \ln \lk \det \( \rho \) \rk 
  = \Sp \lk \ln \( \rho \) \rk \quad \folg \quad
  \ln \( \rho \) = \o^a T^a \quad , \quad
  \rho  = e^{\vcsm \o \vcsm T} \quad .
\ee 
Hence, $\rho$ has $n$ real parameters too, also found in the
exponent. $2n=n+n$ \ --- \ end of counting. From this point
of view, $M=V\rho$ is nothing but a special way of booking
down the elements of SL(N,C). 


\sec{ The spaces  \quad \lower .6cm\vbox{
     \hbox{\boldmath$\;\; \cl A \,\quad | \quad \cl C 
                     \;\quad |$ } \vskip -.4cm
     \hbox{--------------------------} \vskip -.4cm
     \hbox{ \boldmath$\cl M \quad | \quad \cl H 
                     \quad |\quad \cl G_* $ }} }

Here we relax a bit, to develop our state of mind and
our strategy. The horizontal in the head line separates upper 
and underworld. All the five spaces, now provided with names, 
are more or less known already$\,$:  
\\[4pt] 
$\cl A \;$: \parbox[t]{15cm}{
   the space of \ a l l  \  gauge fields (nevermind, whether 
   we think in terms of the real $A_j^a(\vc r)$ or its
   elegant combination to the complex traceless 
   matrix field $A$).}
\\[8pt] 
$\cl M \;$: \parbox[t]{15cm}{ 
       the space of all ($\vc r$  dependent)
       matrices out of SL(N,C).}
\\[4pt] 
$\cl H \;$: \parbox[t]{15cm}{
    the physical subspace of the underworld $=$
    the space of all ($\vc r$ dependent) hermitean 
    $N\times N$ matrices $H$ with determinant 1.}
\\[8pt] 
$\cl G_* \;$: \parbox[t]{15cm}{
    the gauge group $=$ the space of all ($\vc r
    $ dependent) unitary matrices $U$ $=$ the elements of SU(N).}
\\[4pt] 
$\cl C \; $: \parbox[t]{15cm}{the space of only such $A$  
      fields, which are not connected by gauge transformations
      $=$ the space of gauge--invariant field configurations
      $=$ {\bf the}\, interesting physical subspace, in which
      quantization is allowed.}
\\[4mm]
No grey hairs might grow by considering the following relation,
\be{6.1}
  \hbox{ space } \; \cl C \; = \; { \hbox{ space } \; \cl A \0
  \hbox{ gauge group } \;  \cl G_* \; } \quad , 
\ee 
because \eq{6.1} just \ d e f i n e s \ the meaning of a quotient 
in the group chinese language. KKN  give some amount of references 
for the geometry of the space $\cl C\,$ (ref. [4] there).
One could ask whether perhaps the spaces $\cl H$ and $\cl C$ are
identical. Well, the $H$'s live in the underworld. If intergration
over $\cl A$ differs from integration over $\cl M$, because
there is a Jacobian in between, then we expect something to be
between $\cl H$ and $\cl C\,$, too.

By the term ``Schr\"odinger wave function'' it is commonly made clear
that ordinary $\psi$--function quantum mechanics is going on \ --- \ 
instead of working with creators and annihilators. The latter are
good for perturbation theory, but here we like to do it better.
There is an other famous example for the uselessness of creators,
namely the exact solutions by Bethe ansatz of a few special
1D many--particle systems as e.g. the Hubbard model. For a 1D 
oscillator we need an x--axis, the wave function attains complex
values on. In field theory, each of the $\infty$ many points 
$\vc r$ of the discretized space is the origin of a few (here $2*n$) 
``field axes'' (specific to $\vc r$). A point on the $j$--$a$--th 
axis gives the real value of $A_j^a\,$, and on these $A$--axes the 
wave function attains complex values: $\psi \!\lk A_j^a(\vc r) 
\hbox{ or~? }\rk$. Now, ``Schr\"odinger wave--functional'' is the 
appropriate term, indeed. 

For a moment, let us forget about the gauge freedom. $j$, $a$, 
$\vc r$ number variables. Correspondingly, a scalar product 
$\lw 1 | 2 \rw$ between two states $\psi$ contains $2*n*\infty$
integrals$\,$:
\be{6.2}  \!
  \int \psi_1^* \psi_2  =  \!\int \! dA_1^1(1) 
  \, \ldots \, dA_2^n(\infty)\bigg|_? \; 
  \psi_1^* \!\lk H (\vc r) \rk 
  \psi_2 \!\lk H (\vc r) \rk \, \glr  
  \!\int\! d\mu (\cl A)\bigg|_? \,\psi_1^* \psi_2 \;\; . \;
\ee 
To the right in \eq{6.2}, the product of differentials has been
given a name$\,$: volume element $d\mu (\cl A)$ in the space
$\cl A$. 

The question marks in \eq{6.2} refer to the problem. \eq{6.2}
makes sense only if before the integration has been restricted to
physical variables. $\psi$ depends on only these, and we know 
them~: $H(\vc r )\,$. At this point, our rough strategy (``ask 
the underworld'') can be a bit detailed. Relate
the measure $d\mu(\cl A)$ with that (called $d\mu( \cl M)$) of the 
underworld, i.e. obtain the Jacobian. Split off the gauge volume
(in the underworld, of course, with best regards from Faddeev
and Popov), and than turn back in upward direction$\,$: 
\bea{6.3}  
   & & \nonu \\[-6mm]
   & & \lower .4cm\vbox{
  \hbox{$d\mu \(\cl A \) \hspace*{2cm}  d\mu \(\cl C \)$}
  \vskip -.1cm
  \hbox{$\;\;\; \downarrow \hspace*{3cm} \uparrow$}
  \vskip -.1cm
  \hbox{$ d\mu \(\cl M \) \hspace*{.5cm}
         \longrightarrow \hspace*{.5cm}
          d\mu \(\cl H \) \;\; \cdot \;\; \lb \; 
          d\mu (\cl G_*) \; \rb$} }
\eea 
In the next four sections we shall drown in details of measure 
and volume elements. But then, the $\psi$'s will call for a 
Hamiltonian. According to \eq{2.9} and \eq{2.10}, and going
to quantum mechanics by $\P \to -i\d_A$, the kinetic energy will 
turn into a $2*n*\infty$ dimensional Nabla operator. This,
in turn, when applied to $\psi (H)$, will become the Laplacian
on $\cl C$. 

A new volume element is expected to be the product of 
differentials of the new variables times a Jacobian. The latter
is the absolute value of the ``Jacobi matrix'' $\Im \gll 
\6({\rm old\; variables})/\6({\rm new\; variables})\,$. For 
warming up,  divide $ds^2$ in the second line by $dt^2$ and 
remember the kinetic energy, i.e. $v^2$, of a particle. Obviously,
spherical coordinates well illustrate that and how $\Im$ is 
obtained from the metrics $ds^2$ (let them smile, who 
are experienced with general relativity: $g\omn$ from 
$ds^2$ and $\wu {\rm det (g)}$ in the action).
Our starting point is the space $\cl A$. Its metrics in the
third line is Euclidian and rather trivial (read $\int\! 
d^2r$ as $\sum_{\vcsm r}$). The other lines are outlook.
{\ft
\bea{6.4} 
   & &  \nonu  \\[-.2cm]  
   & &  \hspace{-8mm}
   \begin{tabular}{ r | c | c | l | l | }\hline
  & space & elements & \qquad metrics & 
           \ \ \ volume element \\ \hline\hline
{\ft 1} & $ R^3 $  &  $\vc r $  
        &  $ ds^2 = dx^2 + dy^2 + dz^2 $ 
        &  $ d^3r = dx\, dy\, dz $  \\ \hline 
{\ft 2} & $ R^3 $  &  $r$, $\ta$, $\ph$  
        & $ ds^2 = dr^2 + r^2 d\ta^2 + 
            r^2 \sin^2 (\ta ) d\ph^2 \!$ 
        & $ d^3r = dr r^2 \, d\ta \sin (\ta )\, d \ph $ \\  
        & &  $ \matrix{ \Im\,\gll\, \cr
               {\6 (x,y,z) \0 \6 (r,\ta ,\ph )} \cr } $  
        &  $ ds^2 = \hbox{\ft $\( \matrix{ dr \cr d\ta 
             \cr d\ph \cr} \)$} \, \Im^{T} \; \Im 
   \hbox{\ft $\( \matrix{ dr \cr d\ta \cr d\ph \cr} \)$} $ 
   & $ d^3r = dr \, d\ta \, d \ph \, |\det (\Im )| $ \\ \hline 
{\ft 3} &  $\cl A$   &  $A$  
        &  $ ds^2 = \int\! d^2r \;\d A_j^a \d A_j^a $ 
        & $ d\mu \( \cl A \) = d\mu \( \cl M \) \, 
           \det ( D^\dagger D ) $ \\ \hline
{\ft 4} & $\cl M $  &  $M$  
        & $ ds^2 = 8 \int\!d^2r \, \hspace{2.9cm} $ 
        & $ d\mu \( \cl M \) = \hspace{.8cm} $ \\ 
   & {\scriptsize SL(N,C) }    & 
        & $ \quad  \Sp \lk ( \d M M^{-1} )
            ( M^{\dagger\, -1} \d M^\dagger ) \rk $ 
        & $ \qquad  d\mu(\cl H ) \cdot d\mu(\cl G_*)$ \\ \hline
{\ft 5} & $\!\cl H = {{\rm SL(N,C)} \0 {\rm SU(N)} }\!$ 
           \rule[-.3cm]{0pt}{.9cm} &  $H$  
        &  $ds^2 = 2 \int\! d^2r \; 
            \Sp (H^{-1} \d H H^{-1} \d H) $
        & $ d\mu \( \cl H \)$  \\ \hline 
{\ft 6} &  $\cl G_* =${\ft SU(N) }  &  $U$  
        & & $ d\mu \( \cl G_* \)$  \\ \hline
{\ft 7} & $\cl C$   &  $A_{\rm phys}$  
        & & $ d\mu \( \cl C \) = 
        d\mu (\cl H )\, \det ( D^\dagger D) $ \\ \hline
\end{tabular}   \hspace{1cm}
\eea 
}

\noindent
Finally, let the variety of $A$'s be put in a scheme$\,$:
\bea{6.5}
     A_j = - iT^a A_j^a \quad \longleftarrow 
   & \hbox{\ft $2n$ real} \; A_j^a & 
     \longrightarrow \quad A^a = {1\02} \( A^a_1 
         + i A^a_2 \) \nonu \\[-.2cm]
     \downarrow \hspace*{3.4cm} 
   & & \hspace*{1.3cm} \downarrow \nonu \\[-.15cm]
     {1\02}\(A_1+iA_2 \) \;\; = \; 
   &\hbox{\ft one traceless\,}\, A& = \; - i T^a A^a
   \quad . \quad
\eea 


\sec{ \boldmath$ d\mu(\cl A) \to d\mu(\cl M) $ : 
      Jacobian determinant }

At a point $\vc r = (x,y)$ in space, and at a position on its
$j$--$a$--th field axis, let $\d A_j^a (\vc r)$ a small change
of this position. Of course, this change affects each of the 
linear relations \eq{6.5} as well$\,$:
\bea{7.1}  
  \d A_1^a \d A_1^a + \d A_2^a \d A_2^a  
  & = & (\d A_1^a + i \d A_2^a)\, (\d A_1^a - i \d A_2^a) 
  \; = \; 4\; \d A^a\; \d A^{a*} \nonu \\
  & = & 8\,\Sp \( T^a \d A^a \, T^b \d A^{b*} \) 
  \; = \; 8\,\Sp \( \d A \,\d A^\dagger \) \quad .\quad
\eea 
\pagebreak[2]

\subsection{ \boldmath$ d s_{\cl A}^2$ and $\d M$} 
 
\nopagebreak[3]
Certainly, using the relation $A=-(\6 M) M^{-1}$, the 
expression \eq{7.1} can be written in terms of $\d M$.
Just
$$ M\, M^{-1} = 1 \; \folg \; 
   \d M^{-1} = - M^{-1} \d M M^{-1} \;\; \hbox{and} \;\;
   \6 M^{-1} = - M^{-1} (\6 M) M^{-1}\, $$
has to be used repeatedly$\,$:
\bea{7.2}
 \d A &=& - (\6\d M ) M^{-1} - (\6 M) \d M^{-1} \nonu \\
      &=& \; - \6 \lk \d M M^{-1} \rk + \d M \6 M^{-1}
           + (\6 M) M^{-1} \d M M^{-1} \nonu \\
      &=& \; - \6 \lk \d M M^{-1} \rk - \d M M^{-1} (\6 M ) 
          M^{-1} + (\6 M) M^{-1} \d M M^{-1}  \nonu \\
      &=& \; - \lb \; \6 \lk \d M M^{-1} \rk \; + \; 
          \lk A \, , \,\d M M^{-1} \rk \; \rb  \nonu \\
      &=& \; - \; \dod \;\,\d M M^{-1} \qquad 
     \hbox{with} \quad \dod \,\gll 
                \6 + \lk A \, , \, \;\; \rk
\eea 
Probably, we could write $d$ in place of $\d$, as well, but then
parantheses might limit how far $d$ acts. Let $\d$ only refer
to the quantity immediately following. The covariant derivative
comes in several versions. In fundamental representation we have 
$D_j = \6_j + A_j$ and may combine them to $D \gll {1\02} 
( D_1 + i D_2) = \6 + A\,$. $\dod$ in \eq{7.2} is the commutator
version in adjoint representation. Its index version $D^{ab}$
comes into play when $\dod$ is applied to a matrix field 
$\,\L^a\, T^a\,$:
\bea{7.3}
  \dod\, \L^a T^a &=& T^a \6 \L^a 
      - i A^b \lk T^b\, ,\, T^c \rk \L^c  \nonu \\
      &=& T^a D^{ac} \L^c \qquad \hbox{with} \qquad
      D^{ac} \, \gll \, \d^{ac} \6 + f^{abc} A^b \quad .
\eea 
$A^b$ means ${1\02}\( A_1^b + i A_2^b\)$, of course. For
\eq{7.1} we also need $\d A^\dagger$. Starting from 
$A^\dagger = - M^{\dagger\, -1} \,\ov{\6} M^\dagger$ with 
$\ov{\6} ={1\02} \( \6_1 - i\6_2 \)$, every step of \eq{7.2}
appears daggered$\,$:
\be{7.4}
  \d A^\dagger = \;\ldots\; = - \; \ov{\dod} \; 
  M^{\dagger\, -1} \d M^\dagger \quad \hbox{with} \quad 
  \ov{\dod} \,\gll \ov{\6} - 
       \lk A^\dagger \, , \, \;\; \rk \quad .
\ee 
Inserting into \eq{7.1} and summing by $\int\! d^2r$ over
space points the intermediate result is
\be{7.5}
   ds^2_{\cl A} = \int\! d^2r \; \d A_j^a \,\d A_j^a  
   \; = \; 8 \int\! d^2r \;\Sp \( \lk \dod \; 
  \d M M^{-1} \rk \lk \ov{\dod} \; 
  M^{\dagger\, -1} \d M^\dagger \rk \) \quad ,
\ee 
This was just the first of three steps. In a second step
(\S~7.2) we study $\d M$ and find \ --- \ independently of
\eq{7.5} \ --- \ the volume elment $d\mu (\cl M )\,$. It needs
a third step (\S~7.3) to establish the relation between the measures
$d\mu (\cl A )$ and $d \mu (\cl M )\,$. 
\pagebreak[2]

\subsection{ \boldmath$d \mu (\cl M )$ } 

\nopagebreak[3]
Elements of a group (here $M$'s) are related by multiplication. To 
study the metrics, N. Dragon said, one starts from the 1--element.  
An infinitesimal deviation from 1 can be parametrized by 
$ 1 + {1\02} \vc \e \vc T\,$ . The components $\e^a$ of $\vc \e$ 
are complex. Linear in $\vc \e$ $\det (M)=1$ is guaranteed.
Now we settle down in the middle of the group and like to formulate
a small difference between $M$ and its neighbour $M + 
\d M$ \ --- \ by multiplication$\,$:
\bea{7.6}
  M + \d M  &=& \big( \, 1 + \2 \vc \e \vc T \, \big) \; 
  M \quad \folg \quad \d M = \2 \vc \e \vc T M 
      \nonu \\
  \d M \, M^{-1} &=& \2 \vc \e \vc T \quad , \quad 
  \e^a \; = \;  4 \;\Sp \( T^a \,\d M \, M^{-1} \)\quad .
\eea 
The prefactor \2, seemingly unnecessary, will keep the results simple.
$d \mu (\cl M )$ follows from the metrics, and the latter needs
quadratic infinitesimal quantities. Already the first tempting idea
works well$\,$:
\be{7.7}
  8\; \Sp \( \d M M^{-1} \, M^{\dagger\, -1} \d M^\dagger \)
  \; = \; 2\; \Sp \( \e^a T^a \; \e^{b\, *} T^b \)
  \; = \;  \e^a \e^{a\, *} \; = \; \e_1^a \e_1^a + \e_2^a \e_2^a 
  \quad , \;\;
\ee 
where of course $a$ is summed over, and $\e_1^a \gll \Re e 
\(\e^a\)\,$, $\e_2^a \gll \Im m \(\e^a\)\,$. Hence, the metrics
in the space of $\vc r$--dependent fields $M$ is
\be{7.8} 
     ds^2_{\cl M} = 8 \int\! d^2r \;
     \Sp \( \d M M^{-1} \, M^{\dagger\, -1} \d M^\dagger \)
      \; = \; \int\! d^2r \;  \e^a \e^{a\, *} \quad .
\ee 
The left half of this line is \ek{2.12} and the fourth line in 
the table \eq{6.4}, the right half is something own. By construction,
the metrics is kartesian. So, in our $\e$--language,
\be{7.9}
  d\mu (\cl M ) \; = \;\prod_{\vcsm r}\;
   \prod_{a,\, j} \,\; \e_j^a  \; = \;
   \prod_{\vcsm r}\; \prod_a \,\; \e_1^a \e_2^a \quad
\ee 
is the volume element (the {\sl Haar measure}) in the space 
$\cl M\,$. \eq{7.9} is the starting point for splitting off
the volume in \S~8. 
\pagebreak[2]

\subsection{ \boldmath$d \mu (\cl A )\;= \; d \mu 
     (\cl M )\;$ times Jacobian } 

\nopagebreak[3]
Only now a big Jacobian may be announced. We start from
the metrics \eq{7.5}, express $\d M$ through \eq{7.6} by $\e$'s
there and make use of \eq{7.3}, \eq{7.4}$\,$: 
\bea{7.10} 
   ds^2_{\cl A}\! &=&  2 \!\int\! d^2r \;\Sp \( 
       \lk \dod \,\e^a T^a \rk 
       \lk \ov{\dod} \,\e^{b\, *} T^b \rk \) 
   \, = \,  2 \!\int\! d^2r \; \Sp \Big( 
      \lk T^a \, D^{ac} \e^c \rk \lk T^d \, D^{db \,*} 
      \e^{b\, *} \rk \Big) \;\; \nonu \\
   &=&  \!\int\! d^2r \, 
     \lk D^{ac} \e^c \rk \lk D^{ab\, *} \e^{b\, *} \rk
     \qquad \hbox{where} \quad D^{ab\, *} \gll \,\lb \, 
      \d^{ab} \, \ov{\6} + A^{c\,*} f^{acb} 
      \,\rb    \nonu \\
   &=& \!\int\! d^2r \,\lk D \vc \e \rk \cdot 
       \lk D \vc \e \rki^*  \; = \; 
       \int\! d^2r \; \( \lk \Re e ( D \vc \e ) \rki^2
	       \, + \, \lk \Im m ( D \vc \e ) \rki^2 \)  
       \nonu \\ 
  & &  \qquad\quad  D \glr D_1 + i D_2 \;\; , \;\;
       \vc \e = \vc \e_1 + i \vc \e_2 \;\; : \nonu \\
  &=&  \!\int\! d^2r \, \( 
       \lk D_1 \vc \e_1 \rki^2 + \lk D_2 \vc \e_2 \rki^2 
        - 2 \lk D_1 \vc \e_1 \rk \lk D_2 \vc \e_2 \rk  
        \right. \nonu \\[-9pt]
   & & \hspace*{2cm} \left.  + \;
       \lk D_1 \vc \e_2 \rki^2 + \lk D_2 \vc \e_1 \rki^2 
       + 2 \lk D_1 \vc \e_2 \rk \lk D_2 \vc \e_1 \rk 
        \) \nonu \\
   &=&  \!\int\! d^2r \,  \lk 
   \( \matrix{ D_1 & -D_2 \cr D_2 & D_1 \cr } \)
   \( \matrix{\vc \e_1 \cr \vc \e_2 \cr} \) \rk \cdot \lk
   \( \matrix{ D_1 & -D_2 \cr D_2 & D_1 \cr } \)
   \( \matrix{\vc \e_1 \cr \vc \e_2 \cr} \) \rk \,\glr\, 
    \ueb{\rightharpoonup}{\hbox{\small\bf [\ ]}} \cdot 
    \ueb{\rightharpoonup}{\hbox{\small\bf [\ ]}} \;\; .
\eea 
In the third line ``$D$'' is shorthand for $D^{ab}$ of course 
(sorry, too many $D$'s, simply omit the former meaning 
$\6+A$ now). The operator $D$ is not only a $n\times n$ matrix, 
but also contains differentiation. In \eq{7.10}, their action is 
limited by square brackets. 

Turning to the expression $\ueb{\rightharpoonup}{\hbox{\small\bf [\ 
]}} \cdot \ueb{\rightharpoonup}{\hbox{\small\bf [\ ]}}$ 
to the right in the last line of \eq{7.10}, the integral $\int\! d^2r$ 
was omitted by extending the sum convention. At the same moment 
$\ueb{\rightharpoonup} {\hbox{\small\bf [\ ]}}$ must be viewed as 
a big $\vc r$--indexed vector. Correspondingly, the operators $D$
become big matrices {\bf D}, carrying the index pair $\vc r$, 
$\vc r'$. {\bf D} acts on $\e$ by $\int' \!\! D_{r r'} \e_{r'}\,$. 
We encountered this language already in \eq{4.5} and therefore 
maintain the boldface--Notation. These changes of mind are 
required for correctly reading off the Jacobi--matrix $\Im$ from 
\eq{7.10}. The $\cl A$--space volume element is now obtained as 
\be{7.11}
  d \mu (\cl A ) \; = \; d \mu (\cl M ) \; \left| 
      \det (\Im ) \right| \; = \;\; d \mu (\cl M ) \;
  \left| \det \( \matrix{ {\bf D}_1 & -{\bf D}_2 
   \cr {\bf D}_2 & {\bf D}_1 \cr } \) \right| \quad .
\ee 
The matrix $\Im$ is real. But $\det (\Im )$ has not yet 
the desired form $\det ( {\bf D}^\dagger 
{\bf D} )\,$. The latter is achieved by the following nice
derivation (diagonalize \I, J.~Schulze said)$\,$: 
\bea{7.12}
  \Im &=& \( \matrix{ {\bf D}_1 & -{\bf D}_2 \cr 
   {\bf D}_2 & {\bf D}_1 \cr } \) \; = \; 
   {\bf D}_1 \, 1 + {\bf D}_2 \, \I \quad , \quad
   \I = \( \matrix{ 0 & -1 \cr 1 & 0 \cr } \) \quad ,
     \nonu \\
  W &=& {1\0 \wu 2 } \( \matrix{ 1 & i \cr i &
       1 \cr } \) \;\; ,  \;\;
   W \I W^\dagger \; = \; \( \matrix{ i & 0 
      \cr 0 & -i \cr } \) \;\; , \;\; \det(W) 
      = \det (W^\dagger) = 1 \quad , \nonu \\
  \det ( \Im ) &=& \det \( W \Im W^\dagger \) \; = \; 
  \det \( \matrix{ {\bf D}_1 + i {\bf D}_2 & 0 \cr 
          0 & {\bf D}_1 -i {\bf D}_2 \cr } \) \; = \;
  \det \( {\bf D}\, {\bf D}^* \) \quad .
\eea 
The $\cl A$--space volume--element now reads $\, d\mu (\cl A ) 
= d \mu(\cl M ) \left| \det ({\bf D} {\bf D}^* ) \right|$. 
Fine~? ~Is there a need for those absolute value bars~?

To the hell with all the errors ever produced in physics 
papers by missing absolute value bars around Jacobians.
Remembering \eq{7.3} let us book down the matrix ${\bf D}$ 
with all indices,
\be{7.13}  \;\;
  {\bf D}\, : \quad  D^{ab}_{r r'} = \d^{ab} \6_{r r'} + 
  A^c (\vc r ) f^{acb} \d_{r r'}  \quad , \hspace*{5cm}
\ee 
and notice that 
\bea{7.14}
  {\bf D}^T\, : \quad  (D^T)^{ab}_{r r'}  = \d^{ab} 
     \( - \6_{r r'}\)  +  A^c (\vc r ) f^{bca} \d_{r r'}  
  & & \nonu \\
  \folg\qquad {\bf D}^T = - {\bf D} \quad , \quad 
  \hbox{hence} \quad   {\bf D}^* = - {\bf D}^\dagger 
         \quad , & &  \hspace*{4.6cm} 
\eea 
because $\,A^c = \2 \( A_1^c + i A_2^c \)\,$, 
$\,A_j^c$ real, and $\6_{r r'} = -\6_{r' r}$, see the 
line below \eq{4.5}. Vector arrrows over $r$--indices are suppressed 
for simplicity. But, please, add them by mind. If the hermitean 
matrix $\,{\bf D}^\dagger {\bf D}\,$ has no zero eigenvalue (we 
endure the risk), then we may state that
\be{7.15} 
  - {\bf D} {\bf D}^* = {\bf D} {\bf D}^\dagger  \quad
  \lower 4pt\vbox{\hbox{{\ft is positive}} \vskip -.26cm
                  \hbox{{\ft \ \ definit}}} \;\quad
  \folg \;\quad  \big| \;\det \( {\bf D} {\bf D}^* 
  \)\,\big| \; = \; \det \( {\bf D}^\dagger  {\bf D} \) 
  \;\; \glr \; e^\G \quad ,
\ee 
and
\be{7.16}
  d \mu \( \cl A \) \; = \; d \mu \( \cl M \) 
  \; \det \( {\bf D}^\dagger  {\bf D} \) \quad
\ee 
is the result of this section 7. It completes line 3 in the table 
\eq{6.4}. The quantity $\G$ defined in \eq{7.15} is real. Aside, 
through \eq{7.10} to \eq{7.15} we proceeded by two lines 
text in KKN below \ek{2.12}.


\sec{ \boldmath$ d\mu(\cl M) \to d\mu(\cl H) $ : splitting off 
      the ``volume'' }

To the present job of making path for pedestrans, it was
very helpful to have Jens Reinbach being engaged too, in 
particular in this section and the next. But do not ask 
``who--what'', because all has been actually done by KKN. 

The ``separation ansatz'' for decoupling unphysical degrees
of freedom is known from \S~5 to be 
\be{8.1}
    M \; = \; U \, \rho \qquad \folg \qquad
    \d M = \d U \,\rho \, + \, U \,\d\rho \quad .
\ee 
Now preferring $U$ rather than $V$ signals a slight change in
philosophy$\,$: all possible gauge transformations $U$ are 
started from a physical ``point'' $\rho$. As the content
of this section can be well stated in words, we do something
strange, start with the summary (next paragraph up to the 
figure) and go into the details only afterwards.

On one hand, we know the $\cl M$--space measure $\,d \mu (\cl M )
=\prod_{\vcsm r} \prod_a \e_1^a \e_2^a\,$ from \eq{7.9}, and
the relation of $\e$'s with $\d M$ from \eq{7.6}, on the other. 
Through the split \eq{8.1} it will turn out that 
$\e_1^a$ is a pure $\rho$--ic expression (depending on only
$\rho$ and $\d \rho$). But $\e_2^a$ will become a sum of a 
$U$--ic and a $\rho$--ic piece. If so, the latter must be a 
linear combination of the $\e_1^a$'s. At this point, new variables 
$h$ and $u$ are in order, 
\bea{8.2}
  & & \e_1^a \; = \; d h^a \quad , \quad \e_2^a \; = \; 
  Q^{ab} d h^b \; + \; d u^a \quad : \nonu \\[6pt]
  \Big[ \prod_a \e_1^a \e_2^a \Big]
  \!\! &=&  \Big[ \prod_a dh^a du^a \Big] \;
  \left| \, \det \( \matrix{ 1 & 0 \cr Q^{ab} & 1 \cr } \) 
  \,\right| =  \Big[ \prod_a d h^a \Big] 
  \Big[ \prod_a du^a \Big] \quad , \quad
\eea 
to reach the desired decomposition. \\
\parbox[t]{8.8cm}{Integration over $du$ will give the
gauge orbit $\cl G_*$, and the remaining measure 
$d\mu (\cl H )$ might be related to the $d h$'s. Other than in 
\ek{2.13}, \ek{2.14} there appear no {\sl wedge} products or 
$\sim$ signs in the present derivation. But, admittedly, it 
is high time to see a figure. \vskip .2cm } 
\hspace*{.3cm} \parbox[t]{5.3cm}{\unitlength .96cm 
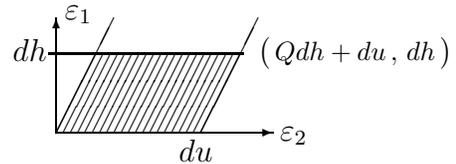
\begin{picture}(5,0)
   \put(1,-1.8){\vector(1,0){3}} 
   \put(1,-1.8){\vector(0,1){1.6}}
   \put(.9,-.7){\line(1,0){2.7}}  
   \put(1,-1.8){\line(1,2){.8}} 
   \put(3,-1.8){\line(1,2){.8}}
   \put(1.12,-.22){$\e_1$} \put(4.1,-1.9){$\e_2$}
   \put(.4,-.8){$dh$}     \put(2.7,-2.2){$du$}
   \put(0,-5.3){$ $}  
   \put(3.8,-.79){\ft $\big(\, Q dh + du\, ,\, dh\, \big)$}
   \multiput(1,-1.8)(.1,0){21}{\line(1,2){.54}}
\end{picture}} 

For detailing the above, we have from \eq{7.9}, \eq{7.6} 
and \eq{8.1} that
\bea{8.3}
  \e^a &=& 4 \, \Sp \( T^a \lk \d U \rho + U \d\rho \rk 
  \rho^{-1} U^\dagger \) 
  \; = \; 4\,\Sp \( \ov{T}^a U^\dagger \d U \)
  \; + \; 4\,\Sp \( \ov{T}^a \d \rho \,\rho^{-1} \) 
     \qquad \nonu \\
  \e^{a\, *} &=& {} - 4\,\Sp \( \ov{T}^a U^\dagger \d U \)
  \; + \; 4\,\Sp \( \ov{T}^a \rho^{-1} \d\rho \) \qquad ,
  \qquad \ov{T}^a \gll U^\dagger T^a U \quad ,
\eea 
where $\lk \Sp(A) \rki^* =\Sp (A^\dagger)$ and $\d U^\dagger U 
= - U^\dagger \d U$ have been used. One can see already that,
in the real part, $U$-ic terms will compensate$\,$:
\bea{8.4}
  \e_1^a = {\e^a + \e^{a\,*} \0 2} 
  &=& 2 \,\Sp \( \ov{T}^a \rho^{-1} \lk \rho \,\d\rho 
   + \d\rho \,\rho \rk  \rho^{-1} \) \nonu \\
  &=& 2\, \Sp \( \ov{T}^a H^{-1/2} \d H H^{-1/2} \) 
     \;\; \glr \;\; dh^a \quad .
\eea 
But there remain two terms in the imaginary part$\,$:
\bea{8.5}
  \e_2^a = {\e^a - \e^{a\,*} \0 2 \, i } &=& - 2\, i 
  \,\Sp \( \ov{T}^a \lk \d\rho \,\rho^{-1} 
   - \rho^{-1} \,\d\rho \rk \) \; + \; du^a \nonu \\
  \hbox{with} \qquad  du^a &\gll& - \; 4\, i \, 
  \Sp \(\ov{T}^a U^\dagger \,\d U \)  \; = \; - \; 4\, 
  i \, \Sp \( T^a \d U\, U^\dagger \) \quad . \quad
\eea 
Below \eq{5.8} it was understood that $\rho$ is determined by
$n$ real parameters. Hence, every single--infinitesimal
pure $\rho$--ic quantity \ --- \  such as the corresponding term 
in \eq{8.5} \ --- \ can be linearly combined from $dh^a$'s$\,$:
\be{8.6}  - 2\, i \,
  \Sp \( \ov{T}^a \lk \d\rho \,\rho^{-1} - \rho^{-1} 
  \,\d\rho \rk \) \;\glr\; Q^ {ab} dh^b \quad .
\ee 
We have thus obtained that $\e_2^a = Q^{ab} dh^b + du^a$,
hence \eq{8.2} is valid. One could object that there is
still some dependence on $U$ in \eq{8.8}, hidden in the 
gererators $\ov{T}^a \gll U^\dagger T^a U$. Yes. But the whole
matrix $Q$ drops out in \eq{8.2}$\,$! 

It remains to relate the physical part $\prod_a dh^a$ of the
volume with the measure $d\mu (\cl H )$ in the space of
hermitean unit--determinant matrices $H$. In order that the 
neighbour matrix $H+\d H$ is hermitean too, we must write
\be{8.7}
  H + \d H = H^{1/2} \( 1 + \eta^a \ov{T}^a \) H^{1/2} 
  \quad , \quad \eta^a \;\; {\rm reell} \quad .
\ee 
Linear in $\eta^a$ we even have $\det(H+\d H)=1$ due to
$\Sp(\ov{T}^a)=0$. Solving \eq{8.7} for $\eta$ and comparing 
with \eq{8.4}, 
\be{8.8}
  \eta^a = 2\, \Sp \( \ov{T}^a H^{-1/2} \d H 
   H^{-1/2} \) \; = \; dh^a \quad , 
\ee 
exhibits the simplest possibility to be true. The measure
in the space $\cl H$ is thus given by 
\be{8.9}
    d \mu (\cl H ) = \prod_{\vcsm r} \prod_a d h^a \quad ,
    \hbox{ and} \quad d \mu ( \cl M ) =  \prod_{\vcsm r} 
    \prod_a d u^a  \,\cdot\,  d \mu ( \cl H ) \quad
\ee 
states the connection to $d\mu(\cl M)$. In passing, the
metrics in the $\cl H$ space can be written as 
\be{8.10}
  ds^2_{\cl H} = \int\! d^2 r \; \eta^a \eta^a 
  = \int\! d^2 r \; 2 \,\Sp \(\eta^a \ov{T}^a 
    \ov{T}^b \eta^b \) \, = \, 2 \int\! d^2 r \; 
    \Sp \( H^{-1} \d H H^{-1} \d H \) \quad .
\ee 
\eq{8.10} agrees with \ek{2.17}. But KKN state something
else in place of \eq{8.8}, namely  $2\Sp(T^a H^{-1}\d H)\,$,
and call it the {\sl Haar measure}. Wether there is a real 
difference or not~(!) we come back to in \S~14.3.  

With view to \eq{8.8} one may again object that there is
reminescent $U$ dependence hidden in $\ov{T}^a$. This time it 
is removed by observing that  $\eta^a_{\rm non}\,\gll\, 2\Sp ( 
T^a H^{-1/2} \d H H^{-1/2} )$ and $\eta^a$ are related by an 
ordinary real rotation matrix. To realize this, we define $\cl D$ 
by $\eta^a \glr {\cl D}^{ab}\eta^b_{\rm non}$ and read off 
from \eq{8.10} that
\be{8.11} 
  \eta^a_{\rm non}  \eta^a_{\rm non} =  \eta^a \eta^a 
   = \eta^b_{\rm non} ({\cl D}^T)^{ba} {\cl D}^{ac}  
    \eta^c_{\rm non} \quad \folg \quad
     {\cl D}^T {\cl D} =1 \quad . 
\ee 
To make use of this, we start from \eq{8.9}, put integrals in 
front of the relation to the right, rotate in the integrals over 
$d\mu(\cl H)$ until their memory on $U$ hase gone, and finally 
shift the integrals over $du^a$ to the right~: 
\be{8.12}
  \int\!\!\int\; d \mu (\cl M )\,\; = \,\int d\mu (\cl H ) \;\,
  \int \prod_{\vcsm r} \prod_a d u^a \;\; = \; \int 
   d \mu (\cl H ) \;\cdot\, \int d \mu ( \cl G_* ) \quad .
\ee 
There it is, the gauge volume $\int d \mu ( \cl G_* )$,
to be removed before quantizing.

Let us summarize the essence of the last three sections and 
combine the equations \eq{6.1}, \eq{7.16} and \eq{8.12}$\,$:
\be{8.13}
  \fbox{\rule[-7mm]{0pt}{18mm}\qquad $\dis
  d \mu (\cl C )  \;\; = \;\; { d\mu (\cl A ) \0 
  d \mu ( \cl G_* ) } \;\; = \;\; { d \mu (\cl M ) \, 
  \det \( {\bf D}^\dagger  {\bf D} \) \0 
  d \mu ( \cl G_* )} \;\; = \;\; d\mu ( \cl H ) \;
  \det \( {\bf D}^\dagger  {\bf D} \)  \;\;\quad . $ 
  \quad}
\ee 
\eq{8.13} is \ek{2.19}~: {\sl the problem is thus
reduced to the calculation of the determinant of the
two--dimensional operator $\;{\bf D}^\dagger {\bf D}\,$.}

Perhaps, while recapitulating the present section, one could be 
dissatisfied with the argument leading to \eq{8.6} which merely 
states the existence of the matrix $Q$. But $Q$ can be made 
a bit more explicit also, e.g. by using the $\rho$ representation 
\eq{5.8}~:
\bea{8.14}
 \rho \; = \; e^{\vcsm \o \vcsm T} \; &,& \;
  \d\rho = d\o^a \,\6_{\o^a} \, e^{\vcsm \o \vcsm T} 
    = d\o^a \int_0^1\! ds \; e^{s\,\vcsm \o \vcsm T} 
     \big[ \6_{\o^a} \;\vc \o \vc T \big]
      e^{(1-s)\,\vcsm \o \vcsm T} 
      \quad \nonu \\
  \d\rho\, \rho^{-1} 
  &=& d\o^a \int_0^1 \! ds\; \tau^a(s) \qquad , \qquad  
    \tau^a(s) \;\gll\;  e^{s\,\vcsm \o \vcsm T} 
    \, T^a \, e^{-s\,\vcsm \o \vcsm T} \nonu \\
  \rho^{-1} \d\rho &=& \big[ \d\rho\, \rho^{-1} 
    \big]^\dagger \; = \; d\o^a \int_0^1 \! ds \;
    \tau^{a\, \dagger} ( s ) \quad . \quad
\eea 
With these details at hand ($\ov{T}^a$ is now called $T^a$),
\eq{8.4} and \eq{8.6} turn into
\be{8.15}
  \matrix{ \,\; dh^a = 2 \, S^{ab}\, d\o^b \cr
   Q^{ab} dh^b = - 2 \, R^{ab}\, d\o^b\cr} \qquad 
  \hbox{with} \qquad 
 \matrix{S^{ab} \cr R^{ab} \cr } \Bigg\} \;\gll\; 
 \int_0^1 \! ds\; \Sp \( T^a  \lb
   \matrix{ [\tau^b + \tau^{b\,\dagger} ] \cr
   i \, [ \tau^b - \tau^{b\,\dagger} ] \cr} \rb\) \quad .
 \ee 
Eliminating $d\o^b$ we have~:
\be{8.16}
  2 \, d \vc \o = S^{-1}\, d\vc h \quad : \qquad 
  Q \; = \; - \, R \; S^{-1} \quad . \quad
\ee 
A quantity, which can be booked down, does exist.  


\sec{ Functional differential equations for \boldmath$S[H]$ }

The material of this long section strongly defended itself 
to be understood. Sometimes, our way out is probably not
the best.   

Following the KKN guide \cite{kkn} the somewhat delicate 
relation \ek{2.20} is in order,
\be{9.1}
  e^\G \, = \,\det \( {\bf D}^\dagger {\bf D} \) \,
  = \,\s^n \, e^{2N\, S}  \qquad \hbox{with} \quad \s \; 
  = \; {\det' \( - \ov{\6} \6 \) \0 \int \! d^2r } \quad , 
\ee 
which presumably needs a regularization philosophy before any
closer look at (see \S~9.4 and \S~11). For the stage 
being\footnote{\ So far (April 18, 2000), the value 
    of $\s$ given to the right in \eq{9.1} is not understood. 
    May be that it leads into higher spheres \cite{bn,gaw}. The 
    determinant of a positive definit operator is, of course, 
    the product of its eigenvalues$\,$: $\det(P) =  
    \exp\(\Sp[\ln(P)]\) = \exp\(\sum \ln (\l )\) = \prod \l\,$. 
    Especially, $ - \ov{\6} \6 = - \D/4$ has $\l = \vc k^2/4$, 
    and  $\vc k =0$ might be excluded. Also, $\det'$ 
    needs a large--$k$ regularization. The power $n$ is clear, too,  
    \ -- \ but that area in the denominator of $\s$
    remains a mystery.}, let the factor $\s$
contain all that is left for vanishing fields $A$. Then, all 
non--trivial $A$ dependence is in the exponential, i.e. in $S$, 
and we have the condition
\be{9.2}
   \lim_{A\to 0} S \; = \;\lim_{H \to {\rm const}} 
   S\lk H \rk \; = \; 0 \quad .
\ee 
For the details of this line note that, as $\,\det \( 
{\bf D}^\dagger {\bf D} \)\,$ might be a gauge invariant, $S$ 
should depend on the physical degrees of freedom $H$ only. This 
will be seen more explitly (see the headlines) and, admittedly, had 
been used already in splitting off the gauge volume. Due to
$A=-(\6 M)M^{-1}$ a vanishing $A$ corresponds to a constant matrix 
$M$, and this in turn (due $H = M^\dagger M$) to a constant $H$. 
\eq{9.2} may be viewed to give initial values to the functional 
first order differential equation to be derived. The reader may 
look forward to \eq{10.1} below to see the condition respected.

``Differentiate, regularize and integrate up again'' is the
general rule for obtaining $S\,$. It seems to be familar 
to those experienced in anomaly calculations \cite{powi}. 
Under variation with respect to $A$ fields,
\be{9.3}
  \G = \ln \lk \det \( {\bf D}^\dagger {\bf D} \) \rk
     = n\,\ln (\s ) + 2NS \quad \folg \quad
    \d\,\G = 2N \,\d\,S \quad ,
\ee 
the stressy factor $\s$ in \eq{9.1} becomes irrelevant.
\pagebreak[2]
      
\subsection{ \boldmath$\d \,$ ln ( det ) }
     
\nopagebreak[3]
For calculating $\d \G$, a true pedestrian recalls \eq{7.13} 
and goes ahead\footnote{\ 
   Through $\d \Sp (\ln(X))=\Sp(\d X/X)\,$, 
   the second line of \eq{9.5} can be reached 
   immediately, of course. Bute we like to get 
   used to the giant matrices.}~: 
\bea{9.4}
  \G &=& \ln \lk \det \( {\bf D}^\dagger{\bf D} \) \rk 
     = \fhsp \lk \ln \( {\bf D}^\dagger{\bf D} \) \rk 
       \quad , \quad
       1- {\bf D}^\dagger{\bf D} \,\glr \,\mho  \nonu \\
   &=& \fhsp \( - \sum_{n=1}^\infty \,{1\0n} \;\mho^{\; n}\) 
   \; = \; - \sum_{n=1}^\infty \,{1\0n} \; \( \;\;\;
   \unitlength 1cm \parbox[t]{3cm}{\begin{picture}(2,1.1)
    \put(1,.6){$\mho_{12}$}     \put(2,.3){$\mho_{23}$}
    \put(-.1,.3){$\mho_{n1}$}   \put(2,-.3){$\mho_{34}$}
    \put(-.1,-.3){$\mho_{n-1\, n}$}
    \put(1,-.08){\vbox{\hbox{\tiny $\;\;\; n$} 
       \vskip -.36cm \hbox{\tiny factors}}}
    \put(.7,-.8){$\bullet$} \put(1.2,-.92){$\bullet$}
    \put(1.7,-.8){$\bullet$} \end{picture}}\;\) \quad.
\eea 
Let a boldface {\bf Tr} refer to $r r'$ indexed giant 
matrices, and the hat to their $ab$ indices. E.g. the 
index 1 in \eq{9.4} represents $b, \vc r'$, 2 stands for 
$c, \vc r''$ and so forth. A variation $\d$ grasps 
into each of the $n$ buckets $\mho\;$:
\bea{9.5}
  \d\,\G &=& - \sum_{n=1}^\infty \lk \d\,\mho_{12} \rk 
          \mho_{23} \ldots \mho_{n1} \; = \;
  - \lk \d\,\mho_{12} \rk \( {1\0 1 - \mho} \)_{21}  \nonu \\
  &=& \fhsp \( \lk \d ( {\bf D}^\dagger {\bf D} ) \rk
      {\bf D}^{-1} {\bf D}^{\dagger -1} \)
  \; = \; \fhsp \( {\bf D}^{-1} {\bf D}^{\dagger -1}
          \d ( {\bf D}^\dagger {\bf D} ) \) \quad .
\eea 
If only one {\bf D} changes under the variation $\d$ (${\bf D}$ 
itself, say), then the other (which is ${\bf D}^\dagger\,$)
drops out in \eq{9.5}. Hence, when differentiating with respect 
to $A^a (\vcsm r )$, the result is
\bea{9.6}
  \d_{A^a (\vcsm r )} \G 
  &=& \fhsp \( {\bf D}^{-1} \d\, {\bf D} \) 
    \; = \; (D^{-1} )_{12} (\d\, D)_{21}
    \; = \; \int' \int'' (D^{-1})^{bc}_{r' r''} 
    \d_{A^a (\vcsm r )} D^{cb}_{r'' r'}  \nonu \\
  &=& \int' \int''  (D^{-1} )^{bc}_{r' r''} \;
    \d (\vc r - \vc r'' ) \; f^{abc} \; 
    \d (\vc r'' - \vc r' ) \nonu \\
 \d_{A^a (\vcsm r )} \G
 &=& f^{abc} \, (D^{-1} )^{bc}_{r' \, r} \;\hbox{\Large 
     \boldmath$|$}_{\vcsm r' \to  \vcsm r} \quad , \quad
 \d_{A^{a*} (\vcsm r )} \G = f^{abc} \, 
    (D^{* -1} )^{bc}_{r' \, r} \;\hbox{\Large 
     \boldmath$|$}_{\vcsm r' \to  \vcsm r} \quad . \quad
\eea 
In the second line, there are two delta functions. The first 
arised by functional differentiating the $A$ field in 
$D^{cb}_{r'' r'} = \d^{cb} \6_{r'' r'} + A^\bullet 
(\vc r'') f^{c \bullet b} \d_{r'' r'}\,$. 
It is harmless$\,$: one may just integrate over 
$\vc r''\,$. The second delta is the one which
was already present in $D^{cb}_{r'' r'}\,$. 
If integrating, it would make the two spatial indices on
$D^{-1}$ equal. In the third line, following KKN, we postpone 
this dangerous coincidence. It must wait for the regularizaition
of \S~11. In the last line, since $\G$ is real, the second 
differential equation was added as the c.c. of the first.
\pagebreak[2]

\subsection{ The inverse of {\bf D} }
     
\nopagebreak[3]
The inverse of {\bf D}, required in \eq{9.6} and now denoted 
by (?), is the unique solution of the $n^2$ equations
\be{9.7}
  D^{ac}_{r r''} \;  (\,\hbox{\bf ?}\, )^{cb}_{r'' r'}
  \; = \; \Big(\d^{ac} \6 + A^{ac}(\vc r )\Big) \;
  (\,\hbox{\bf ?}\, )^{cb}_{r r'}  \;  = \; \d^{ab} 
  \,\d (\vc r - \vc r' )  \quad \hbox{with} \;\; 
  A^{ac} \gll A^\bullet f^{a \bullet c} \quad . \quad
\ee 
We like inventing. Due to \eq{4.3}, i.e. $\6 G_{r r'} 
= \d (\vc r - \vc r' )\,$, the quantity (?) might
be anyhow related to $G\,$. If we insert $\d^{cb} G_{r r'}$ 
for (?), a $\d \d$ arises to the right but also an additional 
term $A G$. To compensate the latter, we make the next attempt
with (?)$=M^{cb} (\vc r ) G_{r r'}\,$. This time the r.h.s.
becomes $(\6 M^{ab} ) G + M^{ab} \d_{r r'} + A^{ac} M^{cb} G\,$. 
No. But we see that a further matrix $(M^{-1})^{db}
(\vc r' )$ would restore the $\d\d\;$:
\be{9.8}
  \Big(\d^{ac} \6 + A^{ac}(\vc r )\Big) \; M^{cd}_r 
  G_{r r'} (M^{-1})^{db}_{r'}  = \lk (\6 M)_r^{ad} 
  + A^{ac}_r M^{cd}_r \rk  G_{r r'} 
  (M^{-1})^{db}_{r'} + \d^{ab} \d_{r r'} \quad . 
\ee 
The so far arbitrary matrix $M^{ab}$ can now be fixed
to make the square bracket vanish. If we are able
to construct the ``adjoint'' matrix $\hat{M}$ having
elements $M^{ab}$ with the property announced, i.e. if
the equation $\hat{A}\hat{M} = - \6 \hat{M}$ can be solved, 
then the solution to \eq{9.7} reads~:
\be{9.9}
  (D^{-1})^{ab}_{r r'} =  M^{ac}_r G_{r r'} 
  (M^{-1})^{cb}_{r'} \quad , \hbox{ \ or} \quad 
  {\bf D}^{-1}  = \hat{M} \, G \, \hat{M}^{-1} \quad .
\ee 
The elements of the matrix $\hat{A}$ are those given to the 
right in \eq{9.7}. The equation $\hat{A}\hat{M} = - \6 \hat{M}$ 
to be solved may be called the adjoint version of the 
familar relation $AM= - \6 M\,$. By the way, do not integrate 
over pairs of spatial indices in \eq{9.8} and \eq{9.9} 
(the last pair to be integrated ocurred to the very left
in \eq{9.7}).
\pagebreak[2]

\subsection{ The adjoint matrix \boldmath$\hat{M}$ }
    
\nopagebreak[3]
We start from the fundamental representation, i.e.
from $A= - i T^a A^a = - (\6 M) M^{-1}\,$, and put this 
in the commutator $[ T^b\, , \,  \ldots ]$ to obtain
\be{9.10}
  A^{bc} T^c = - T^b (\6 M ) M^{-1} + (\6 M) M^{-1} T^b
  = - T^b (\6 M ) M^{-1} - M (\6 M^{-1} ) T^b  \quad .
\ee 
Multiplication by $M^{-1}$ from the left and by $M$ from
the right gives
\bea{9.11}
 A^{bc} M^{-1} T^c M  &=& - M^{-1} T^b \6 M 
 - (\6 M^{-1} ) T^b M \; = \; - \6 \; M^{-1} T^b M 
  \quad ,\nonu \\
 \hbox{i.e.} \quad A^{bc} \,\Sp\( M^{-1} T^c M T^d \)  &=&
     - \6 \; \Sp \(M^{-1} T^b M T^d \) \quad , 
\eea 
which allows for reading off the solution to $\hat{A}\hat{M} 
= - \6 \hat{M}$ only up to normalization. For the latter we 
require $M^{ab}\to \d^{ab}$ for $A\to0$ in order to make 
\eq{4.4} or \eq{4.5} valid even adjointly. To summarize~:
\be{9.12}
 \hat{M} \, : \;\;  M^{ab} = 2 \, \Sp 
 \Big( T^a M T^b M^{-1} \Big) \qquad \hbox{solves} 
 \quad \hat{A} \hat{M} = - \6 \hat{M} \quad . \qquad\,
\ee 
Because of $(\hat{M}^\dagger )^{ab} = M^{ba\; *} = 
2\, \Sp ( M^{\dagger -1} T^a M^\dagger T^b )$
one may add that
\be{9.13}   \hspace*{.5cm}
 \hat{M}^\dagger \, : \;\,  (M^\dagger )^{ab} = 2\,\Sp 
 \Big( T^a M^\dagger T^b M^{\dagger -1} \Big) \quad 
 \hbox{solves} \quad \hat{M}^\dagger \hat{A}^\dagger  
  = - \ov{\6} \hat{M}^\dagger \quad . \quad
\ee 
As $\hat A$ has the elements $A^{ab} = A^\bullet f^{a\bullet b}$, 
$\hat{A}^\dagger$ has the elements $(A^\dagger)^{ab} 
=  A^{\bullet *} f^{b\bullet a} = - A^{\bullet *} 
f^{a\bullet b}\,$.
\pagebreak[2]

\subsection{ Jump into the regularized version }
    
\nopagebreak[3]
Disgraceful~! But guided by KKN, and because things are worked 
out in \S~11, let us be allowed to anticipate the result of 
regularization here. Remember from \eq{9.6} and \eq{9.9} that
it is the Greens function $G_{r r'}$ suffering under the 
coincidence limit $\vc r' \to \vc r\,$. $G$ has to be
replaced by its regularized version $\cl G_{r r'}\,$.
Then the coincidence limit can be performed, see \eq{11.28},
to give
\be{9.14}
  \lk D^{-1}_{r' r} \rki_{\vtop{\hbox{\ft \ reg}
  \vskip -.24cm \hbox{\ft $\vcsm r'\!\to\!\vcsm r $}}}
  \, = \, - \, { \hat{A}^\dagger - ( \ov{\6} \hat{M} ) \, 
     \hat{M}^{-1}  \0 \pi } \;\;\; , \;\;\; 
  \lk D^{* -1}_{r' r} \rki_{\vtop{\hbox{\ft \ reg}
  \vskip -.24cm \hbox{\ft $\vcsm r'\!\to\!\vcsm r $}}}
  \, = \; { \hat{A} - \hat{M}^{\dagger -1} \, 
    \6 \, \hat{M}^\dagger \0 \pi } \;\;\; . \quad
\ee 
Herewith the ill equations \eq{9.6} turn into well behaved, 
local (but still adjointly formulated) functional differential 
equations, namely$\,$:
\be{9.15}
  \d^a \G = {- 1\0\pi}\, f^{abc} 
  \( \hat{A}^\dagger - ( \ov{\6} \hat{M} ) \, 
   \, \hat{M}^{-1} \)^{bc} \quad , \quad 
  \d^{a*} \G = {1\0\pi}\, f^{abc} 
   \( \hat{A} - \hat{M}^{\dagger -1} \, 
    \6 \, \hat{M}^\dagger \)^{bc} \quad . \quad
\ee 
Here, of course, $\d^a$ stands for $\d_{A^{a} (\vcsm r )}$
and $\d^{a*}$ for $\d_{A^{a*} (\vcsm r )}\,$. The right 
equation is the c.c. of the left one. Our next task is
the return to the more familar $N\times N$ matrices $M\,$. 
Adjoint traces might become fundamental ones. 
\pagebreak[2]

\subsection{ Trace times trace \ -- \ back to fundamental }
    
\nopagebreak[3]
There is a nice technical detail on traces, which can be
discovered by searching the inverse of $\hat M$. It has to be
determined from $(M^{-1})^{ac} M^{cb} = \d^{ab}\,$. We 
\ g u e s s \ the result,
\be{9.16}
   ( M^{-1} )^{ac} = 2 \, \Sp \( T^a M^{-1} T^c M \)
   \; = \;  M^{ca} \quad ,
\ee 
and start checking. Two fundamental traces are to be combined~: 
\bea{9.17}
  ( M^{-1} )^{ac} M^{cb} &=& 2 \, \Sp \( T^a 
   M^{-1} T^c M \) \; 2 \, \Sp \( T^c M T^b M^{-1} \) \nonu \\
  &=&  2 \, \Sp \( \lk M T^a M^{-1}\rk T^c \) \; 2 \, \Sp \( T^c 
      \lb M T^b M^{-1}\rb \)  \nonu \\
  &=& 2 \lk  M T^a M^{-1} \rki_{\ell\, m} \; 2 \, T^c_{m\, \ell}   
      \, T^c_{p\, q} \; \lb  M T^b M^{-1} \rb_{q\, p} \nonu \\[-.4cm]
  & & \hspace*{3.08cm} \unt{1.8}{.4}{.3}{\d_{m\, q} \; 
      \d_{\ell \, p} \, - \, {1\0N}\; \d_{m\, \ell}\; 
      \d_{p\, q} }   \nonu \\
  &=& 2 \,\Sp\( \lk  M T^a M^{-1} \rk \; \lb  M T^b M^{-1} \rb \)
      \; = \; \d^{ab} \quad \hbox{q.e.d.} \quad
\eea 
The $1/N$ term dropped out since $\Sp ( M T^a M^{-1} ) = \Sp ( 
M T^b M^{-1} ) = 0\,$. The trick will be repeatedly used in the 
sequel. Having Miss Maple in mind, we call it {\sl concatenation}. 
To summarize, $\Sp\(A T^a\)\,2\, \Sp \( T^a B\) = \Sp \( A\,B \)$, 
{\sl if} either $\Sp\(A\)=0$ or $\Sp \( B\) = 0\,$ or both.

Already in the next step, namely fundamentalizing \eq{9.15},
{\sl concatenation} is at work~:
\bea{9.18}
  (\ov{\6} M^{bd} ) \, (M^{-1})^{dc} 
  &=&  2\, \Sp \( T^b (\ov{\6} M)  T^d M^{-1} + T^b M 
       T^d \ov{\6} M^{-1} \)\, 2  \, \Sp \( T^d M^{-1} 
       T^c M \)   \nonu \\
  &=& 2\, \Sp \( M^{-1} \lk T^b \ov{\6} M - 
      (\ov{\6} M ) M^{-1} T^b M \rk T^d \) \,  
      2 \, \Sp \( T^d M^{-1} T^c M \)  \nonu \\
  &=& 2 \, \Sp \( \lk T^b (\ov{\6} M ) M^{-1} 
        - (\ov{\6} M ) M^{-1} T^b \rk T^c \) \nonu \\
  &=& 2\,\Sp \( \lk T^c , T^b \rk (\ov{\6} M ) M^{-1} \) 
      \; = \; - \, 2 \, i \, f^{abc} \,\Sp \( T^a 
      (\ov{\6} M) M^{-1} \) \quad , \quad \nonu \\
  (M^{\dagger -1} )^{bd}\, \6 \, (\hat{M}^\dagger )^{dc}   
  &=& \quad \ldots \quad = \; - \, 2 \, i \, f^{abc} \,\Sp 
      \( T^a M^{\dagger -1} \6 M^\dagger \)  \quad . 
\eea 
But it is rather trivial to fundamentalize the $A$ terms in 
\eq{9.15}~: 
\be{9.19}
   A^{bc} =  A^a f^{bac} =  - 2i f^{abc} \,\Sp 
   \( T^a A \) \;\; , \;\; (A^\dagger )^{bc}  
   = - A^{\bullet *} f^{b \bullet c} =  - 2i 
     f^{abc} \,\Sp \( T^a A^\dagger \) \quad .
\ee 

The $f$ factors in \eq{9.18}, \eq{9.19} are highly welcome,
because, since \eq{9.3}, $\d\G \,\glr\, 2NS$, we mused
on the factor $N\,$. Now it turns about through $f^{abc} 
f^{bc\bullet}=N\d^{a\bullet}\,$. Using \eq{9.19} and 
\eq{9.18} in \eq{9.15} one obtains
\be{9.20}
 \d^a \G = 2N {i\0\pi} \,\Sp \( T^a \big[ A^\dagger 
     - (\ov{\6} M ) M^{-1} \big] \) \; , \;
 \d^{a*} \G = - 2N {i\0\pi} \,\Sp \( T^a \big[ A 
  - M^{\dagger -1} \6 M^\dagger \big] \) \; . \;
\ee 
\eq{9.20}, right equation, is \ek{2.23} and the c.c.
of the left. Of course, the functional differential equations
for $S$ are given by \eq{9.20} by simply omitting the 
prefactors $2N\,$. But they have by far not yet the
appropriate form. 
\pagebreak[2]

\subsection{ \boldmath$\, S\, =\, S_1 + S_2\;$: $\;$the 
             differential equations for $\,S_2$}
    
\nopagebreak[3]
Reformulation, one more reformulation \ --- \ to which end?
Let us have a look ahead to the first equation of \S~10.
The solution should be a functional of $H$ only, $S[H]$, and
anyhow this might be seen already in the differential
equations. Things have been as in a chess game~: three good
moves, each one followed by three further good moves, an so on.
May be, the actual way gone (with some temporal inflation of 
notations, sorry for) reflects a bit of these difficulties. 

In \eq{9.20} there occur quantities with ``good'' 
differentiation ($\6$ likes $M$, and $A=-(\6 M) M^{-1}$ is 
good) and with a ``false'' one (therefore $F$)~: 
\be{9.21}
   F \,\gll\, - \( \ov{\6} M \) M^{-1} \quad , \quad  
   F^\dagger \, = \, - M^{\dagger -1} \6 M^\dagger 
   \quad . \quad
\ee 
To repeat \eq{9.20} it is very convenient to write the generators
as integrated delta functions $\, T^a = i \int \d^a A\,$ 
and $\,T^a = - i \int \d^{a*} A^\dagger\,$. Then
\be{9.22}
  \d^a S  =  - {1\0\pi} \int \Sp \Big( \lk A^\dagger 
             + F \rk \d^a A \Big)  \quad , \quad
  \d^{a*} S  = - {1\0\pi} \int \Sp \Big( \lk A 
  + F^\dagger \rk \d^{a*} A^\dagger \Big) \quad . \quad
\ee 
This shows $2*n$ functional differential equations for $S$. 
$\int$ stands for $\int\! d^2r$ over the whole 2D plane. $\vc r$ 
is integration varable. Hence (to distinguish variables) we
read $\d^a$ as $\d_{A^a (\vcsm r_0)}\,$. In \eq{9.22} we may
replace $\d^a A$ by $\d^a [ A + F^\dagger ]$, and $\d^{a*} A^\dagger$ 
by $\d^{a*} [ A^\dagger + F ]$, because
$$  A \; , \; F \quad \hbox{depend on only} \quad A^a \qquad 
   \;\; \hbox{and} \qquad \;\; A^\dagger \; , \; 
   F^\dagger \quad \hbox{on only} \quad A^{a*} \quad . $$
To realize this, remember \eq{4.5} showing the expansion of 
$M$ (hence $\ov{\6} M$ aswell) in powers of $A^a\,$. The 
next abbreviaion, which is $\,\O \gll A+F^\dagger\,$, 
shortens \eq{9.22} to
\be{9.23}  \hspace*{-.2cm}
  \d^a (2\pi S) = \int \Sp \( - 2\,\O^\dagger \d^a \O \)
  \quad , \quad  \d^{a *} (2\pi S) = \int \Sp \( - 2\,\O 
  \d^{a*} \O^\dagger \) \quad . \quad
\ee 
It is tempting here to begin with guessing. But the attempt 
with $\int\Sp( \O \O^\dagger)$ only leads to a reasonable 
decomposition~:
\be{9.24} 
  2\pi S \; = \; 2\pi S_1 + 2\pi S_2 \qquad , \qquad 
       2\pi S_1 \,\gll\, \int \Sp \( - \O \O^\dagger 
  \) \quad . \quad
\ee 
As will be seen, $S_1$ remains the harmless first term of the 
action $S$, while $S_2$ advances to become the volume term 
of the WZW action. Conveniently we study the functional
differential equations for the part $\,S_2 = S - S_1\,$
separately. Hence $\d (2\pi S_1) = \int \Sp \( - \O^\dagger \d \O 
- \O  \d \O^\dagger \)$ has to be subtracted from \eq{9.23}
to give
\bea{9.25}
 \d^a (2\pi\, S_2) = \int \Sp \Big( \O \d^a \O^\dagger
  - \O^\dagger  \d^a \O  \Big) \quad , \quad
 \d^{a*} (2\pi\, S_2) = \int \Sp \Big( \O^\dagger 
 \d^{a*} \O - \O \d^{a*} \O^\dagger  \Big) \quad . \quad
\eea 
\pagebreak[2]

\subsection{ Variables \boldmath$H$ only }
   
\nopagebreak[3]
If $S$ depends on only $H=M^\dagger M$ then we expect the 
differential equations to depend on only $H\,$, $\d^a H$ and 
$\d^{a*} H$~:
\bea{9.26}
  \O &=& A + F^\dagger = - \lk (\6 M) M^{-1} 
       + M^{\dagger -1} (\6 M^\dagger ) \rk \nonu \\
     &=& - M^{\dagger -1} \lk M^\dagger \6 M  + 
         ( \6 M^\dagger ) M  \rk M^{-1} = - M^{\dagger -1} 
         (\6 H) M^{-1} \quad , \quad \nonu \\
  \O^\dagger &=&  - M^{\dagger -1} (\ov{\6} H) M^{-1} 
  \; = \; M (\ov{\6} H^{-1} ) M^\dagger \quad . \quad
\eea 
Herewith, the part $S_1$ in \eq{9.24} becomes a $H$--ic object 
immediately~:
\be{9.27}
  2\pi\, S_1 = \int \Sp \Big( (\6 H) \, \ov{\6} 
               H^{-1} \Big) \quad . \quad
\ee 
But for $S_2$ we must keep track with the differential equations,
e.g. with the left equation \eq{9.25}. It becomes $H$--ic through
\bea{9.28}
  \d^a (2\pi\, S_2) &=& \int \Sp \( M^{\dagger -1}
  (\6 H) M^{-1} \d^a \lk M^{\dagger -1} (\ov{\6} H)
  M^{-1} \rk \; - \; \hbox{ditto}_{\6 \gdw \ov{\6}} \;\)
  \nonu \\
  &=& \int \Sp \( (\6 H) H^{-1} \d^a \lk (\ov{\6} H) 
      H^{-1} \rk \; - \; \hbox{ditto}_{\6 \gdw 
      \ov{\6}} \;\)  \nonu \\
  &=& {i\02} \int \Sp \( H_{\prime 2} H^{-1} \d^a 
      \lk H_{\prime 1} H^{-1} \rk \; - \; 
      \hbox{ditto}_{1 \gdw 2} \;\)  \nonu \\
  &=& {i\02} \int \Sp \( H^{-1} H_{\prime 2} H^{-1} 
      \d^a H_{\prime 1} - H^{-1} H_{\prime 2}
      H^{-1} H_{\prime 1} H^{-1} \d^a H \; - \; 
      \hbox{ditto}_{1 \gdw 2} \;\)  \nonu \\
  &=& {i\02} \int \Sp \Big( \ph^a \lk X_1 X_2 - X_2 X_1 
      \rk \, + \6_1 (X_2 \ph^a) - \6_2 (X_1 \ph^a) \Big) 
      \quad , \quad
\eea 
where the abbreviations in the last line are
\be{9.29}
   X_j \gll H^{-1} H_{\prime j} \quad (j=1,2)
   \quad \hbox{and} \quad
   \ph^a \gll H^{-1} \d^a H \quad .
\ee 
In the first line of \eq{9.28} we were allowed to
commute $M^{\dagger -1}$ with $\d^a\,$. The third line was
obtained through ``$\,\6\ov{\6} 
- \ov{\6}\6 = {1\04} ( \6_1 + i \6_2) ( \6_1 - i \6_2) 
- {1\04} ( \6_1 - i \6_2) ( \6_1 + i \6_2) = {i\02} \lk 
\6_2\,\6_1 - \6_1\,\6_2 \rk\,$''. In the last line of \eq{9.28}, 
finally,
\bean
 X_2 H^{-1} \d^a H_{\prime 1}  
 &=&  X_2 H^{-1} \6_1 \d^a H 
      \; = \; \6_1 ( X_2 \ph^a ) - ( H^{-1} 
      H_{\prime 2} H^{-1})_{\prime 1} \d^a H \nonu \\
 &=&  \6_1 (X_2 \ph^a) - H^{-1} H_{\prime 2 \prime 1} 
      \,\ph^a + \lk X_1 X_2 + X_2 X_1 \rk \ph^a 
      \quad , \qquad \nonu
\eea
was used, whereof 1-2--symmetric terms drop out.  
Clearly, the desired $H$--ic version is reached. 
\pagebreak[2]
 
\subsection{ Stokes \ --- \ a common \boldmath$\d$ }
   
\nopagebreak[3]
In the result \eq{9.28}, last line, one recognizes the third
component of a curl operator. We now add the equation
for $\d^{a*}S$, which follows from an analoguous calculation or
simply as the conjugate complex of $\d^a S_2\,$~:
\bea{9.30} \!
  \d^a S_2 \! &=& \!
  {i\04\pi} \int\! \Sp\,\bigg( \,\ph^a [ X_1 X_2 
       - X_2 X_1 ] + \left[ \nabla \times 
        \hbox{\Large\bf (}  X_1 \ph^a 
        \hbox{\Large\bf\ , } X_2 \ph^a  
        \hbox{\Large\bf\ , } \ldots  
        \hbox{\Large\bf\ )} \right]_3 \bigg) \nonu \\
  \d^{a*} S_2 \! &=& \!
  {i\04\pi} \int\! \Sp\,\bigg( \,\psi^a [ X_1 X_2 
       - X_2 X_1 ] - \left[ \nabla \times 
        \hbox{\Large\bf (}  X_1 \psi^a 
        \hbox{\Large\bf\ , } X_2 \psi^a  
        \hbox{\Large\bf\ , } \ldots  
        \hbox{\Large\bf\ )} \right]_3 \bigg) \;\; , \qquad 
\eea 
where $\psi^a \gll H^{-1} \d^{a*} H\,$, which is just $\ph^a$ with
$\d^{a*}$ in place of $\d^a\,$. Is there any other small difference?
For the first term together with the l.h.s. of each line we may in 
fact introduce a common $\d$, which may be freely chosen to be 
$\d_{A^a (\vcsm r_0)}$ or $\d_{A^{a*}(\vcsm r_0)}\,$. 
In place of $\ph^a$, $\psi^a\,$ a comon $\ph \gll H^{-1} \d H\,$ 
would be sufficient. But in the curl term the difference in sign
prevents from such unification. 

$\nabla \times\,$ calls for Stokes theorem (for a plane, in the case 
\eq{9.30} at hand). If there is nothing left at the 
border of the infinite 2D plane, the curl terms in
\eq{9.30} may be omitted, and we arrive at
\bea{9.31}
  \d\, S_2 &=& {i\04\pi} \int \Sp \Big( \, \ph \,
           [ X_1 X_2 - X_2 X_1 ] \,\Big) 
      \quad , \quad \ph = H^{-1} \d H  \\
  &=& {i\04\pi} \int \Sp \Big( \, H^{-1} \d H \lk H^{-1} 
  H_{\prime 1} H^{-1} H_{\prime 2} - H^{-1} H_{\prime 2}   
  H^{-1} H_{\prime 1} \rk \Big) \quad , \quad \nonu 
\eea 
which is the final result of this lengthy \S~9.

For safety, however, let us look at possible border terms,
Stokes could have left. Note that $\ph^a=H^{-1}\d^a H$ is a 
function of two spatial variables. One is $\vc r$, $H$ depends on
and $\nabla\times$ acts upon. The other is $\vc r_0$, the position
mark in $\d^a =\d_{A^a(\vcsm r_0)}\,$. In addition, as $H$ is
also a functional of the fields $A$, there is also the 
integration variable $\vc r'\,$, see \eq{4.4}, \eq{4.5}. 
Through $\d^a$  a $\vc r'$ is forced at the position 
$\vc r_0$. If now $\vc r$ runs far away from $\vc r_0\,$, the
Greens functions $G$ in \eq{4.5} might care for $\ph^a$ vanishing  
at infinity. \\[.3cm]
\hspace*{.6cm}
\vrule depth -3pt height 3.5pt width 1.2cm 
\ \ $\vc r$ \ border \ 
\vrule depth -3pt height 3.5pt width 2cm 
\ \ $A(\vc r' )$ \
\vrule depth -3pt height 3.5pt width .6cm 
\ \ $\d_{A^a(\vcsm r_0)}$ \ 
\vrule depth -3pt height 3.5pt width 2cm 
\ \ border \ $\vc r$ \
\vrule depth -3pt height 3.5pt width 1.2cm \\[.3cm]
If, in addition, $A\to0$ at $r\to\infty$ ($H$ and 
$M\to {\rm const}$), then even the factors $X_j$ support
the omission of border contributions due Stokes.

For later use, also the $H$--ic functional differential
equations for $S_1$ and for $S=S_1+S_2$ should be
noticed. If starting from \eq{9.23} and using \eq{9.26}  
we obtain
\be{9.32}
  \d S_1 = {1\04\pi} \int \Big\{ \; \Sp \Big( \,\ph 
     \lk \6_1 X_1  + \6_2 X_2 \rk \Big)\,  
     + \6_1 \,\Sp \( \ph X_1 \)  + \6_2 \,\Sp \( 
     \ph X_2 \) \;\Big\} \quad , \quad
\ee 
i.e. a common $\d$ immediately. This time it is Gauss, who
removes the derivative terms in \eq{9.32}. To summarize,
we have
\bea{9.33}
  \d S_1 &=& {1\0 2\pi} \int \Sp 
           \Big( \, \ph \lk \6 \ov{X} + \ov{\6} X \rk \big)
     \qquad ,\qquad X \gll H^{-1} \6 H \;\; , \;\; 
     \ov{X} \gll H^{-1} \ov{\6} H \nonu \\
  \d S_2 &=& {1\0 2\pi} \int \Sp \Big( \, \ph \lk \ov{X} X 
             - X \ov{X} \rk \Big) \nonu \\
  & & \hspace*{-2cm} 
      \vrule depth -2pt height 2.2pt width 9.3cm \nonu \\
  \d S &=& {1\0 2\pi } \int \Sp \Big( \, \ph \lk \6 \ov{X} 
         + \ov{\6} X + \ov{X} X - X \ov{X} \rk \Big) 
    \; = \; {1\0 \pi } \int \Sp \Big( \;\ph  \;\6 
         \ov{X} \,\Big) \quad , \quad
\eea 
because of the identity
\be{9.34}
   \ov{\6} X + \ov{X} X = \ov{\6} \lk H^{-1} \6 H \rk 
   + \ov{X}\, X \, = \, H^{-1}  \6 \ov{\6} H \, 
    = \, \6 \ov{X} + X \ov{X} \quad . \quad
\ee 

Assume that all the nice (?) above notations survive the \S~12,
but are forgotten in \S~14 of part II. So, here is equation 
\eq{9.33} in normal life~:
\be{9.35}
   \d \, S \; = \;  {1\0 \pi } \int \Sp \Big( \; 
   H^{-1} (\d H) \;\6\;  H^{-1} \ov{\6} H \; \Big) 
   \quad . \quad
\ee 


\sec{ Solution \boldmath$S[H]$ to the differential equations }

Well, we like inventing. But at this moment the view to the 
alleged result \ek{2.21} is really irresistible. The solution 
$S$ to the above functional differential equations is the 
hermitean WZW action 
\be{10.1}  \!\!   
    \fbox{\rule[-.6cm]{0pt}{1.4cm} $\,\dis      
    S = {1\0 2\pi} \int \Sp 
           \Big( ( \6 H ) \ov{\6} H^{-1} \Big)
     + {i\0 12\pi} \int_V \epsilon^{j k \ell} 
       \Sp \Big( H^{-1} ( \6_j H ) H^{-1} ( \6_k H )
           H^{-1} ( \6_\ell H ) \Big) \,$ } 
\ee 
The first term is $S_1$, and is well known from \eq{9.27}.
Thus, the second term is $S_2\,$, and it is purely $H$--ic, too. 
It has to solve the differential equation \eq{9.31}. Obviously, 
both terms respect the condition \eq{9.2} separately, as they 
vanish for $H\to {\rm const}\,$.  
\pagebreak[2]

\subsection{ The volume term }

\nopagebreak[3]
$S_2$ is a volume integral. How this~?? ~$H$ is defined on the
xy plane. We live on this plane and can't any other. Obviously,
someone mad has given our $H$'s an additional dependence on the
variable $z\,$. In fact, in her TFT'98 proceedings article 
\cite{dika}  Dimitra Karabali states that \   
{\sl ``the integrand thus requires an extension of the 
matrix field $H$ into the interior of $V$, but physical 
results do not depend on how this extension is done. 
Actually for the special case of hermitian matrices the 
second term can also be written as an integral over} [2D] 
{\sl spatial coordinates only.''} \ --- \ aah \ --- 
\ ? \ --- . 

Here, there is a pitfall to run into with ease. The volume 
integral makes sense, one could think, only if it is transformed 
into a surface integral by means of the Gauss theorem (a scalar 
integrand can always be written as $\nabla\cdot \vc C\,$). Then, 
the values of $S_2$ might lie on this surface, and now one can 
proceed with functional differntiation. But this is only 
one possibility of going ahead with checking the $S_2$ differential 
equation (we still believe that one can go this way, anyhow and at 
least formally). The other way was learned from \cite{efna}~:
functionally  differentiate \ f i r s t , i.e. the content
of the volume, and use Gauss only afterwards. Here we
do what we can, and go the easy way.

As mentioned, let the hermitean matrices $H$ anyhow depend 
on $z\,$. Now, apart from $X_1$, $X_2\,$, there is also a 
$X_3 = H^{-1} H_{\prime 3}$. In the $\e$--tensor language of
\eq{10.1} there are six terms under the trace. But cyclic
permutations reduce to two$\,$: 
\be{10.2} 
   S_2 = {i\0 4\pi} \int_V \Sp \Big( \, X_3 \lk X_1 X_2 
         - X_2 X_1 \rk \Big) \quad . \quad 
\ee 
Remember the common $\d$, and let it act at a position 
$\vc r_0$ anywhere in the volume $V\,$. Recalling also 
$\ph = H^{-1} \d H\,$, we are going to check whether 
\eq{10.2} solves the differential equation \eq{9.31}~:
\bea{10.3}
  \d\, S_2 
  &=& {i\0 4\pi} \int_V \Sp \Big(\,\lk X_1 X_2 - X_2 X_1\rk 
      \d X_3  \;\; + \;\; \hbox{cyclic}
      \,\;\Big) \nonu \\[2pt]
  & & \hspace*{1.4cm}
      \d X_3 = \d ( H^{-1} H_{\prime 3} ) \, = \, X_3 \ph 
       - \ph X_3 + \6_3 \ph \quad , \quad \nonu \\[2pt]
  & & \hspace*{1.4cm}
      \Big[\, [ X_1 , X_2 ]\, ,\, X_3\, \Big] \; + \; 
      \hbox{cyclic} \; = \; 0  \quad
      \hbox{(Jacobi identity)} \nonu \\[2pt]
  &=& {i\0 4\pi} \int_V \Sp \Big(\,\lk X_1 , X_2 \rk \6_3 
      \ph \;\; + \;\; \hbox{cyclic}\;\Big) \nonu \\[2pt]
  &=& {i\0 4\pi} \int_V \Sp \Big(\,\6_3\, \ph \lk X_1 , 
       X_2 \rk  \;\; + \;\; \hbox{cyclic} \; - \; \ph 
       \,\hbox{\large\boldmath$\{\,$} \6_3 \lk X_1 , X_2 
       \rk \; + \; \hbox{cyclic} 
       \,\hbox{\large\boldmath$\,\}$}\;\Big) \nonu \\[2pt]
  & & \hspace*{1.4cm}
       X_1 X_2 - X_2 X_1 \, = \,\6_2 X_1 
       - \6_1 X_2 \qquad \folg \quad 
      \hbox{\large\boldmath$\{ \qquad \}$}\; = \; 0  
      \nonu \\[2pt]
  &=& {i\0 4\pi} \int_V \nabla \cdot \,
      \hbox{\Large\bf ( }  \Sp \( \ph [ X_2 , X_3 ] \) 
      \hbox{\Large\bf\ , } \Sp \( \ph [ X_3 , X_1 ] \) 
      \hbox{\Large\bf\ , } \Sp \( \ph [ X_1 , X_2 ] \) 
      \hbox{\Large\bf\ )} \nonu \\[4pt]
  & & \hspace*{3cm} \hbox{{\bf Gauss but now :}} 
      \nonu \\[3pt]
  &=& {i\0 4\pi} \int_{\6 V} d\vc f \cdot \; 
      \hbox{\Large\bf ( }  \Sp \( \ph\,[ X_2 , X_3 ] \)
      \hbox{\Large\bf\ , } \Sp \( \ph\,[ X_3 , X_1 ] \)
      \hbox{\Large\bf\ , } \Sp \( \ph\,[ X_1 , X_2 ] \)
      \hbox{\Large\bf\ )} \nonu \\[4pt]
  \d\, S_2 
  &=& {i\0 4\pi} \int \Sp \Big(\,\ph \lk X_1 X_2 
       - X_2 X_1 \rk \Big) \quad \hbox{on top of the surface ,} 
      \;\;\qquad \hbox{q.e.d.} \qquad
\eea 
The differential equations are fulfilled. Hence, all is done what
the headline announces. The question mark at the end of the above 
Karabali text refers to the claim that $S_2$ can be written as a 
plane integral, too. But we had not the strength to verify this 
detail.
\pagebreak[2]

\subsection{ Recapitulation }
 
\nopagebreak[3]
Five sections have gone just to learn how a wave function $\psi$ 
of the Schr\"odinger wave functional quantum mechanics might be 
normalized. The variables, $\psi$ depends on, vary on ``field axes'' 
($2n$ axes, at the beginning). Already in \eq{6.2} the scalar product 
was noticed. But the question marks there were overcome while 
changing from the measure $d\mu (\cl A)$ to $d\mu (\cl C )$ in 
\S~8, thereby removing the unphysical gauge volume, nothing 
depends on, even not the remaining weight exp$\lk 2NS[H] \rk$. 
To make this weight explicit, we were led into all the trouble 
with the ``giant'' Jacobian $\det({\bf D}^\dagger{\bf D})$ in 
\S~9. After all, the scalar product now reads
\be{10.4}
  \int \psi_1^* \psi_2 \; = \; \int\! d\mu (\cl C)
  \;\,\psi_1^*[H] \, \psi_2 [H] 
  \; = \; \s^n \int\! d\mu (\cl H) \; e^{2NS[H]} \;
  \,\psi_1^*[H] \, \psi_2 [H] \quad .
\ee 
\eq{10.4} ist \ek {2.25}. KKN$\,$: {\sl This formula shows 
that all matrix elements in (2+1)--dimensio\-nal SU(N) gauge 
theory can be evaluated as correlators} of the hermitean 
WZW--Modell. Note that the number of field axes has reduced to
$n$, because with \eq{8.14} $H=\rho^2 =e^{2\vcsm \o \vcsm T}\,$ 
carries the $n$ real parameters $\o^a$. The normalization
prescription is inherent in \eq{10.4}. But a general
wave functional $\psi$ can never be normalizable, as we 
know well from non--relativistic quantum mechanics.
At minimum, we will have to learn on this selection.
And than there is the equation of motion $i \hbar \p 
\psi = {\bf H} \psi\;$. How might the Hamiltonian ${\bf H}$
look like in Schr\"odinger wave field theory~?
\pagebreak[2]

\subsection{ Polyakov--Wiegmann identity }

\nopagebreak[3]
To work with the action $S[H]$, one has not necessarily to 
go into the details of \eq{10.1}. The simpler property is 
its variation \eq{9.35}. There are even other properties, 
as its conformal invariance (see \S~12.2) or the following
{\sl remarkable identity} \cite{powi2} concerning a product 
argument. Its three--lines reasoning in \cite{powi2} is 
hardly understood, and the result given there is wrong\footnote{\
    It is equation (7) there. The relation is stated in 
    terms of a functional $W$, which is nowhere defined. 
    If (5) is meant (with or without the prefactor 
    $1/(2\pi)\,$), then (7) is wrong. Some authors might 
    repeat their half--time examinations. }.  
By \ek{3.2} the identity is given incorrect even in \cite{kkn}. 
So, let us do it right~:
\be{10.5}
  S[AB] = S[A] + S[B] - {1\0\pi} \int \Sp \Big( \lk 
  \6 B \rk B^{-1} A^{-1} \,\ov{\6} A \,\Big) \quad .\quad
\ee 
To derive \eq{10.5}, we write $S = S_1 + S_2$ again and obtain
\bea{10.6}
  S_1[AB] &=& {1\0 2\pi} \int \Sp \Big( \lk \6 A B \rk 
              \ov{\6}\, B^{-1} A^{-1} \Big) \nonu \\
    &=& {1\0 2\pi} \int \Sp \Big( \lk (\6A) B + A\, \6 B \rk 
       \lk  (\ov{\6} B^{-1} ) A^{-1} + B^{-1} \ov{\6} 
        A^{-1} \rk \Big) \nonu \\
    &=& S_1[A] + S_1[B] - {1\0 2\pi} \int \Sp \Big( \;
        (\ov{\6} B ) B^{-1} A^{-1} \6 A + (\6 B) B^{-1} 
        A^{-1} \ov{\6} A \;\Big) \nonu \\[3pt]
    &=& S_1[A] + S_1[B] - {1\0 4\pi} \int \Sp 
        \Big( \, b_1 a_1 + b_2 a_2 \,\Big) \quad ,\quad 
   \lower 10pt\vbox{\hbox{$a_j \gll A^{-1} A_{\prime j}$}
     \hbox{$b_j \gll B_{\prime j}\,B^{-1}$}} \quad . \quad
\eea 
Turning to $S_2$, using \eq{10.2} and again replacing $H$ by $AB\,$, 
we have $X_j = B^{-1}A^{-1} (AB)_{\prime j} = B^{-1} \big( a_j 
+ b_j \big) B\,$ to be inserted there~: 
\bea{10.7}
 S_2[AB] &=& {i\0 4\pi} \int_V \Sp \Big( \, X_3 \lk X_1 
             X_2 - X_2 X_1 \rk \Big) \nonu \\ 
  &=&  {i\0 4\pi} \int_V \Sp \lk \, (a_3 + b_3 ) 
       (a_1+b_1)(a_2+b_2) \;\; - \;\; 
       \hbox{ditto}_{\rm anticyclic} \rk  \\
  &=&  S_2[A] + S_2[B] + {i\04\pi} \int_V \Sp 
       \,\Big[\; \hbox{mixing terms}\; \Big] \quad . 
       \quad \nonu 
\eea 
Among the six mixing terms there are three with one $b$ 
times a commutator of two $a$'s, and three with one $a$ and 
a commutator of $b$'s. The three $a$--commutators, for instance, 
are the components of  $\vc a \times \vc a\,$. Altogether we have
$[${\small mixing terms}$] = \vc b (\vc a \times \vc a ) 
+ ( \vc b \times \vc b ) \vc a\,$ under the trace. This is
very welcome, because through 
\bea{10.8}
 \nabla \times \vc b = \nabla \times \lk  (\nabla B ) 
    B^{-1} \rk &=& \vc b \times \vc b  \quad , \quad 
    \nabla \times \vc a = \nabla \times \lk A^{-1} \nabla A 
    \rk = - \, \vc a \times \vc a \quad , \nonu \\
   \,\Big[\; \hbox{mixing terms}\; \Big] 
   &=& (\nabla \times \vc b ) \vc a - \vc b (\nabla \times 
   \vc a ) = \nabla (\vc b \times \,\vc a ) \quad , \quad
\eea 
Gauss helps us to get rid of the unwanted last volume integral.
On the surface on top of the volume $V$ it yields
$ ( \vc b \times \,\vc a )_3 = b_1 a_2 - b_2 a_1 $, and in total
\be{10.9}
  S_2[AB] = S_2[A] + S_2[B] - {1\0 4\pi} \int \Sp
  \Big( \, i \, b_2 a_1 - i \, b_1 a_2 \, \Big) 
  \quad . \quad
\ee 
Adding \eq{10.9} to \eq{10.6} gives the announced result
\be{10.10}
 S[AB] = S[A] + S[B] - {1\0 4\pi} \int \Sp \Big(\; 
   (b_1 + i b_2 ) \, (a_1 - i a_2 ) \; \Big) \;\; 
   \equiv \;\; \hbox{\eq{10.5} \quad ,\quad q.e.d.} \quad
\ee 
Reporting him of this success, Sergei Ketov smiled~: {\sl yes, 
a pedestrian needs Gauss, but the elegant derivation
is that in \cite{powi2}}.

For a nice check of the identity \eq{10.5} one may use it to
obtain the differential equation \eq{9.35} again. 
\bea{10.11}
 \d S &=& S[H+\d H] - S[H] = S\lk H\, \(1+H^{-1} \d H\) \rk 
          - S[H] \nonu \\
      &=& S\lk 1+H^{-1} \d H \rk - {1\0\pi} \int \Sp \(
          (\6 H^{-1} \d H ) (1- H^{-1} \d H ) H^{-1} \ov{\6} H \)
             \nonu \\
      &=& {1\0\pi} \int \Sp \( H^{-1} \d H\, \6\, H^{-1} \ov{\6} H \)
           \; \equiv \; \hbox{\eq{9.35}} \quad
\eea 
using partial integration and neglecting terms $\sim (\d H)^2\,$.

In the next section we even need $S[ABC]\,$. But this is merely
one more exercise on \eq{10.5}~:
\bea{10.12}
  S[ABC] &=& S[A] + S[B] + S[C] - {1\0\pi} \int \Sp 
            \bigg( \nonu \\ 
  & & \hspace{-2.8cm} (\6 C) C^{-1}B^{-1}A^{-1} 
      (\ov{\6} A ) B \; + \; (\6 C) C^{-1}B^{-1} 
       \,\ov{\6} B   \; + \; (\6 B) B^{-1} 
    A^{-1}\,\ov{\6} A  \;\;\bigg) \quad . \quad
\eea 
While checking \eq{10.12} with e.g. $\, A B C = H H^{-1} H\,$,
one becomes aware of two more strange relations, namely
\be{10.13}
   S_1 [ H^{-1} ] = S_1 [H]  \qquad , \qquad   
   S_2 [ H^{-1} ]\, =\, -\, S_2 [ H ]
   \quad , \quad 
\ee 
and obtains $\, 3 S_1 + S_2 - 2 S_1 = S\,$ to the right of
\eq{10.11}, as expected.


\sec{ Regularization }

In section 9.4, the step to the regularized version \eq{9.14}
of the inverse matrix $D^{-1}$ was done by citation. Here
this painful gap will get closed. In the equations for
$\G$, \eq{9.6}, the limit $\vc r' \to \vc r$ could not be
performed, not naively at least, because according to \eq{9.9} 
the Greens function comes across its pathological argument zero.
 
First of all, a regularisation has to respect gauge invariance, 
i.e. it might favour $H$'s rather than $M$'s. But there is
still one more redundancy to be respected, namely the ``very old'' 
one in fixing the mapping from space $\cl A$ to space $\cl M\,$. 
\pagebreak[2]

\subsection{ Holomorphic invariance }
   
\nopagebreak[3]
The term could be due to KKN ({\sl we shall refer to ... as 
...}). We encountered the corresponding freedom while fixing 
it especially in \eq{4.4}. An arbitrary $N\times N$ matrix 
$\ov{V}$, depending on $\ov{z}=x+iy\,$, could have been used
there in place of the inhomogenity 1. Of course, if allowing 
for all these inhomogenities, then a special element $A$ of space
$\cl A$ is mapped to some subspace $M\,\ov{V}\,$. If $M$ is 
in SL(N,C) than $M\ov{V}$ is not. Nevertheless, changing $\ov{V}$ 
leaves the $A$ fields invariant,
\be{11.1} 
  M \to M\, \ov{V} \;\quad \folg \;\quad
  A=-(\6M)\, M^{-1} \;\to\; - \(\6 M \ov{V} \) \; 
  \ov{V}^{-1}  M^{-1} = A \quad  \;\; , \quad
\ee 
because of $\,\6\ov{V}(\ov{z})=0\,$, see~\eq{3.12}. No physics
depends on the special $V$, people on earth ($\ov{V}=1$), moon 
or neptun work with to fix their mapping. 
\\[12pt]
{\bf The action {\boldmath$S[H]$} is holomorphic invariant}.
$M\to M\ov{V}$ makes $H=M^\dagger M$ turning into $H \to 
VH\ov{V}$, where $V\gll \ov{V}^\dagger$ depends on only $z=x-iy$ 
and $\ov{\6} V(z)=0\,$. Hence, the question is, whether
$S[VH\ov{V}]$ agrees with $S[H]$, and this is clearly a matter
of the $S[ABC]$ relation \eq{10.12} as prepared in the last 
subsection. The three terms under trace vanish all due to $\6 C 
= \6 \ov{V} = 0$ and/or $\ov{\6} A = \ov{\6} V = 0\,$. The 
remaining unwanted terms are $S[V]$ and $S[\ov{V}]\,$. They 
vanish, as is seen next, because either $V$ or $\ov{V}$ 
suffer under the ``false differentiation''. For $S_1[V]$ this
is seen in \eq{9.27} directly. Concerning $S_2[V]$ consider
the square bracket in \eq{10.2}~:
\bea{11.2}
   \lk X_1 X_2 - X_2 X_1 \rk &=&  V^{-1} V_{\prime 1}
   V^{-1}V_{\prime 2} - V^{-1} V_{\prime 2}
   V^{-1}V_{\prime 1}  \nonu \\
   &=& {2\0i} V^{-1} (\ov{\6} V) V^{-1} \6 V 
   - {2\0i} V^{-1} (\6 V) V^{-1} \ov{\6}V = 0 \quad ,
\eea 
since $\ov{\6} V(z)=0\,$. Similarly $S_2[\ov{V}]\,$ is shown to 
vanish. Thus, $S[VH\ov{V}]= S[H]\,$, q.e.d.
\\[12pt]
{\bf The Greens function becomes a matrix.} Let the people
on the moon (L for Luna) represent the field $A$ by
$A= - (\6 L ) L^{-1}$ and use the inhomogenity $\ov{V}$ when 
solving it for $L$. They write (using $G_{r r'}= -G_{r' r}$)
\be{11.3}
  L = \ov{V} + \int' (AL)_{r'} G_{r' r} \quad 
  \hbox{in place of} \quad M = 1 + \int' (AM)_{r'} 
  G_{r' r} \quad .
\ee 
We (living on earth) multiply our integral equation (the right 
one in \eq{11.3}) by $\ov{V}\,$ from the right,
$$ M \ov{V} = \ov{V} + \int' (AM\ov{V})_{r\prime} \;\;
   \ov{V}_{r'}^{-1} G_{r' r} \ov{V}_r  \quad , \quad $$
and recognize the two manipulations in parallel 
\bea{11.4}
  M \;\to\; M\,\ov{V} \quad  &\hbox{and}& 
  \quad G_{r r'} \;\to\; \ov{V}^{-1}_r 
  G_{r r'} \ov{V}_{r'} \;\glr\, 
     \schl{G}_{r r'} \hspace{1cm} \nonu \\
  \hbox{or} \quad M^\dagger \;\to\; V\, M^\dagger\quad 
  &\hbox{and}& \quad \ov{G}_{r r'} \;\to\; 
  V_r \;\ov{G}_{r r'} V_{r'}^{-1}
\eea 
by which one is beemed to Luna. In passing, $\schl{G}_{r r'}$
is a Greens function as well~: $\6 \schl{G}_{r r'} = \d(\vc r 
- \vc r')\,$. 

Just for fun, let as also turn to neptun, where people write
$A = -(\6N)N^{-1}$ and work with inhomogenity $\ov{W}$ and Greens
function $\ov{U}_r G_{r r'} \ov{U}_{r'}^{-1}\,$, say~: 
$$ N = \ov{W} + \int' (AN)_{r'} \, \ov{U}_{r'} G_{r' r} 
   \ov{U}_r^{-1}  \quad . \quad $$ 
Now, multiplication with $\ov{U}$ from the right leads back
to Luna equations with $L=N\ov{U}$ and inhomogenity
$\ov{V} = \ov{W}\ov{U}$. Allright.

Finally, for later use, we add the adjoint representation
of $\schl{G}_{r r'}$, as defined in \eq{11.4}~: 
\be{11.5}
 G_{r r\prime} \,\d^{ab} \;\to\; ( \ov{V}_r^{-1} )^{ac}
 G_{r r\prime} (\ov{V}_{r'})^{cb} \; \glr\, 
  ( \schl{G}_{r r\prime})^{ab} \quad . \quad
\ee 
To derive this, simply repeat the steps leading from \eq{11.3} 
to \eq{11.4}, but in adjoint representation.
\\[12pt]
{\bf The integration measure $d\mu(\cl H)$ is holomorphic invariant}. 
According to KKN this is {\sl easily checked}. To do so, we look at
\eq{8.10}, i.e. at $ds_{\cl H}^2 = 2 \int \Sp ( H^{-1} \d H H^{-1} 
\d H)\,$. While varying $H$ the inhomogenity $\ov{V}$ has to be kept
fixed: one either lives on moon or neptun or just here. But
under $H\to VH\ov{V}$ and $\d H \to V\d H\ov{V}$ the metrics
remains clearly unchanged, and so does the measure.
\\[12pt]
{\bf The operators \boldmath$p^a$ and $\ov{p}^a$ split off
a matrix}. Two functional differential operators, called $p^a$ and 
$\ov{p}^a$, are particular useful for the present task (see 
\eq{11.15} below). But they enjoy still an other nice property as 
detailed in \S~14.3. $p^a$ and $\ov{p}^a$ are defined through
\be{11.6}
  \d^a \,\glr\, M_r^{ab} \int' G_{r r'}  p^b_{r'} 
  \quad , \quad \d^{a*} \,\glr\, - (M_r^\dagger)^{ba} 
  \int'  \ov{G}_{r r'} \ov{p}^b_{r'} \quad , \quad
\ee 
which is KKN's \ek{2.34}. Remember $\,\d^a=\d_{A^a(\vcsm r)}$, 
and note that $\ov{p}^a = -p^{a*}\,$. \eq{11.6} can be solved for
$p$ by operating with $\6_r \hat{M}^{-1}$ from the left~:
\be{11.7}
  p_r^a = \6_r \, (M_r^{-1})^{ac} \; \d_r^c \quad , 
  \quad  \ov{p}_r^a = - \ov{\6}_r \, (M_r^\dagger)^{ac} 
  \; \d_r^{c*}  \quad . \quad
\ee 
Note the position marks: $\6$ or $\ov{\6}$ act on both. Turning 
to Luna, the $p$'s change according to~:
folgendes$\,$:
\be{11.8}
  p^a \;\to \; (\ov{V}^{-1})^{ab} p^b \qquad , \qquad
  \ov{p}^a \;\to \; V^{ab} \ov{p}^b \quad . \quad
\ee 
To derive this, note that $A^a$, or $\d^a$, knows of no
earth--moon difference. Hence
\bea{11.9}
  p^a \;\to\; \6 \,\; 2 \,\Sp \Big( T^a \ov{V}^{-1} 
     M^{-1} T^c M \ov{V} \Big) \; \d^c  
  &=& \6 \; 2\Sp \Big( \ov{V} T^a \ov{V}^{-1} T^b \Big)\, 
   2\Sp \Big( T^b M^{-1} T^c M \Big)\; \d^c \qquad \nonu \\
     = \,\6 \; ( \ov{V}^{-1})^{ab} (M^{-1})^{bc} \,\d^c 
  &=&  (\ov{V}^{-1} )^{ab} p^b \quad . \quad
\eea 
The step within the first line used {\sl concatenation} 
in backward direction.
\pagebreak[3]
\\[12pt]
{\bf The Hamiltonian density is holomorphic invariant}. 
This is trivial for the potential term in \eq{2.10}, because
it is made up of the holomorphically insensitive $A$ fields.
The same is true for the kinetic energy density $\cl T 
= - {e^2\02} \d^{a*} \d^a$ \ --- \ sorry for looking
forward to \eq{12.2} below. But, using \eq{11.6},  $\cl T$ 
may be also written as 
\be{11.10}
 \cl T = - {e^2\02} \d^{a*} \d^a \; = \; 
    {e^2\02} H^{ab} \,\big( \ov{G} \ov{p} 
     \big)^a\, \big( G p \big)^b  \quad ,
\ee 
where $H^{ab}=2\Sp(T^a H T^b H^{-1})\,$. While checking
the invariance with the expression to the right, we
enjoy the harmony among the previous equations
\eq{11.5} to \eq{11.8}~: 
\be{11.11}
  \big( G p \big)_r^b \,\gll\, \int' G_{r r'} p_{r'}^b 
  \; \to \; \int' \Big( \ov{V}^{-1}_r G_{r r'} \ov{V}_{r'} 
  \Big)^{bc} \( \ov{V}_{r'}^{-1} \)^{cd} p_{r'}^d 
  = \big( \ov{V}_r^{-1} \big)^{bc} \big( G p \big)^c_r 
  \quad , \quad
\ee 
and similarly $(\ov{G} \ov{p})^a \to V^{ab}(\ov{G} \ov{p})^b\,$. 
From $H\to VH\ov{V}$, and using concatenations and $V^{ac} 
=(V^{-1})^{ca}\,$, one obtains
\bea{11.12}
  H^{ab} &\to& V^{ac} H^{cd} \,\ov{V}^{db} \qquad \folg 
                  \quad \nonu \\
  \cl T &\to& {e^2\02} \, V^{ac} \, H^{cd} \, 
   \ov{V}^{db}\, V^{ae} \, \big( \ov{G} \ov{p}\big)^e \,
   (\ov{V}^{-1})^{bf} \, \big( G p \big)^f  \,\; = \,\; 
   \cl T \quad , \quad \hbox{q.e.d.} \quad . \quad
\eea 
\pagebreak[2]

\subsection{ Point splitting}
     
\nopagebreak[3]
Usually, the infinitites of a field theory are recognized
and regularized as UV catastrophes in momentum space ($\L$, 
$M$, $d-\e$). If this is done in real space, we expect that
some short distances must be washed out, as e.g. when going
from the delta function to one of its representations. KKN's 
smooth delta function is
\be{11.13}
   \s (\vc r ) \gll {1\0\pi\eta} e^{-r^2/\eta}
   \quad , \quad \int \! d^2r \;\s (\vc r ) = {1\0\pi} 
   \int e^{-r^2} = 1
\ee 
The inverse momentum cutoff $\eta$ is small but non--zero.
Compared to this the small parameter $\e$ in the Greens 
function \eq{4.2} is an icecold $+0\,$. 

In a manner that respects holomorphic properties, the
smoothening $\s$ is first buildt in the operators $p\,$.
KKN then form $(Gp)^a$ to read off from it the fate of $G$
itself. The regularized Greens function, thanks to some
labour of evaluation, exhibits the desired finite value in 
the coincidence limit $\vc r' \to \vc r\,$.

Things start with a notation. Let $H$, if considered as
a function of $z$ and $\ov{z}$, be denoted by $K\,$:
\be{11.14}
  H^{ab}(\vc r) \, \glr\, K^{ab}(z,\ov{z}) \; \to \;
  V^{ac} (z) \, K^{cd}(z,\ov{z}) \,\ov{V}^{db}(\ov{z}) 
  \quad . \quad
\ee 
For $\eta\to0$, i.e. $\s_{rr'}\to \d_{rr'}\,$, the two
expressions
\bea{11.15}
  p_{\rm reg}^a \,&\gll& \int' \s_{r r'}  \lk K^{-1} 
  (z' , \ov{z} ) \, K(z' , \ov{z}' ) \rki^{ab} \, 
  p_{r'}^b  \nonu \\
  \ov{p}_{\rm reg}^a \,&\gll& \int' \s_{r r'} 
  \lk K(z, \ov{z}' ) \, K^{-1}(z' , \ov{z}' ) \rki^{ab} 
  \, \ov{p}_{r'}^b  \quad
\eea 
clearly turn into $p^a\,$ and $\ov{p}^a$, respectively.
But even for $\s_{rr'} \neq \d_{rr'}\,$ they have the right 
holomorpohic behaviour \eq{11.8}. This is seen by means of 
\eq{11.14}: $\;\big[\, K^{-1}(z',\ov{z}) \,\big]^{ac} 
\to \,\big[\, \ov{V}^{-1}(\ov{z}) K^{-1}(\ldots ) V(z')
\,\big]^{ac}\;$, and the right $V$ matrix at unprimed argument 
survives at the left end. Anyhow clever.     

As announced, the replacement $p^a \to p^a_{\rm reg}$ is now
performed in $(Gp)^a\,$, 
\be{11.16}
   \int' \! G_{r r'} p_{r'}^a  \,\to\, \int' \! G_{r r'} 
   \int''\! \s_{r' r''} \lk K^{-1} (z'' , \ov{z}' ) \, 
   K(z'' , \ov{z}'' )  \rki^{ab}\,  p_{r''}^b 
   \, \glr \int'' \cl G^{ab}_{r r''} \, p_{r''}^b \quad , 
\ee 
to read off the regularized Greensfunktion $\cl G$ as 
\be{11.17}
  \cl G^{ab}_{r r''} = \int' G_{r r'} \s_{r' r''}
  \lk K^{-1} ( z'' , \ov{z}' ) \,  K ( z'' , \ov{z}'' ) 
  \rki^{ab} \; \glr \; \int'  G_{r r'} \s_{r' r''} \; 
  f^{ab}(\ov{z}' ) \quad . \quad
\ee 
Denoting $\lk K^{-1} ( z'' , \ov{z}' ) \, K 
( z'' , \ov{z}'' ) \rki^{ab}\,$ by $f^{ab}(\ov{z}')\,$, 
we suppress the variables irrelevant in the next subsection. 
But let $f^{ab} (\ov{z}'') = \d^{ab}\,$ be kept in mind.
\pagebreak[2]

\subsection{ Performing the integration in \boldmath$\cl G$ }
   
\nopagebreak[3]
At first glance, it appears absurd to perform the $d^2r'$ 
integration in \eq{11.17}, because,  though $G$ and $\s$ 
are known funtions, we \ \hbox{c a n n o t} \  specify 
$f(\ov{z})$. The integration is nevertheless possible \ --- \ 
thanks to the fact that $G$ is a Greens function of $\6\,$.

There is a test, the result will have to pass. According to 
\eq{11.17} it is
\be{11.18}
  \6 \, \cl G^{ab}_{r r''}  = \s_{r r''}\, f^{ab} (\ov{z}) 
  = {1\0 \pi \eta} e^{- (\vcsm r - \vcsm r''  )^2 / \eta} 
    f^{ab} (\ov{z}) \quad ,
\ee 
and after evaluation this must be still valid, of course. The 
reader might look forward to \eq{11.26} and do the test right 
now. 

It is convenient to introduce an integral $S$ of $\s\,$: 
\be{11.19}
 \6 S (\vc r ) = \s (\vc r ) \quad , \quad \hbox{\ft ansatz } 
 \; S = {z\0 r^2} h(r)  \quad  \folg \quad   S(\vc r ) = 
 {1\0 \pi\, \ov{z} } \( 1 - e^{-r^2/\eta} \) \quad , \quad
\ee 
where a constant of integration was chosen to be 1 at will.
By means of $S$, and omitting the indices $a$, $b$ for brevity, 
we may write
\be{11.20}
  \cl G_{r r''} = \int'\! G_{r r'} 
  \,\lb \,\6' S(\vc r' - \vc r'' ) \,\rb\, f (\ov{z}') 
  \, = \, \int'\! \6' (GSf) - \int' \!(\6'G) Sf
  \, \glr \, \cl G_1 + \cl G_2  \quad 
\ee 
since $\6'f=0\,$. The second part is readily evaluated~:
\bea{11.21}
 \cl G_2 &=& S(\vc r - \vc r'' ) f(\ov{z})
  = {1\0\pi} {1\0 \ov{z}- \ov{z}'' } \( 1 - e^{(\vcsm r 
   - \vcsm r'' )^2 / \eta} \)  f(\ov{z}) \nonu \\
  &=& \; G_{r r''}   \( 1 -  e^{(\vcsm r - \vcsm r'' )^2 
    / \eta} \) f(\ov{z}) \quad . \quad
\eea 
The use of $G$ in the last line was allowed due to the zero
of the round bracket at the pole of $G$. 
The other part may be written as
\be{11.22}
  \cl G_1 = \int' \! \6' \,G_{rr'} {1\0\pi (\ov{z}'- \ov{z}'')}\,
  \(1-e^{-\ldots}\) f(\ov{z}')  \; = \; {-1\0\pi (\ov{z}
  - \ov{z}'')}\, f(\ov{z}) \, + \, \cl G_{1e} \quad
\ee 
with
\be{11.23}
  \cl G_{1 e} = - \int\! d^2 r' \;\, \6^{\,\prime} 
  \lk {1\0\pi} \, { z-z' \0 (z-z' )\, (\ov{z} - \ov{z}' ) 
  + \e^2 } \;\, {1\0\pi}\, {1 \0 \ov{z}' - \ov{z}'' } \;
  e^{- (\vcsm r' - \vcsm r'' )^2 / \eta} 
  \; f(\ov{z}' ) \rk \quad , 
\ee 
For the first step in analysing this expression, the reader might
realize that the limit $\e\to0$ may be performed here. The second
step is the shift $\vc r' \to \vc r' + \vc r''$ of the
integration variable. The third step introduces the variables 
$u$, $v$ by
$$ 
  x' = {1\02} (u-iv) \;\; , \;\; y' 
  = {1\02} (-iu+v)  \;\; , \;\; \hbox{Jacobian} 
  = {1\02}  \;\;\; , \;\;\; \ov{z}' = u 
    \;\; , \;\; z'  = -iv \;\; ,  \quad  $$
and leads to
\be{11.24}
  \cl G_{1\, e} = - {i\0 2} \, {1\0 \pi} \int \! du' \; 
  {f(u' + u'') \0 \; u-u'' - u' \;} \;\,  {1 \0 \pi\, u'} 
  \int \! dv' \; \6_{v'} \; e^{i\, {u'\0\eta}\,v' }\,
  \; = \; {1\0\pi} \, {1\0 \ov{z} - \ov{z}''} 
   \; f(\ov{z}'') \quad . \quad
\ee 
To realize this, perform $\6_{v'}\,$, obtain $\d(u')$
from the $v'$ integration and return to $u''= \ov{z}''\,$. 
Using \eq{11.24} in \eq{11.22} we arrive at
\be{11.25}
 \cl G_1 = {1\0 \pi} \, {1\0 \ov{z} -  \ov{z}'' }
  \lk f( \ov{z}'' ) - f( \ov{z} ) \rk
  \;=\; G_{r r''} \lk f( \ov{z}'') - f( \ov{z} ) \rk
\ee 
and together with \eq{11.21} at the final result
\be{11.26}
  \cl G^{ab}_{r r''} = G_{r r''} \( \d^{ab} 
  - e^{-(\vcsm r - \vcsm r'' )^2  / \eta} \; 
  f^{ab} (\ov{z} ) \;\) \quad . \quad
\ee 
\eq{11.26} is \ek{3.8}. It was obtained with the special
embedding \eq{11.13} of the delta function, but we bet that 
any other delta representation would lead to the same result. 
\pagebreak[2]

\subsection{ The coincidence limit}
     
\nopagebreak[3]
After all, the limit $\vc r' \to \vc r$ can now be performed
without troubles. We look back at $\cl G_2$, \eq{11.21}, and 
realize that the round bracket vanishes $\sim (z-z'')\,
(\ov{z} - \ov{z}'')\,$, i.e. faster than the denominator. 
In short, in the coincidence limit $\cl G_2$ turns to zero. 
For $\cl G_1$, the inner expression of \eq{11.25} shows a
differential quotient$\,$:
\bea{11.27}
  \cl G_{r r}^{ab} \; = \;\cl G_{1 \; r r}^{ab} &=& - 
    \, {1\0\pi} \, 
  \apf{\ov{\6}} \lk K^{-1} (z, \apf{\ov{z}} ) \, 
  K(z,\ov{z}) \rki^{ab} = {1\0\pi} \lk H^{-1} \ov{\6} 
  H \rki^{ab} \nonu \\ 
  & & \hspace*{-3.7cm} {} =  
  {1\0\pi} \lk M^{-1} M^{\dagger -1} \Big( (\ov{\6} 
  M^\dagger ) M + M^\dagger  \ov{\6} M \Big) \rki^{ab}
  = {-1\0\pi} \lk M^{-1} \Big( A^\dagger - (\ov{\6} M)
  M^{-1} \Big) M \rki^{ab} . \qquad  
\eea 
\eq{11.27} is \ek{3.10}. In regularizing \eq{9.9}, just $G$ 
has to be replaced by $\cl G\,$:
\be{11.28}
  \Big[ \; D^{-1}_{\rm reg} \; \Big]^{ab}_{rr} 
   = M^{ac}\, \cl G^{cd}_{rr} \, (M^{-1})^{db} 
   =  -\,{1\0 \pi} \,\lk   A^\dagger 
   - (\ov{\6} M) M^{-1}  \rki^{ab} \quad , 
   \quad \hbox{q.e.d.} \quad 
\ee 
and hooray, because this is the desired result \eq{9.14}.


\sec{ Kinetic energy , mass gap and CFT }

Anywhere, far behind, we had noticed the classical Hamiltonian
density of the 2+1~D YM system, \eq{2.8} to \eq{2.10}$\;$:
\be{12.1}
  \cl L = {1\0 2 e^2} \p A{\!}^a_j \p A{\!}^a_j - \cl V
          \quad , \quad 
    \P^a_j = {1\0 e^2} \p A{\!}^a_j \quad , \quad
  \cl H = {e^2\02} \P^a_j \P^a_j + \cl V \quad   
\ee 
with $\,\cl V \gll {1\0 2 e^2} B^a B^a\,$ and $\,B^a = 
\6_1 A^a_2 - \6_2 A^a_1 + f^{abc} A^b_1 A^c_2\,$. Were there 
no gauge freedom, the quantum mechanics \ \hbox{w o u l d} \
be initiated by $\P^a_j \;\to\; (1/i)\,\d_{A^a_j}\,$:
\be{12.2}
 \cl H = \cl T + \cl V \qquad \hbox{with} \qquad
  \cl T \, = - {e^2\02} \d_{A^a_j} \d_{A^a_j} 
  \; = \; - {e^2\02} \d^{a*} \d^a \quad
\ee 
as already announced in \eq{11.10}. The question, how
$\cl T$ can be restricted to the physical subspace $\cl C$,
has a rather simple answer$\,$: apply $\cl T$ to functionals
of only $H\;$! KKN make some efforts in constructing the
{\sl Laplacian on $\cl C$} by splitting off the unphysical part
from $p^a$. The use of these details will be realized in \S~14.4.
But they are irrelevant for the particular question of reducing 
$\cl T$. It restricts itself by working with in the space 
$\cl C$. Thereby the above ``would be'' turns into
``is'', automatically.
\pagebreak[2]

\subsection{ \boldmath$ T\; J^a \; = m \; J^a $}
     
\nopagebreak[3]
Let $\cl T$ be applied to special functionals $\psi[H]\,$. 
Which ones~? ~At this point we jump and servile follow the text 
of KKN around \ek{2.28}. There one finds the statement, which is
crucial and is as interesting as (at first glance) mysterious~:
\bea{12.3} 
  \hbox{To be normalizable~, ~the } \psi\hbox{'s must be 
  functionals $\psi[J^a]$ of the ``currents'' } & & 
  \nonu \\[1mm]
  J^a \,=\, {2N\0\pi} \, \Sp \Big( \, T^a \,(\6H) 
  \, H^{-1}\, \Big) \quad . \hspace*{4cm} & & \quad 
\eea 
This is the decomposition of $\psi[H]$'s expected at the end 
of \S~10.2. The statement means that the dependence on $H$ must
be indirect as shown, otherwise the norm $\int\!|\psi|^2$ 
diverges. The objects $J^a$ are indeed the currents of the
WZW action $S$ as detailed at the end of \S~12.2. To justify 
\eq{12.3}, apparently (see \cite{kkn}, this time really), 
there are several possibilities. Probably, conformal field theory 
(CFT) provides with the most convincing reasoning. But let CFT be 
outside of the scope of the present treatise (apart from the 
next subsection). Some weaker argument in favour of \eq{12.3}
rests on fact that the Wilson loop, hence any (normalizable)
physics, can be shown to depend on only $J\,$. This will be 
detailed in \S~16.5.

For ending up part I, consider the most simple functional of $J\,$, 
namely \cite{vor}
\be{12.4}
  \psi_{\rm sp} \lk J \rk \;\gll\; \int c^a (\vc r)
  J^a (\vc r ) \nonu \\
\ee 
(with arbitrary c number functions $c^a(\vc r)\,$) and 
apply $\int \cl T\,$ to it. Using
\bea{12.5}
  J^a (\vc r) &=& {N\0\pi}\, 2 \, 
  \Sp \( T^a \lk (\6 M^\dagger) M^{\dagger -1}
  +  M^\dagger iT^c A^c M^{\dagger-1} \rk \) \nonu \\
  & & \folg \qquad  \d^a_r J^d_{r'} 
  = i {N\0\pi} (M^\dagger_r)^{da} \d(\vc r-\vc r') \quad
\eea 
we are led to the following second line. But for the details 
$(*)$, $(**)$ (to be commented afterwards) we are much to 
excited~:
\bea{12.6}
 {\bf T} \;\psi_{\rm sp} &=& - {e^2\02} \int \d^{a*}_r \, 
    \d^a_r \, \int' c_{r'}^d \, 
     J^d_{r'} \nonu \\
  &=& - {e^2\02} \int \,\d_r^{a*} \; {N\0\pi}\, i\, c^d_r 
        (M_r^\dagger)^{da}  \; = \;
     - i \; {e^2N\0 2\pi} \int c^d \lk \d^{a*}_{r'}
      (M_r^\dagger)^{da} \rki_{\vcsm r' \to \vcsm r} 
        \nonu \\  
  &=& - i \; {e^2N\0 2\pi} \int c^d \lk  
      (M_{r'}^\dagger)^{db} f^{bae} \( 
       D^{* -1}_{r r'} \)^{ae} \rki_{\vcsm r' 
       \to \vcsm r}          \qquad\qquad (*) \nonu \\
  &=&  - i \; {e^2N\0 2\pi} \int c^d \, (M^\dagger)^{db}
       \;{iN\0\pi}\, 2\,\Sp \(\, T^b \lk M^{\dagger -1} 
       \6 M^\dagger + (\6 M) M^{-1} \rk \) 
                        \qquad (**) \qquad \nonu \\  
  &=&  {e^2N\0 2\pi} \int c^d \lk 
       {N\0\pi} \, 2 \, \Sp \( T^d (\6 H ) H^{-1} \) \rk   
       \nonu \\  
  &=& {e^2 N \0 2\pi} \int c^d J^d \;\; = \;\; 
       m  \;\; \psi_{\rm sp} \qquad , \qquad
       m \gll {e^2 N \0 2\pi}  \quad . \quad
\eea 
Special eigenfunctions of ${\bf T}\,$ are found. Due to the 
arbitrarity of the weight $c^a(\vc r )\,$ the corresponding
eigenvalue $m\,$ is infinitly degenerated.

In the third line $(*)$, one recognized the same coincidence 
limit as already in \eq{9.6}. Moreover, the factors 
$f^{^{\ldots}} D^{*-1}$ in the square bracket \ a r e \ 
$\,\d^{b*} \,\G\,$, i.e. they can be replaced by the regularized 
expression \eq{9.20} directly. This explains the step from line 
$(*)$ to line $(**)$. Then, by just a {\sl concatenation} one 
reaches the next-to-last line. There remains the problem of 
verifying the step from the second to the third line. It states 
the equality of two square brackets. For simplicity, consider 
the c.c. of the first one. $\,\d^a_{r'} \hat{M}_r\,$ must 
derive from $\, 0 = \(\6 + \hat{A}\)\, \hat{M}\,$, anyhow 
(see \eq{9.12} and remember $A^{ab}=A^\bullet f^{a\bullet b}\,$). 
We apply $\d^a_{r'}\,$, use \eq{7.13} and obtain  
\be{12.7}
  \int''\! D^{eb}_{r r''} \; \d^a_{r'} M_{r''}^{bd} 
  = \d_{r r'} f^{aeb} M_{r'}^{bd}  \quad \folg \quad  
  \d_{r'}^a M_r^{cd}  = (D^{-1}_{r r'})^{ce} f^{aeb} 
  M_{r'}^{bd} \quad . \quad
\ee 
One may set $a=c$ here and sum over $a\,$. Since $M^{db\, *} 
= (M^\dagger)^{bd}\,$ we may even add the c.c. with ease~:
\be{12.8}
  \d_{r'}^a M_r^{ad} = (D^{-1}_{r r'})^{ae} f^{aeb} 
  M_{r'}^{bd} \quad , \quad  \d^{a*}_{r'} 
  (M_r^\dagger)^{da} = (M_{r'}^\dagger)^{db} f^{bae} 
  \( D^{* -1}_{r r'} \)^{ae} \quad , \quad
\ee 
q.e.d. But there is danger to anyone going his own way.
Instead of \eq{12.7} he could derive the very similar
relation 
\be{12.9}  
  \d^a_{r'} M_r^{cd} = (D^{-1}_{r r'})^{ea}
  f^{ecb} M_r^{bd} \quad . \quad
\ee 
Now, by $c=a$ and summation, he obtains \eq{12.7} only nearly, 
because $M^{bd}$ carries the index $r$ (not $r'$). Hence, he 
condamnes \eq{12.6} and is searching for his or our error till 
tomorrow. The answer: both are right. \eq{12.7} and \eq{12.9} can 
be derived from each other by means of \eq{9.9}. In the special 
combination \eq{12.8}, one is allowed to switch the $M$--index 
from $\vc r'$ to $\vc r\,$.
\pagebreak[2]

\subsection{ \boldmath$S[H]\,$ is conformally invariant}
  
\nopagebreak[3]
Already since section 2 we live in a 2D Euclidian world. This is 
the case in which the call for conformal invariance is
particular restrictive \cite{gins,ket,difr}, because
in the coordinate transformation
\bea{12.10} 
    z \;\to\; f(z) \quad & &, \;\; \hbox{i.e.} \;\;\;
    z_{\rm new} = u(x,y)-iv(x,y) \;\ueb{!}{=}\; f(x-iy) \\ 
    & & \quad \folg \quad u_{\prime x} = v_{\prime y} 
    \quad , \quad  u_{\prime y} = - v_{\prime x} \nonu
\eea 
the function $f(z)\,$ remains arbitrary. All these nice facts as
e.g. the conservation of angles (explaining the word ``conform'')
under the transformation \eq{12.10}, or the change of flat metrics
from {\scriptsize $\( \matrix{ 1 & 0 \cr 0 & 1 \cr} \)\,$} to 
$ f' f^{\prime *} ${\scriptsize $ \(\matrix{ 1 & 0 \cr 0 & 
1 \cr} \) $} \cite{ket}, may be bypassed here.

As \eq{12.10} is a pure coordinate transformation, i.e. the 
values $H(z, \ov{z}\,)$ of the fields remain unchanged. 
Infinitesimally ($\e\to0 \,$, $\, g(z)\,$ arbitrary) this means 
\bea{12.11} 
 z\to z' &=& z + \e\, g(z) \nonu \\
 H' (z' , \ov{z}') &=& H(z,\ov{z})
   = H(z' - \e g , \ov{z}' - \e \ov{g}) \nonu \\
 &=&  H(z', \ov{z}' ) - \e g \6 H - \e \ov{g} \, 
 \ov{\6} H \; \glr \; H(z', \ov{z}' ) + \d H \quad . \quad
\eea 
To demonstrate the conformal invariance of $ S[H]\, $ we have 
to show that
\bea{12.12} 
  S' - S &=& \int d^2 r' \; \cl L \lb H' (z' , \ov{z}' )
  \, , \, \6'  \, , \, \ov{\6}' \, \rb - S
  = \int d^2 r \; \cl L \lb H+ \d H \, , \, \6 
  \, , \, \ov{\6}\, \rb - S \nonu \\
  & & \hspace*{-20mm} =\, S\lk H+\d H\rk - S\lk H\rk 
  = \d S \quad \hbox{\ft with especially} \quad 
  \d H = - \e g \6 H - \e \ov{g} \ov{\6} H  \quad
\eea 
does vanish.
For a variation $\d S\,$ of $S\,$ \ --- \ due to whatsoever
$\d H$ \ --- \ we have the ready formula \eq{9.33}~:
$$ \d S = {1\0 \pi } \int \Sp \Big( \;\ph  \;\6 
           \ov{X} \,\Big) \qquad  \hbox{with} \qquad
  \lb \ph , X , \ov{X} \rb = H^{-1} 
  \lb \d , \6 , \ov{\6} \rb H \quad . \quad  \eqno{(9.33)}
$$  
Just $\d H\,$ from \eq{12.12}, i.e. $\ph = H^{-1}\d H
= -\e gX - \ov{g}\ov{X}\,$, has to be inserted here~:
\bea{12.13} 
  \d_{\rm confo}\;\, S &=& - {\e\0\pi} \int 
     \bigg[ g  \,\Sp \( X  \, \6 \ov{X} \) \; + \: 
        \ov{g} \,\Sp \( \ov{X} \, \6 \ov{X} \) \bigg]  
        \nonu \\
  &=& - {\e\0\pi} \int \bigg[ g\,\Sp \( X  
      \lk \ov{\6} X + \ov{X} X  - X \ov{X} \rk \) \; + \; 
      \ov{g} \, \Sp \( \ov{X} \6 \ov{X} \) \bigg]  
      \;\;\; \nonu \\
  &=& - {\e\02\pi} \int \bigg[ \ov{\6}\; g \;
       \Sp \( X X \)  \; + \;  \6\; \ov{g} \;
       \Sp \( \ov{X}\,\ov{X} \) \bigg] \;\; 
      = \;\; 0 \quad , \quad \hbox{q.e.d.} \quad                 
\eea 
In the second line, \eq{9.34}, i.e. $\ov{\6} X + \ov{X}X 
= \6\ov{X} + X \ov{X}\,$, was used, and $\,\ov{\6} g = \6 \ov{g} 
= 0\,$ in the third. The conformal invariance of $S\,$ is thus an 
immediate consequence of its functional differential equations.
One more reason for remembering the above $\d S\,$ formula \eq{9.33}
is realized next.

The WZW equations of motion follow from $\d S=0\,$ under independent
variation of the $n$ elements of the matrix $H\,$. They can be 
read off from \eq{9.33} as
\bea{12.14}  
  \6 \;\ov{X} = 0 \qquad & & \qquad \hbox{or} \;\;\;\qquad 
    \6 \, J^\dagger = 0 \quad \hbox{with} \quad 
    J^\dagger\, \gll \, {N\0\pi}\,H^{-1} \ov{\6} H \nonu \\
   & & \hbox{as well as} \qquad 
    \ov{\6}\, J = 0 \quad \hbox{with} \quad 
     J \;\gll \, {N\0\pi}\, (\6 H)  H^{-1} \quad . \quad 
\eea 
As \eq{12.14} exhibits equations of continuity, $J\,$ and 
$J^\dagger\,$ are matrix versions of the currents of the WZW model.

The currents $J^a$ as given in \eq{12.3} are now recognized
to be the components $J^a=2\,\Sp\(T^a J\)\,$ of its matrix 
version. By concatenation one obtains $J=T^a J^a\,$. But 
are we really allowed for expanding $J\,$, \eq{12.14}, into 
the traceless generators~? ~For this, $(\6H)H^{-1}\,$ has to be
traceless, too. To verify it, O. Lechtenfeld had the right
idea. Remembering \eq{5.8} or \eq{8.14}, we may use the 
representation $\, H = \rho^2 = e^{2\vcsm \o\vcsm T} \,
\glr\, e^{\vcsm \ph \vcsm T}\;$ to get 
\be{12.15} 
  \Sp \Big( (\6 H)H^{-1} \Big) = \Sp \( \int_0^1 ds \; 
   e^{s \vcsm \ph \vcsm T} (\6 \ph^a )T^a 
   e^{- s \vcsm \ph \vcsm T} \) =  (\6 \ph^a ) 
   \;\Sp \( T^a \) = 0 \qquad 
\ee 
indeed. 

Looking back to \eq{12.14}, one could ask, why not the other 
version $J^\dagger\,$, hence $J^{a*}={N\0\pi} 2\Sp\(T^a H^{-1} 
\ov{\6} H\)\,$, has been declared to be the WZW current. May one 
work with $\psi[J^{a*}]\,$ as well? Yes. Then one would derive the 
c.c. of the Hamiltonian of \S~13.2. But decide to work with
\ e i t h e r \ $\psi[J^a]\,$ \ o r \ $\psi[J^{a*}]\,$. This rule 
would be violated, for instance, by absorbing $e^{NS}\,$ into the 
wave functionals. 
\pagebreak[2]

\subsection{ End }
     
\nopagebreak[3]
Some first period of efforts ends up here. Just the first 
piece of physics has been grasped, namely the mass gap 
$m=e^2N/(2\pi)$. But its stability under the inclusion of the 
potential term is not yet studied. There is still no functional 
Hamiltonian, no ground state wave functional, no Wilson loop, 
no confinement, no application to the 4D gluon plasma. 

The reason for breaking the notes into two parts has a 
psychological background. For the first time feeling good 
with the matter, one finds reasons for relaxation \ --- \ more
than half a year, actually. So, have at least a nice weekend. 
First part, first run.


\newpage

\addtocontents{toc}{ $ $ \vskip 4mm}

\addtocontents{toc}{\hspace*{-8mm} 
\vrule width 5cm height .2pt depth 0pt 
 \hfill PART II \hfill \vrule width 5cm height .2pt depth 0pt 
\quad \vskip 2mm}

\noindent
\rule[2mm]{4cm}{1pt}
    \hfill {\Huge P A R T \ II} \hfill\rule[2mm]{4cm}{1pt} 

\vskip 22mm 
One year after ending up the first run. Meanwhile, there 
is a first attempt of using KKN's results for thermal 3+1~D 
physics \cite{tau}. In the static limit ($\o\to 0$), the 
transverse selfenergy could be identified with the squared mass 
gap $m^2 =e^4N^2/(4\pi^2)\,$ through $e^2=g^2T$. In other words, 
we (Reinbach and Schulz) verified this strong conjecture by 
ruling out potential other contributions at order $g^4\,$.
Regulators for the 3D Theorie were derived from its embedding in 
the 4D setup. One may state that, for the static selfenergy, 
Lindes infinite numerical series has been summed up. The relation
to 4D TFT is enforced in a recent report of Nair \cite{dubna} 
(conference at Dubna).

In the following second part of the notes, they reach the end 
of \cite{kkn}, in essence, and continue with KKN's paper 
\cite{nach}~: {\ft\sl On the vacuum wavefunction and string 
tension of Yang--Mills theories in (2+1) dimensions}. Equations 
in \cite{nach} are referred to as {\bf\{xx\}}.


\sec{ The full Hamilton in \boldmath$\psi [J]\,$ space } 

In the retrospect, at least, we become aware of the most
painful break in \S~12, namely as we were satisfied with 
the first rough look at the spectrum: $m$. \ A l l \ about 
2+1~D YM has to be worked out, and this ``all'' is contained 
in the functional Hamiltonian ${\bf H}=\int\!\cl H 
= \int\!\cl T + \int\!\cl V$ \ \ p l u s \ \ 
the associated exotic scalar product $\lw 1 | 2 \rw\,$. The latter 
is taken up again in \S~14. We shall continue with keeping distance 
to CFT. But let the Wilson--loop argument as detailed in \S~16.4 
give some confidence in the strange statement that the wave 
functionals $\psi$ become a finite norm through the indirect 
dependence on $H$ via $J^a={N\0\pi}2\Sp\(T^a (\6 H)H^{-1}\)\,$. 
\pagebreak[2]

\subsection{ Kinetic energy in \boldmath$\psi [J]\,$ space } 

\nopagebreak[3]
It is obvious how to overcome the extremely special wave 
functionals  $\psi = \int\! c^a J^a\,$ of \S~12.1. Apply the 
operator ${\bf T} = \int\! \cl T = -{e^2\02} \int\! \d^{a*} \d^a\,$ 
of the kinetic energy to a general functional $\psi [J^a]\,$. 
Thereby, {\bf T} will turn out as an expression in $J^a\,$ and 
$\d_{J^a}\,$, automatically. As the symbolic line
\be{13.1}
  \d^* \d \,\psi = \d^* \int (\d J)\; \d_J \psi
  = \int (\d^* \d J ) \;\d_J \psi \, + \,\int (\d J)
    \;\d^* \d_J \psi
\ee 
shows, two terms  will arse, which we denote by {\bf T}$_1$ 
and $({\bf T}_2+{\bf T}_3)$, respectively. KKN's headline
of \S~4 reads {\sl An expression for $\cl T$ in terms of currents}. 
But no, not ``An'', the one unique $J$ version of {\bf T} is in search. 
We overlook all the epsilontic care in KKN, find a direct way
to the more aesthetic final result \ek{4.13} $\equiv$ \ekk{17} 
$\equiv$ \eq{13.9} below, and enjoy giving details~: 
\bea{13.2}
  {\bf T}_1 \,\psi  [J^a] &=&  - {e^2\02} \int_r \, 
  \int_{r'} \( \d^{a*}_r \d^a_r J_{r'}^d \) \d_{J_{r'}^d} 
  \psi \, = \int_{r'} \( \d_{J_{r'}^d} \psi  \) {\bf T}\, 
  J_{r'}^d  \qquad \nonu \\[1mm]
  \hbox{(12.6) : \ \ } &=& m \int J^a \d_{J^a} \, \psi [J^a] 
  \qquad \folg \qquad {\bf T}_1 = m \int J^a \,\d_{J^a} 
  \quad . \quad
\eea 
This could have been guessed. For $({\bf T}_2+{\bf T}_3)$, however, 
such a quick tracing back is no more available, because after 
commuting the differentiations $\d_r^{a *}\,$ and 
$\d_{J_{r'}^d}\,$, 
\bea{13.3}
  \({\bf T}_2 + {\bf T}_3\) \,\psi [J^a] &=&  - {e^2\02} 
  \int_r \, \int_{r'} \( \d^a_r J_{r'}^d \) \d_{J_{r'}^d} 
  \;\d^{a*}_r \psi  [J^a] \nonu \\[-4.6mm]
  & & \hspace*{3.8cm} 
     \unt{1.52}{.2}{1.4}{\int_{r''} \( \d_{J_{r''}^c}
                \psi \) \, \d_r^{a*} J_{r''}^c } \\
  &=& - m\, i \int_r \( M^\dagger_r \)^{da} \d_{J^d_r} 
      \int_{r'} \( \d_{J_{r'}^c} \psi \) \, 
      \d_r^{a*} J_{r'}^c  \quad , \quad
\eea 
we become aware of the new combination $\,\d^{a*}_r J^c_{r'}\,$. 
For the round bracket in the first line, instead, we remembered 
\eq{12.5} and were led to the second line with ease. 

To study the object $\,\d^{a*}_r J^c_{r'}\,$ we turn to its 
c.c. and interchange the indices, for temporarily getting the 
familar Greens function $G_{r r'}$ (not $\ov{G}$) involved. 
At first we obtain
\bea{13.4}
   \d^a_{r'}\, J^{c*}_r &=& \,\d^a_{r'}\,{N\0\pi}
   2\,\Sp \( T^c H^{-1} \ov{\6} H \) \nonu \\
   &=& {N\0\pi}2\,\Sp\( T^c \Big[ - H^{-1} (\d^a_{r'} H)
   H^{-1} \ov{\6} H + H^{-1} \ov{\6} (\d^a_{r'} H) \Big] \) 
   \quad . \quad
\eea 
Remember that $\d^a_{r'} H = M^\dagger \d^a_{r'} M\,$. To 
continue with $\d^a_{r'} M\,$, only the adjoint version \eq{12.8} 
is available in the previous. So, let us obtain $\d^a_{r'} M\,$ 
anew, namely from $0=(\6+A)M$. Applying $\d^a_{r'}$ leads to the 
differential equation $(\6+A)(\d^a_{r'}M) = iT^a M_r \d_{r r'}\,$. 
Its solution is $\d^a_{r'}M = M_r G_{r r'} M_{r'}^{-1} i T^a 
M_{r'}\,$ and is easily verified (do \ n o t \ sum over $\vc r$ 
or $\vc r'\,$). This gives
\be{13.5}
  \d^a_{r'} H = i H G_{r r'} M_{r'}^{-1} T^a M_{r'} 
  = i H G_{r r'} T^b 2\,\Sp\( T^b  M_{r'}^{-1} T^a M_{r'} \)
  = H T^b \; i G_{r r'} M_{r'}^{ab} \;\; 
\ee 
under suppression of un--primed $r$ indices, as already in 
\eq{13.4}. Hence, the last $\ov{\6}$ in \eq{13.4} will produce 
two terms, one with $\ov{\6}H$ and one with $\ov{\6} G_{r r'}\;$:
\bea{13.6}
  \d^a_{r'}\, J^{c*}_r &=& {N\0\pi} 2\,\Sp\( \lk T^b, 
  T^c\rk H^{-1} \ov{\6}H \) i G_{r r'} M_{r'}^{ab} \,
   +\, {N\0\pi}\, \d^{bc} \,i\,\ov{\6} G_{r r'}
  M_{r'}^{ab}  \nonu \\
  &=&  M_{r'}^{ab} \lk - {1\0 \pi (\ov{z} - \ov{z}') }
       f^{bce} J_r^{e*} - {N\0\pi}\, i\, \d^{bc} 
       {1\0 \pi \,( \ov{z} - \ov{z}')^2} \rk \quad . \quad
\eea 
Somewhat dangerous happened in turning to the second line 
by inserting the un--embed\-ded Greens function $G_{r r'} 
= 1/ \lk \pi \,(\ov{z} - \ov{z}')\rk$ (and even operating
with $\ov{\6}= \6_{\ov{z}}$ on it). Ultimately, the corresponding
doubts will have gone in the Fourier subsection \S~13.3.
We now return to what we need in \eq{13.3}~:
\be{13.7}
  \d^{a*}_r\, J^c_{r'} = (M^{\dagger -1}_r)^{ab} 
  \lk {1\0 \pi (z-z')} f^{bce} J_{r'}^e + {N\0\pi}\, i\, 
  \d^{bc} {1\0 \pi \,(z-z')^2} \rk \quad , \quad
\ee 
because $M^{ab*}= (M^\dagger)^{ba} = (M^{\dagger -1})^{ab}\,$.
Clearly, the objects $M^\dagger$ will recombine in
\eq{13.3}. There we are. Just the unspecified $\psi[J]$ now might
be omitted on both sides~:
\be{13.8}
 {\bf T}_2 + {\bf T}_3 =  m \int_r  \int_{r'} \d_{J^b_r}
 \lk  {N\0\pi^2}\, { \d^{bc} \0 (z-z')^2}  - {i\0 \pi}{f^{bce} 
 \0  (z-z')} J^e_{r'} \rk  \d_{J_{r'}^c} \quad . \quad
\ee 
Doesn't it smell to CFT a bit? Due to the $f$ antisymmetry 
the functional derivative $\d_{J^b_r}$ may be shifted through
the square bracket. Thus, the functional operator of the
kinetic energy in the space of $\psi [J]$ is given by
\bea{13.9} \!
 {\bf T} \!&=& {\bf T}_1 + {\bf T}_2 + {\bf T}_3 \nonu \\
         \!&=&\!  m \int\!\! d^2r \, J^a \,\d_{J^a}
    + \; m \int\!\! d^2r \int\!\! d^2r' \,
   \lb \, {N\0\pi^2}\, { \d^{ab} \0 (z-z')^2} 
   - {i\0 \pi}{ f^{abc} \0  (z-z')} J^c_{r'} \rb   
   \d_{J^a_r} \d_{J_{r'}^b} \;\; . \qquad
\eea 
As already announced, the seemingly dangerously singular 
denominators will turn out to be harmless in \S~13.3. 
\eq{13.9} ist \ekk{17}. Doesn't the last term vanish due to
$f$ antisymmetry~? \ No, because $\vc r$, $\vc r'$ 
would have to be interchanged, too, under change of sign of 
the denominator.
\pagebreak[2]

\subsection{  Potential {\bf V} in \boldmath$\psi [J]$ space } 

\nopagebreak[3]
The term ${\bf V}=\int\! \cl V$ with $\cl V={1\02e^2} B^a B^a$
and $B^a=\6_1 A_2^a-\6_2 A_1^a + f^{abc} A_1^b A_2^c$ was
noticed long ago in \eq{2.9} and nearly forgotten afterwards. It 
is real and contains no functional derivatives. So, as it stands, 
it might turn out to be a functional of only $J$. Via 
$B^a \to A \to M \to H \to J$, where each arrow stands for
``eliminate in favour of'', it gets thrilling with the last two
arrows. For the first, one can clearly write
\bea{13.10}
 \cl V &=& {1\02e^2} B^a B^a = {1\0e^2}\,\Sp\(B B \)
 \quad , \quad B\gll B^a T^a = B^\dagger \nonu \\
 B &=& T^a \( \6_1 A_2^a - \6_2 A_1^a + f^{abc} A_1^b A_2^c \)
       = i\6_1 A_2 - i\6_2 A_1 + \lk A_1 , A_2 \rk \nonu \\
   &=& 2 \( \ov{\6} A - \6 \ov{A} + \lk \ov{A} , A \rk \)
       = 2 \Big\{\; (\ov{\6} + \ov{A}) A - (\6 +A)\ov{A} 
         \;\Big\} \quad ,
\eea 
where $\ov{A}\gll (A_1-iA_2)/2 = - A^\dagger\,$. Matrices $M$
come into play by $\,A=-(\6M)M^{-1}\,$ and $\,\ov{A}=-A^\dagger 
= M^{\dagger -1} \ov{\6} M^\dagger\,$. Here, it would be easier
to go from the result \eq{13.12} in backward direction. But we
like inventing so much~:
\bea{13.11}
  \( \ov{\6} + \ov{A} \)\, A 
 &=& \ov{\6} A + M^{\dagger-1} (\ov{\6} M^\dagger) A\, 
  = \,M^{\dagger-1}\,\ov{\6}\, M^\dagger A
  = - M^{\dagger-1} \,\ov{\6} \, M^\dagger (\6 M) M^{-1} 
     \nonu \\
 &=& - M^{\dagger-1}\, \ov{\6} \, (\6 H) H^{-1} M^\dagger 
     + M^{\dagger-1} \, \ov{\6}\, \6 \, M^\dagger \nonu \\
 &=& - M^{\dagger-1} \lk \ov{\6} \, (\6H)H^{-1} \rk M^\dagger
     \; + \; M^{\dagger-1} \lk \6 - (\6H)H^{-1} \rk 
       \ov{\6} M^\dagger   \nonu \\[-4.6mm]
 & & \hspace*{5.2cm} \unt{5.2}{.2}{5.1}{
   \raise 1pt\hbox{$\!$\scriptsize : second~term}} \nonu \\[-2mm]
 \hbox{second~term}  &=&  
   M^{\dagger-1} \lk \6 - (\6 M^\dagger) M^{\dagger-1}
   - M^\dagger (\6 M) M^{-1} M^{\dagger-1} \rk 
   M^\dagger \ov{A} \nonu \\
  &=&  \( \6 + A \) \, \ov{A} \qquad . \quad
\eea 
The field $B$ is now seen to be a sandwich with a delicious
$H$ matter in between,
\be{13.12}
 B = 2 \Big\{\; (\ov{\6} + \ov{A}) A - (\6 +A)\ov{A} 
     \;\Big\}  = - 2 M^{\dagger-1} \lk \ov{\6} \, 
     (\6H)H^{-1} \rk M^\dagger \quad , \quad
\ee 
and the dry bred is devoured by the trace~:
\bea{13.13}
   \cl V &=& {4\0e^2}\,\Sp\(
   \lk \ov{\6} \, (\6H)H^{-1} \rk \lk \ov{\6} \, 
     (\6H)H^{-1} \rk  \) \nonu \\
   &=& {2\0e^2}\; 2\,\Sp\( T^a 
     \lk \ov{\6} \,(\6H)H^{-1} \rk \) 
               \; 2\,\Sp\( T^a
     \lk \ov{\6} \,(\6H)H^{-1} \rk \)  \nonu \\
   &=& {2\0e^2}\; \( \ov{\6} {\pi \0N} J^a \)
       \; \ov{\6} {\pi \0N} J^a \;\; = \;\; {\pi\0 m\,N}
       \,\( \ov{\6} J^a \) \, \ov{\6} J^a \quad . \quad
\eea 
{\sl De--concatenation} has led to the second line.
\eq{13.13} is \ek{2.45}$\,\equiv\,$\ekk{18}. Other than the
kinetic term, the potential energy density goes as $\sim1/m\,$. 
Remember that $\cl V$ is real. Had we started from $\cl V
= \Sp\(B^\dagger B^\dagger\)$, then 
$\cl V = {\pi\0mN} \(\6 J^{a*}\) \6 J^{a*}$ would have been
obtained, which is valid too. On the other hand, $\6 J^{a*} 
\neq \ov{\6} J^a$. In fact, for $\cl V$ to be real, the summation 
over $a$ is required, or the trace in the above first line, 
respectively.

Thus, the functional Hamiltonian {\bf H} in the space of 
$\psi [J]$ is given by
\bea{13.14} \hspace*{-6mm}
 \fbox{\parbox[b]{13.6cm}{
 \bean
  {\bf H} = {\bf T} + {\bf V} &=&
     m \int\!\! d^2r \; J^a \,\d_{J^a}  \nonu \\
   & & \hspace*{-4mm} + \; m \int\!\! d^2r \int\!\! d^2r' \,
   \lb \, {N\0\pi^2}\, { \d^{ab} \0 (z-z')^2} 
   - {i\0 \pi}{ f^{abc} \0  (z-z')} J^c_{r'} \rb   
   \d_{J^a_r} \d_{J_{r'}^b}  \nonu \\
   & & \hspace*{-4mm}  + \;{\pi\0m\,N\,} \int\!\! d^2r \;
       \( \ov{\6} J^a \) \, \ov{\6} J^a \quad \nonu
 \eea} } & & \nonu \\[-12mm] & & 
\eea 

\vskip 3mm
By a framed picture one is invited contemplating. What is seen
here? Keep your thumb over the $f^{abc}$ term. Then, a quadratic
form of variables $J$ and derivatives $\d_J$ is noticed.
Omitting integrals and color indices, \eq{13.14} would be a harmonic
oscillator involving a $x\6_x$ term. But the latter could be 
removed by a gauge transformation $e^{\a x^2}\,$. Remove the
thumb. Now, the framed \eq{13.14} shows the Hamiltonian of a system
of infinitely many (field theory!) coupled oscillators with
the $f^{abc}$ term as its only nonlinearity. The $f$ term 
``rotates with strength $J\,$''. One also realizes how {\bf H} 
reduces to the first term, if it is applied to a single 
$J^a(\vc r)$ and if by e.g. $m\to\infty$ the last term is removed.  
${\bf T} J^a = m J^a$ was really special.

Some partial decoupling of the oscillators may be expected
from Fourier transformation~: \S~13.3. One may also suspect
that \eq{13.14} could need some normal ordering prescription
due to oscillator zero--point energies summed up~: \S~15. 
\pagebreak[2]
 
\subsection{ Fourier transform } 
 
\nopagebreak[4]
The following longing for harmony goes a bit ahead over
the KKN material. Fourier transfor\-ma\-tion
\be{13.15}
  J^a(\vc r)= \int\! {d^2k\0(2\pi)^2} \; 
  e^{i\vcsm k \vcsm r} \,\schl J^a (\vc k) \quad , \quad
  \schl J^a(\vc k)= \int\! d^2r \; 
  e^{-i\vcsm k \vcsm r} \, J^a (\vc r) \quad , \quad
\ee 
must be anyhow good to the above Hamiltonian, in particular
to the convolution integral ($N\d^{ab}$ term). With view
to the potential term, even the field
\be{13.16} \!\!\!
  I^a (\vc r) \;\gll\, \ov{\6} J^a(\vc r)
    \quad , \quad  \schl J^a (\vc k) 
  = {2\0i}\; {1\0 k_1 - i k_2 }\;  \schl I^a (\vc k)
  \quad , \quad
\ee 
comes into mind, where the second relation rests on
$\,\schl I^a (\vc k) = \int\! d^2r \; e^{-i\vcsm k \vcsm r}
 \; \ov{\6} \; \int\! {d^2q\0(2\pi)^2} \; $ $ 
  e^{i\vcsm q \vcsm r} \,\schl J^a (\vc q)
 = {i\02} (k_1 - i k_2) \schl J^a (\vc k)\, $.
Using \eq{13.15} (right relation) the chain rule of 
differentiation reads
\be{13.17}
  \d_{J^a(\vcsm r)} = \int\! d^2k \; \( \d_{J^a(\vcsm r)} 
  J^b (\vc k ) \) \; \d_{\schl J^b( \vcsm k )} 
   = \int\! d^2k \; e^{-i\vcsm k \vcsm r}\,
    \d_{\schl J^a (\vcsm k)} \quad . \quad
\ee 
The ``wrong'' sign is OK, and the ``missing of $\pi$'s'' as 
well. The factor between $\schl J$ and $\schl I$ drops out in
$\schl J \,\d_{\schl J}\,$. Hence, so far, two of the four $H$ 
terms can be booked down~:
\be{13.18}
  {\bf T}_1 = m \int\! d^2k \,\; \schl I^a (\vc k ) \, 
  \d_{\schl I^a (\vcsm k )} \;\quad \hbox{and} \;\quad
  {\bf V} = {1\0 4\pi m N} \int \! d^2k \,\; 
  \schl I^a (\vc k ) \, \schl I^a (-\vc k )  
  \quad . \quad
\ee 
The convolution integral 
\bea{13.19}
 {\bf T}_2 = {m N\0 \pi^2} \int_r\int_{r'} \d_{J^a_r}
 \; {1 \0 (z-z')^2} \; \d_{J^a_{r'}}
  \!&=&\! 4mN \!\int\! d^2k \,\; \schl \o (\vc k) \,\;
 \d_{\schl J^a (\vcsm k)} \d_{\schl J^a (-\vcsm k)} 
      \nonu \\
  &=&\! m N \!\int\! d^2k \,\; \schl \o (\vc k) \,
      (k_1-ik_2)^2 \; \d_{\schl I^a (\vcsm k)} 
      \d_{\schl I^a (-\vcsm k)} \qquad\qquad
\eea 
involves the Fourier transform $\schl \o$ of $1/z^2\,$,
and the awkward term ${\bf T}_3$ is infected with
the Fourier transform $\schl \kappa$ of $1/z$~:
\bea{13.20}
 {\bf T}_3 &=& \;\ldots\; \hbox{\ft (some labour)} \; 
  = -i \,{m\0\pi} f^{abc} \int\! d^2q  \int\! d^2p \,\;
    \schl \kappa (\vc q ) \,\schl J^c ( \vc p + \vc q )
    \; \d_{\schl J^a (\vcsm q)} \d_{\schl J^b (\vcsm p)} 
    \nonu \\
 &=& {m\02\pi} \, f^{abc} \!\int\!\! d^2q \!\int\!\! d^2p 
     \,\; \schl \kappa (\vc q ) 
    {(q_1-iq_2)\,(p_1-ip_2) \0 (q_1-iq_2) + (p_1-ip_2)}
    \,\schl I^c ( \vc p + \vc q ) \; \d_{\schl I^a 
    (\vcsm q)} \d_{\schl I^b (\vcsm p)} \;\; . \quad\qquad
\eea 

To complete the Fourier version, there remains some integration
work on $\schl \o$ and $\schl \kappa$. A reader, who likes 
pitfalls with contour integrations, might evaluate $\schl \o$
in kartesian coordinates. But the harmless way uses polar
coordinates (even for $\vc k$ by writing it as $\lk k\cos(\a)\,
,\, k\sin(\a)\rk$) and shows the absence of a singularity 
immediately~:
\bea{13.21}
  \schl \o (\vc k) &=& \int\! d^2r\; {1\0 z^2} \,
    e^{-i\vcsm k\vcsm r}  = \int_0^\infty\! dr\; r 
  \int_{(2\pi)} \! d\ph \; { e^{i2\ph} \0 r^2} \, 
    e^{-ik_1r\cos(\ph) - ik_2r \sin(\ph)} \nonu \\
  &=& \int_0^\infty\! dr\; {1\0r}  \int_{(2\pi)} \! 
      d\ph \;  e^{i \lk 2\ph - kr \cos(\ph-\a)\rk}  
      \quad , \quad \ph \to \ph + \a - \pi/2 \nonu \\
  &=& - e^{i2\a} \int_0^\infty\! dr\; {1\0r}  
      \int_{(2\pi)}\! d\ph \;  e^{i \lk 2\ph - kr \sin(\ph)\rk}  
        \nonu \\
  &=& - e^{i2\a} \int_0^\infty\! dr\; {1\0r} \,\Im \quad 
        \hbox{with} \quad
       \Im = \int_{-\pi}^\pi \! d\ph\; \cos \(2\ph - kr \sin(\ph)\)  
           = 2\pi J_2(kr) 
       \nonu \\
  &=& - e^{i2\a} \, \pi \; = \;\, - \, {k_1+ik_2 \0 k_1-ik_2} 
      \;\pi \quad , \quad
\eea 
where for $\int_0^\infty\! dr\; {1\0r}\,\Im = 2\pi \int_0^\infty\! 
dr\; {1\0r} J_2(r) = \pi$ Abramovitz \cite{abra} was consulted
under No. {\bf 11.4.16}, and for the Bessel function under No.
{\bf 9.1.21}. Finally, to evaluate $\schl \kappa$, the first 
steps can be read off from \eq{13.21} (one $e^{i\ph}$ less, 
one $1/r$ less)~:
\bea{13.22}
  \schl \kappa (\vc k) &=& \int\! d^2r\; {1\0 z} \,
    e^{-i\vcsm k\vcsm r}  = \int_0^\infty\! dr 
    \int_{(2\pi)} \! d\ph \; 
    e^{i \lk \ph - kr \cos(\ph-\a)\rk} 
  = - i e^{i\a} \int_0^\infty\!\! dr\; 2\pi J_1(kr) \nonu \\  
  &=& {k_1+ik_2 \0 i\,k} \; {2\pi\0k} \,\; = \,\;
      {2\pi \0 i} \, {1\0k_1-ik_2} \quad . \quad
\eea 

Herewith the details are completed for the translation of ${\bf H}$ 
into $\schl I$ language. In the space of wave functionals 
$\psi [\schl I^a(\vc k)]\,$, the Hamiltonian reads
\bea{13.23}
 {\bf H}_{\rm Fourier}
  &=&    m \int\! d^2k \,\;\schl I^a (\vc k ) \, 
  \d_{\schl I^a (\vcsm k )} \; - \;
   \pi m N \!\int\! d^2k \; k^2 \;\d_{\schl I^a (\vcsm k)} 
   \d_{\schl I^a (-\vcsm k)} \nonu \\
 & & -\; i\, m \!\int\! d^2q \!\int\! d^2p  \,\; 
    {( p_1-ip_2 )\;\;\schl I^c ( \vc p + \vc q ) \0 
    (q_1-iq_2) + (p_1-ip_2)} \,\; f^{cab} \; \d_{\schl I^a 
    (\vcsm q)} \d_{\schl I^b (\vcsm p)}   \qquad \nonu \\
 & & + \;{1\0 4\pi m N} \int \! d^2k \,\; \schl I^a (\vc k ) 
       \, \schl I^a (-\vc k ) \quad . \quad
\eea 
\vskip -3.6cm
\vrule width .2pt height 3.4cm depth 0pt \hspace{-.7mm}
\vrule width .2pt height 3.4cm depth 0pt \, \\
The suspected danger concerning singularities is banned. One can, 
of course, split off a factor $k=|\vc k|$ from $\schl I^a(\vc k)\,$.
Thereby the factor $k^2$ in the second term is removed, but it
appears in the last one, instead. The nonlinear $f$ term has remained 
as terrible as before. The Fourier version \eq{13.23} will develop
some use in \S~15.


\sec{ More on the scalar product } 

If the Hamiltonian is understood as one half of the truth
on the $J$ quantum mechanics, then the exotic scalar product
\be{14.1}
  \lw 1 | 2 \rw = \int\! d\mu(\cl C)\; \psi_1^* \,\psi_2
  = \s^n \!\!\int\! d\mu(\cl H) \; e^{2NS} \;\psi_1^* \,\psi_2
  \,\equiv\, \s^n \!\!\int\! [dh^a] \; e^{2NS} 
     \;\psi_1^* \,\psi_2 \quad . \quad 
\ee 
is the other half. The energy expectation value
\be{14.2}
  \lw \, {\bf H} \, \rw_{\psi} = \int\! d\mu(\cl C) 
  \;\psi^* \, {\bf H} \,\psi \qquad 
\ee 
must be real. But how to show this? In \S~14.4 it will turn
out, indeed, that {\bf H} is ``$d\mu(\cl C)$--hermitean'' 
(and that $d\mu(\cl C)$ is real). Obviously, we might learn
a bit more on making use of $\lw 1 | 2 \rw\,$. For this,
we must step down on the ladder
~$ \ph^a(\vc r)         \;$, 
~$ h^a (\vc r)          \;$, 
~$ h^a_{\rm KKN}(\vc r) \;$, 
~$ J^a  (\vc r)         \;$, 
~$ \schl I^a (\vc k)    \;$ 
of functional variables, where the ``lowest'' level is noticed at 
the left end, referring to $H=e^{\ph^a T^a}\,$. 

Stepping down, we remember four questions left open in part I. 
Question 1 concerns KKN's differential variables $dh^a_{\rm KKN}\,$ 
(see \eq{14.5} below), by which one leaves the space of hermitean 
matrices $H\,$. This question will be answered immediately in the 
next subsection. Questions 2 to 4 are related to CFT. They will be 
bypassed by exploiting the Wilson loop argument \ --- \ though, 
admittedly, arguing is not a very convincing behave. Question 2
was raised in the footnote below \eq{9.1}. It concerns the
definite (?) expression $\s$ as given by KKN~:
\be{14.3}
  \hbox{\ek{2.20} \ : \quad } \quad 
  \det (D D^\dagger ) = \s^n e^{2NS} \qquad \hbox{with} 
  \qquad
  \s \quad\ueb{\raise 1mm\hbox{\large\bf ?}}{=}\quad 
  {\det' (-\6\ov{\6}) \0 \int\! d^2r}
\ee 
Perhaps, the problem merely is how to extract \eq{14.3} from
the material given in ref. 6 of KKN. Question 3 concerns the
verification of
\be{14.4}
  \hbox{\ek{2.26} \ : \quad } \qquad   
   \s^n\! \int \! d\mu (\cl H) \, e^{2NS} 
   \quad\ueb{\raise 1mm\hbox{\large\bf ?}}{=}\quad 1 
  \qquad\quad 
\ee 
and could be an even less trivial problem of extraction, this 
time from \cite{gaw,bn}. Help~! ~But note that $\s$ is some real 
number, not involving functional variables. It could be absorbed 
into the measure $d\mu ({\cl H})$ or in $\psi\,$, as e.g. done in 
\cite{bn}. Missing $\s$ anywhere, it is absorbed anyhow. Qestion 
4 is the call for reasoning the fact that normalisable wave 
functionals depend on $J^a$, see \eq{12.3} and comments below 
there. Now, here is our bypassing of the questions 2 to 4 by hand 
waving . If the Wilson loop argument (see also \eq{16.21} and text 
around it) makes one to believe in the normalisability of 
functionals $\psi[J^a]$, namely by $\s^n\! \int \! d\mu (\cl H) \, 
e^{2NS} \psi^* \psi\,$, then we are allowed to set $\psi \equiv 1$, 
in particular, thus arriving at the l.h.s. of \eq{14.4}. As this 
l.h.s. is finite, \eq{14.4} may be understood to define $\s$. But 
once $\s$ is fixed this way, the question {14.3} is rather 
irrelevant.
\pagebreak[2]

\subsection{ The Haar measure was right } 
 
\nopagebreak[3]
May it happen that, doing it wrong, the result is right?
Sometimes. In \S~8, around \eq{8.7}, we insisted in varying
$H$ by $H + \d H = H^{1/2} \( 1 +  dh^a \ov{T}^a \) H^{1/2}$
since thereby staying within the space of hermitean matrices.
Correspondingly, our integration measure turned out to be
$\,d\mu (\cl H) = \prod_{\vcsm r, a} dh^a \,\glr\, [dh^a]\;$ 
with
\be{14.5}
  dh^a = 2\,\Sp \big(T^a H^{-1/2} dH H^{-1/2} \big)
  \quad \hbox{\LARGE .} \qquad
  dh^a_{\rm KKN} = 2\,\Sp \big(T^a H^{-1} dH \big) \quad ,
\ee 
on the other hand, is what KKN claim (below \ek{2.15}) to be 
correct and what they call the {\sl Haar measure for hermitean 
matrix--valued filelds}. The only possibility, for the two
statements to differ in results, is a Jacobian $\neq 1$ of
the change from variables $h^a_{\rm KKN}$ to $h^a$. The
corresponding Jacobi matrix $\Im^{ab}$ is obtained from
\be{14.6}
  dh^a_{\rm KKN} = \Im^{ab}\, dh^b \quad , \quad \hbox{i.e. from} 
  \quad \Sp\(T^a H^{-1} dH \) = \Im^{ab} \,\Sp\(
  T^b \rho^{-1} dH \rho^{-1} \) \quad .
\ee 
Remember that $H=\rho^2\,$. With concatenation in mind we may 
guess the solution~:
\be{14.7}
  \Im^{ab} = 2\,\Sp\( \rho T^a \rho^{-1} T^b \)
  = 2\,\Sp\( T^b \rho T^a \rho^{-1}  \) \glr \rho^{ba}
  = (\rho^T)^{ab} \quad . \quad
\ee 
To the right we recognise some adjoint version of $\rho$ (cf. 
$H^{ab}$ below \eq{11.10}). Since $\det(\rho)=\det(\rho^T)=1$ we 
get optimistic to have this property even adjointly. But how to 
show. One night, N. Dragon comes through the door and knows how~: 
\bea{14.8}  
  \rho &=& e^{\vcsm \o \vcsm T} \;\; , \quad \rho T^a \rho^{-1} =
  T^a + \lk \vc \o \vc T , T^a \rk + \;\ldots\; = \;\ldots\; 
  =\;  \( e^{i\o^c f^{c \, . \, .} } \)^{ab} T^b \nonu \\
  \Im^{ab} &=& 2\,\Sp\big(  \big( e^{...} \big)^{ad} T^d T^b \big)  
           = \big( e^{...} \big)^{ab} \nonu \\
  \det(\Im ) &=& e^{{\rm Sp}\( \ln (e^{...})\)} 
             = e^{i\o^c f^{cdd}}
             = e^0 = 1 \qquad \hbox{and hence} \nonu \\
    & & \hspace*{2cm}
   \int\! \lk dh^a \rk \; \ldots \; 
    = \; \int\! \lk dh^a_{\rm KKN} \rk \; \ldots 
   \quad . \quad
\eea 
Nice, all are right. May be, it can be seen more directly. Just
looking at \eq{14.5}, M. Flohr realised it to be a change to new 
generators $\rho^{-1} T^a \rho$. Now obtaining a unit Jacobian, 
he said, one could expect. 
\pagebreak[2]

\subsection{ Variables \boldmath$\ph^a$ } 
     
\nopagebreak[3]
A pedestrian (what does it mean?) likes the elementary tools
in favour of stability and explicity, and he has all time of the 
world for going ahead. Clearly, he favours the ``lowest'' type 
$\ph^a$ of functional variables. They are defined by  
$H=e^{\vcsm \ph \vcsm T}\,$, hence real, found already in 
\ek{2.15} and related to our former parameters $\o^a$ 
($\rho=e^{\vcsm \o \vcsm T}\,$) by $\ph^a=2\,\o^a\,$, simply. 
The linear relation between $d\ph^a$ and $dh^a$ defines
a Jacobian matrix. But $dh^a$ comes in two versions, giving
two linear relations, namely
\be{14.9}
  \hbox{OURS} \quad dh^a = s^{ab} \, d\ph^b \qquad\quad
  \hbox{and KKN's} \quad dh^a_{\rm KKN} =  d\ph^b \, r^{ba} 
  \quad . \quad
\ee 
Using \eq{14.5} and \eq{8.14} the matrices $s\,$ and $r$
become explicit~:
\be{14.10} 
  s^{ab} = \int_{-1}^1 \!\! ds \; \Sp \( T^a 
  e^{s\vcsm \o \vcsm T } T^b e^{-s\vcsm \o \vcsm T } \) 
   \quad , \quad 
  r^{ab} = \int_0^1\!\! ds \; 2\,\Sp\( T^a
   e^{s\vcsm \ph \vcsm T} T^b e^{-s\vcsm \ph \vcsm T } \)
  \quad . \quad
\ee 
Due to \eq{14.6}, i.e. $dh^a_{\rm KKN} = \Im^{ab}\, dh^b\,$,
and $\Im^{ab}=\rho^{ba}$, they are related by
\be{14.11}
  r^{ab} \; = \;  \rho^{cb} \, s^{ca}\; = \; s^{ac} 
  \rho^{cb} \quad , \quad
\ee 
because $s^{ab}=s^{ba}\,$. Perhaps \eq{14.11} is worth to be
checked (it works so well)~:
\bea{14.12}
  r^{ab} &=& \int_0^1 \! ds \; 2\,\Sp \( T^a 
  e^{(2s-1) \vcsm \o \vcsm T }  e^{\vcsm \o \vcsm T }
  T^b e^{-\vcsm \o \vcsm T} e^{(1-2s) \vcsm \o \vcsm T }
   \) \quad , \quad 2s-1 = s'  \nonu \\
  &=& \int_{-1}^1\! ds'\; \Sp\( 
      e^{-s' \vcsm \o \vcsm T } T^a 
      e^{s' \vcsm \o \vcsm T } \;   
      e^{\vcsm \o \vcsm T} T^b e^{-\vcsm \o \vcsm T }
      \) \nonu \\        
  &=& \int_{-1}^1\! ds'\; \Sp\(  
      e^{-s' \vcsm \o \vcsm T} T^a 
      e^{s' \vcsm \o \vcsm T } T^c \) 2\,\Sp\(
      T^c \rho T^b \rho^{-1} \)
       \quad \hbox{, \quad q.e.d.} \quad
\eea 
Organized a bit different the above calculation would have
led to $r^{ab}= \rho^{ac} s^{cb}\;$: $s$ and $\rho\,$ do commute,
indeed. The matrix $s^{ab}$ is symmetric and real, while $r$ and 
$\rho$ are only hermitean. But note that
\be{14.13}
  \det \( r \) \; = \; \det \( s \) \; = \;\; 
  \hbox{real}\qquad \folg \quad d\mu(\cl C) \;\; = \;\; 
  \hbox{real} \quad . \quad  
\ee 
Admittedly, $r$ is the more relevant matrix, as is seen shortly.
\pagebreak[2]
   
\subsection{ Hermitean adjoint with respect 
             to \boldmath$d\mu(\cl H )$} 

\nopagebreak[3]
Some sense behind the above stepping back to the ``lowest''
variable becomes obvious, when now asking for the meaning of a 
dagger in our functional quantum mechanics \ --- \ dagger with 
respect to which integration measure? Given a functional operator 
$Q$, then let its adjoint operator $Q^\dagger$ with respect to 
$d\mu(\cl H ) = [dh^a]$ be defined by
\be{14.14}
 \int \! d\mu(\cl H) \; ( Q\chi_1)^* \chi_2 \;\glr\; 
 \int \! d\mu(\cl H) \; \chi_1^* Q^\dagger \chi_2   
 \quad . \quad
\ee 
To determine $Q^\dagger\,$, the relation $[dh^a] 
= [d\ph^a]\,\det(r)$ will help doing partial integrations with 
$\d_{\ph^a}\,$. It looks bad. So, we might be satisfied with
just one example for an operator $Q\,$. Aside, we now enter 
KKN's \S~3.3~: ``{\sl $p$, $\ov{p}$ as adjoints for $d\mu(H)$}''.

Let the given ($n$--fold) operator $Q$ be
\be{14.15}
  p^a \gll -i (r^{-1})^{ab} \, \d_{\ph^b} \quad
  = \quad \6 (M^{-1})^{ac} \d^c \;= \; \eq{11.7} 
  \quad . \quad
\ee 
So far, the functional operator $p^a$, merely played a minor role
in \S~11.3. Here it is revalued. To derive the left expression
in \eq{14.15} from the known right one, show at first that
\be{14.16}
  \,p^a_r \, H_{r'} \, = \, -i\, H\, T^a 
  \,\d(\vc r - \vc r')\, \qquad
\ee 
by using $\d^cM_{r'}^\dagger=0$ and $\d^{c} M_{r'}$ as
given in the text below \eq{13.4}. The commutator version
of \eq{14.16} is \ek{2.39}. Now, using the ansatz 
$\,p^a=X^{ab} \d_{\ph^b}\,$ (as well as $H=e^{\vcsm \ph \vcsm T}$), 
one obtains the matrix $X$ as $X^{ab}= -i (r^{-1})^{ab}\,$. 

The with--respect--to--$d\mu(\cl H)$ adjoint operator of $p^a\,$,
as claimed by KKN in \ek{3.12}, is $\,\ov{p}^a \gll - p^{a*}
= -i(r^{*-1})^{ab}\d_{\ph^b}\,$. We 
verify\footnote{There is an infinity of $\det (r)$'s in
the functional integrations. In \eq{14.17}, only the one is made 
explicit, which carries the same spatial index as $\d_{\ph^b}$}~:
\bea{14.17}
 \int\! [dh^a] \; \(p^a \chi_1 \)^* \, \chi_2 
 &=& \int \! [d\ph^a] \; \det(r) \, \( - i (r^{-1})^{ab}
     \, \d_{\ph^b} \,\chi_1 \)^* \, \chi_2  \nonu \\
 & & \hspace*{-4cm} = \;
     \int \! [d\ph^a] \; \det(r) \, i \, (r^{*-1})^{ab} 
     \,\chi_2 \,\d_{\ph^b} \,\chi_1^*  
  = \int \! [d\ph^a] \; \chi_1^* \, \d_{\ph^b} \,
  \det(r) \,(-i)\,  (r^{*-1})^{ab} \, \chi_2 \nonu \\
 & & \hspace*{-4cm} = \;
     \int \! [d\ph^a] \;  \det(r) \, \chi_1^* \, 
        (-i)\,(r^{*-1})^{ab}\,\d_{\ph^b} \, \chi_2 \;\;
    -i \int \! [d\ph^a] \; \chi_1^* \, \chi_2 \; 
       \d_{\ph^b} \, \det(r) \, (r^{*-1})^{ab} \nonu \\
 & & \hspace*{-4cm} = \;
     \int \! [dh^a] \; \, \chi_1^* \;\; \ov{p}^a \;\chi_2 
     \;\, -i \int \! [d\ph^a] \; \chi_1^* \;\, {\cl O}^{a *} 
     \;\chi_2  \quad \hbox{with}\quad
\eea 
\be{14.18}
 {\cl O}^a = \; \d_{\ph^b} \, \det(r) \, (r^{-1})^{ab}
  = \lim_{\vcsm r' \to \vcsm r} \, \cl P^a \quad , \quad   
  \cl P^a = \d_{\ph^b_{r'}} \, \det(r) \, (r^{-1})^{ab}
  \quad . \quad
\ee 
If KKN`s statement is right, $\cl O^a$ must vanish. We question, 
whether the limit in \eq{14.18} needs care towards regularization
(as e.g. KKN take), or whether perhaps the following quick
way is sufficient, which obtains $\cl P^a=0$ even before the limit
$\vc r'\!\to\!\vc r$ is performed. Let's see. Could--be
dangerous steps carry a $\gef$ mark~:
\bea{14.19}
 {\cl P}^a &=&  (r^{-1})^{ab} \,\d_{\ph^b_{r'}} 
       \,\det(r) \,  + \det(r) \, \d_{\ph^b_{r'}} \, 
      (r^{-1})^{ab} \nonu \\
  &\ueb{\gef}{=}&  \det(r) \lk (r^{-1})^{ab} \d_{\ph^b_{r'}}
       \,\ln \(\det(r)\) \, + \, \d_{\ph^b_{r'}} \, 
      (r^{-1})^{ab} \rk \nonu \\[-3.4mm]
  & & \hspace*{2.56cm} \unt{2.56}{.5}{.2}{\d_{\ph^b_{r'}}
       \,\Sp\( \,\ln(r)\,\) \;   = \;\Sp \big( 
       r^{-1} \d_{\ph^b_{r'}} r \big) \;\;\;\gef\gef }
       \nonu \\[1mm]
  &=&  \det(r) \, (r^{-1})^{ad} \, (r^{-1})^{cb} \, \lb
       \,\d_{\ph^d_{r'}} r^{bc} -
       \d_{\ph^b_{r'}} r^{dc} \,\rb \quad . \quad 
\eea 
Given this intermediate result, the desired number zero is
derived without further danger~:
\bea{14.20}
 \d_{\ph^d_{r'}} r^{bc} &=& \d_{\ph^d_{r'}}
  \int_0^1\! ds\; 2\,\Sp \( b\, e^s c\, e^{-s} \) \quad\qquad 
  \raise 2mm \vtop{\hbox{\ft the short--hand 
     notation} \vskip -2mm \hbox{\ft might be obvious 
     from \eq{14.10}}} \nonu \\
  &=&\!\int_0^1\!\! ds \!\int_0^1\!\! dt\, s \, 2\,\Sp\( 
       c\, e^{-s} b\, e^{ts} d\, e^{(1-t)s}
       - b\, e^s c\, e^{-ts} d\, e^{-(1-t)s} \) \,
       \d (\vc r' - \vc r ) \nonu \\
  &=&\!\d (\vc r' - \vc r ) \int_0^1\!\! ds 
      \!\int_0^s\!\! dt \, 2\,\Sp\( 
       c\, e^{-s} b\, e^t d\, e^{s-t}
       - b\, e^s c\, e^{-t} d\, e^{t-s} \)  \nonu \\
  &=&\!\d (\vc r' - \vc r ) \int_0^1\!\! ds 
      \!\int_0^s\!\! dt \; 2\,\Sp\( 
       c\, e^{-s} b\, e^s\, e^{-t} d\, e^t
       - \, e^{-t} d\, e^t \, e^{-s} b\, e^s c \) 
      \quad , \qquad
\eea 
where the first term in the last line was obtained
through $t\to s-t$, leaving the borders of $t$--integration
unchanged. The other expression in the wavy bracket of
\eq{14.19}, namely $-\d_{\ph^b_{r'}} r^{dc}$, equals
minus the first one through $b\gdw d$. It can be further 
converted by $\int_0^1\!\! ds \!\int_0^s\!\! dt 
= \int_0^1\!\! dt \!\int_t^1\!\! ds$ and $s \gdw t\,$, finally~:
\bea{14.21}
  - \d_{\ph^b_{r'}} r^{dc} 
  &=& \d (\vc r' - \vc r ) \int_0^1\!\! ds 
      \!\int_0^s\!\! dt \; 2\,\Sp\( 
        e^{-t} b\, e^t \, e^{-s} d\, e^s c 
       - c\, e^{-s} d\, e^s\, e^{-t} b\, e^t \) \nonu \\
  &=& \d (\vc r' - \vc r ) \int_0^1\!\! ds 
      \!\int_s^1\!\! dt \; 2\,\Sp\( 
         c\, e^{-s} b\, e^s \, e^{-t} d\, e^t 
       -  e^{-t} d\, e^t\, e^{-s} b\, e^s c \) 
      \quad . \qquad
\eea 
Constituting the wavy bracket $\lb \,\d_{\ph^d_{r'}} 
r^{bc} - \d_{\ph^b_{r'}} r^{dc} \,\rb\,$, two pieces
of the $t$ integral fit together~:
\bea{14.22}
 \lb \phantom{AA} \rb &=& \d (\vc r' - \vc r )  
   \int_0^1\!\! ds \!\int_0^1\!\! dt \; 2\,\Sp\(
   \lk c\, e^{-s} b\, e^s - e^{-s} b\, e^s c \rk 
    \; e^{-t} d\, e^t \,\) \nonu \\
  &=& \d (\vc r' - \vc r )  
   \int_0^1\!\! ds \!\int_0^1\!\! dt \; 2\,\Sp\(
    e^{-s} b\, e^s \lk h , c \rk \) \,2\,
   \Sp\(  h\, e^{-t} d\, e^t \,\) \;\; , \;\;
    \lk c , h \rk = i f^{hcj} j \nonu \\[1mm]
  &=& \d (\vc r' - \vc r ) \, 
       i \, f^{hcj} \; r^{bj} \;\, r^{dh} \quad . \quad 
\eea 
Inserting this into \eq{14.19} we arrive at
\be{14.23}
  \cl P^a =  \,\det(r) \, \d (\vc r' - \vc r ) \, 
  i \, f^{hcj} \;  \d^{ah} \, \d^{cj} \; 
  \sim \; f^{acc} \; = \; 0 
  \quad ,\quad
\ee 
as announced. Never perform $t \to s-t$ in the wrong term
or de--concatenation with an unfortunate content of trace.
Admittedly, we could not understand one detail in KKN, namely
to get \ek{3.14} obvious immediately. But it is correct, this
damned equation. To summarize,
\be{14.24}
  \int \! [dh^a] \; \, \chi_1^* \;\; \ov{p}^a \;\chi_2
  \; =\;  \int\! [dh^a] \; \(p^a \chi_1 \)^* \, \chi_2 
  \quad . \quad
\ee 
\pagebreak[2]

\subsection{ {\bf H} is self--adjoint with respect to 
             \boldmath$d\mu(\cl C )$} 
 
\nopagebreak[3]
What else?! The headline \ m u s t \ be valid, because all
the former treatment was sound and stable. Hence,
$\lw{\bf H}\rw\,$, \eq{14.2}, is real. End. However, to verify the
headline, is also a very crucial test. So, for confidence, let us
show that 
\be{14.25}
  \lw 1 | \,{\bf H} \;2 \rw = \int\! d\mu(\cl C)\; 
   \psi_1^* \, {\bf H} \,\psi_2  \;\;\;\ueb{?!}{=}\;\;
  \int\! d\mu(\cl C)\,\( \,{\bf H}\,\psi_1 \)^* \psi_2  
  =  \lw \,{\bf H}\; 1 | \,2 \rw \quad . \quad
\ee 
The potential term {\bf V}, being real and containing no functional 
derivatives, drops out in \eq{14.25} immediately. For {\bf T} the 
``lower'' version  ${\bf T}=\int {e^2\02} H^{ab}(\ov{G}\ov{p})^a
(Gp)^b$ of (11.10) is convenient by two reasons: (1) we know
about the $p^a$ operators how they become adjoint with respect 
to the measure $d\mu(\cl H)\,$, and (2) we know from \eq{11.16}
how $(Gp)^a$ is regularized, namely as $\int' \cl G^{ab}_{r r'} 
p^b_{r'}$. By the way, this was the reason for always writing 
already $(Gp)^a \gll \int' G_{r r'} p_{r'}^a$ with an $a$ index 
outside of parantheses. We go our own way~:
\bea{14.26}
  \int\! d\mu(\cl C) \,\psi^*_1 {\bf T} \psi_2 
  &=& \hbox{${e^2\02}\!\int_r$} \;\s^n\!\int\! [dh^a] 
      \, e^{2NS} \, \psi_1^* \, H^{ab}(\ov{G} \ov{p})^a \,
      (Gp)^b \, \psi_2 \;\; , \;\; H^{ab*}\!\!=H^{ba} 
      \;\; , \;\; \ov{G}=G^* \nonu \\
  &=& \hbox{${e^2\02}\!\int_r$} \;\s^n\!\int\! [dh^a] \, 
      \lk (Gp)^a \, e^{2NS}\, H^{ba} \,\psi_1 \rki^* 
      (Gp)^b \,\psi_2  \nonu \\
  &=& \hbox{${e^2\02}\!\int_r$} \;\s^n\!\int\! [dh^a] \, 
      \lk (\ov{G}\ov{p})^b \, (Gp)^a \, e^{2NS}\, H^{ba} 
      \,\psi_1 \rki^* \,\psi_2  \nonu \\
  &=& \hbox{${e^2\02}\!\int_r$} \;\int\! d\mu(\cl C) 
      \lk e^{-2NS} \, (\ov{G}\ov{p})^a \,(Gp)^b \, e^{2NS}
      \, H^{ab} \,\psi_1 \rki^* \,\psi_2  \quad . \quad
\eea 
It is seen what wee need. The whole apparatus $ e^{2NS}\, 
H^{ab}$ might commute with the $(Gp)(Gp)$ block. So, the first 
interesting commutator is
\bea{14.27}
  \lk (Gp)^b \, , \,  e^{2NS} H^{ab} \rk &=&
  \lk (Gp)^b \, , \, e^{2NS} \rk  H^{ab} \, + \,
  e^{2NS} \lk (Gp)^b \, , \,  H^{ab} \rk \nonu \\
  &=& e^{2NS} \lb \; 2 N\, H^{ab}\, (Gp)^b \, S \; + \;
  (Gp)^b \, H^{ab} \;\rb \;\;\ueb{?}{=}\;\; 0 \quad . \quad
\eea 
Obviously, the two terms in the wavy bracket might compensate. 
Show!

First term. Our knees turn to jelly. $S$, isn't that this
awkward WZW action with all its volume term complication?
How to avoid doing explicit calculations with? The neat
aspect of the action $S$ was its functional differential
equation (9.35), i.e. $\d S = {1\0\pi} \int \Sp \(  H^{-1} 
(\d H) \6 H^{-1} \ov{\6} H \)$ with $\6$ acting on all factors
to the right of. Indeed, $\d S$ is all what we need for performing
the $(Gp)$ derivative. When performing $(Gp)^b S\,$, $(Gp)$ 
may remain unregularized, with the reason for that already the 
$S$ differential equations had been regularized~:
\bea{14.28}
   2 N\, H^{ab}\, (Gp)^b \, S &=& 
   2N H^{ab}_r \int'\! G_{r r'} (-i r_{r'}^{-1})^{bc} 
   \; {1\0\pi} \int''\! \Sp \( H^{-1}_{r''} 
   \left[ \d_{\ph^b_{r'}} H_{r''} \right] \6'' H_{r''}^{-1}
    {\ov\6}''\! H_{r''} \) \nonu \\
   &=& 2N H^{ab}_r \int'\! G_{r r'} \; {1\0\pi} \int''\!
    \Sp \( H^{-1}_{r''} \left[ p^b_{r'} H_{r''} \right]
    \6''  H_{r''}^{-1} {\ov\6}''\! H_{r''} \) \nonu \\[-1mm]
    \hbox{\ft \eq{14.16} : \qquad } & &  \nonu \\[-4mm]
   &=& -i\,2N H^{ab}_r \int'\! G_{r r'} \; {1\0\pi} 
    \Sp \( T^b \6' H_{r'}^{-1} {\ov\6}'\! H_{r'} \) 
    \nonu \\
   &=& i\,2N H^{ab}_r \int'\! \; {1\0\pi}\, 
    \Sp \( T^b  H_{r'}^{-1} {\ov\6}'\! H_{r'} \) \6' G_{r r'} 
    \nonu \\
   &=& -i\,{N\0\pi} \, 2\,\Sp \( T^a H T^b H^{-1} \) 
        2\,\Sp \( T^b  H^{-1} {\ov\6} H \)  \nonu \\
   &=& -i\,{N\0\pi} \, 2\,\Sp \(  T^a   
       ({\ov\6} H)  H^{-1}  \) \quad . \quad
\eea 

Second term. One can foresee a spatial delta function from
differentiating $H$, hence a conincidence limit in the Greens
function. So, the finite value \eq{11.27} of $\cl G_{rr}$
will have its use again~:
\bea{14.29}
   (Gp)^b_{\rm reg} \, H^{ab} &=& 
   \int'\! \cl G_{r r'}^{bc}\, p_{r'}^c  \;
   2\, \Sp \( T^a H T^b H^{-1} \) \nonu \\
   &=& \int'\! \cl G_{r r'}^{bc}\,  2\,\Sp \( 
   T^a \big[ p_{r'}^c H \big] T^b H^{-1} 
   - T^a H T^b H^{-1} \big[ p_{r'}^c H \big] H^{-1} \) 
     \nonu \\[-1mm]
    \hbox{\ft \eq{14.16} : \qquad } & &  \nonu \\[-4mm]
   &=& -i \,\cl G_{r r}^{bc}\, 2\,\Sp \(  T^a H T^c T^b 
       H^{-1} - T^a H T^b T^c H^{-1} \) 
     = f^{cbd} \,\cl G_{rr}^{bc}\, H^{ad} \nonu \\[-1mm]
    \hbox{\ft (11.27) : \qquad } & &  \nonu \\[-4mm]
   &=& {1\0\pi} H^{ad}\, f^{cbd}  \lk H^{-1} \ov{\6} H \rki^{bc}
    = {1\0\pi} H^{ad}\, f^{cbd}  H^{eb} \ov{\6} H^{ec} \nonu \\
   &=& {-i\0\pi}\, 2\,\Sp \( H^{-1} a  H \lk c , b \rk \) 
       \,2\,\Sp\( b H^{-1} e H \) 
       \,2\,\Sp\( c\,\ov{\6} H^{-1} e H\)  \nonu \\
   &=& {-i\0\pi}\, 2\,\Sp \( H^{-1} \lk e , a\rk  H  c \)
       \,2\,\Sp\( c\,\ov{\6} H^{-1} e H\)  \nonu \\
   &=& {-i\0\pi}\, 2\,\Sp \( H^{-1} \lk \lk e , a \rk , e \rk
       \ov{\6}H \) \nonu \\
   &=& +\;i\,{N\0\pi} \, 2\,\Sp \(  T^a   
       ({\ov\6} H)  H^{-1} \) \;\; = \;\; 
       \hbox{\ft minus \eq{14.28}} \quad . \quad
\eea 
Thus, the vanishing of \eq{14.27} is obtained, indeed, and 
$(Gp)^b$ commutes with $e^{2NS} H^{ab}$.  Next we need a
vanishing commutator also with $(\ov{G}\ov{p})^a\,$.
But this, after all, is easily shown by taking \eq{14.27}
conjugate complex and at interchanged indices $a$, $b$. 
The long chain of equations \eq{14.26} now comes to its
end~:
\be{14.30} \!
 \int\! d\mu(\cl C)\,\psi_1^* {\bf T}\psi_2 
  = \hbox{${e^2\02}\!\int_r$} \!\int\! d\mu(\cl C) 
    \!\lk H^{ab}\,(\ov{G}\ov{p})^a \,(Gp)^b \,\psi_1 \rki^* 
    \!\psi_2  = \int\! d\mu(\cl C) \( {\bf T} \psi_1 \)^* 
    \psi_2 \;\; . \; 
\ee 
Quod erat demonstrandum. 

Presumably, that statement in \cite{nach} below \ekk{17}, ``{\sl In 
principle, one may also obtain the measure of integration 
for the inner product of the wavefunctions by requiring 
self--adjointness of the above expression}'' may be understood
as follows. In order to get {\bf T} self--adjoint with respect 
to $d\mu (\cl C)\,$, \eq{14.28} must hold true, in particular
its first line, i.e. the functional differential equations
for $S$. But having these, one can even turn to its WZW solution.


\sec{ Vacuum and first excitations } 

The energy spectrum of the 2+1D Yang--Mills system follows from
its functional stationary Schr\"odinger equation
\be{15.1}
  (\, {\bf T} +{\bf V} \, ) \, \psi \; = \; E \, \psi 
  \quad , \quad
\ee 
where we look to e.g. the framed $J$ version \eq{13.14}
of the Hamiltonian. Every {\bf T} term exhibits a functional
derivative at its right end. Were there no part {\bf V}, we had
one solution immediately, namely $\,\psi_{00} \equiv 1\,$ with
eigenvalue $E=0\,$ (let the second index refer to the 
approximation $V\!\approx\!0\,$). Hence, we know the vacuum state 
in this approximation. According to \eq{14.4} it is even normalized 
right. But {\bf V} is not zero. 
\pagebreak[2]

\subsection{ The vacuum wave function \boldmath$\psi_0$} 
     
\nopagebreak[3]
There are several possibilities of extracting one or the other
truth from \eq{15.1}. But, most probably, this equation cannot be 
solved exactly. In the spirit of ``{\bf V} is small'', in a sense 
to be developed, KKN propose a systematic iterative procedure 
(\ekk{19} to \ekk{24}), which results in an expansion of the vakuum 
wave function $\psi_0$ in powers of $1/m^2$. Its zeroth approximation 
is $\psi_{00}\equiv 1$. It is field theory. So, let the vakuum 
continue to have energy zero: $({\bf T} + {\bf V})\, \psi_0=0\,$. 
We follow KKN and seek $\psi_0$ in the form
\be{15.2}
  \psi_0 \lk J \rk = e^{P\lk J \rk} \psi_{00} 
  \; = \, e^P \, 1 \quad . \quad
\ee 
Inserting this into \eq{15.1} (and taking $E=0$ there),
\be{15.3}
   e^{-P} \( {\bf T} + {\bf V}\)\, e^P \psi_{00} = 0 
   \quad , \quad \hbox{i.e. } \;
   \schl H\, 1 = 0 \quad \hbox{with} \;
   \schl H \, \gll\, e^{-P} \( {\bf T} + {\bf V}\)\, e^P 
  \quad , \quad 
\ee 
we enjoy the fact that the series of commutators,
\be{15.4}
   \schl H = 
   e^{-P} \( {\bf T} + {\bf V}\)\, e^P 
   = {\bf T} + \lk {\bf T} , P \rk + {1\02} 
     \lk \rule{0pt}{4mm} \lk {\bf T}\, , \, P \rk , P \rk 
     + {\bf V}  \quad , \quad 
\ee 
breaks off, because {\bf T} contains two functional derivatives,
at highest. The last two terms in \eq{15.4} do no more
differentiate. From the condition $\schl H\, 1=0\,$, i.e..
\be{15.5}
    0 \;\;\ueb{!}{=}\;\, \schl H \, 1 \; = \; 
   \lk {\bf T} , P \rk 1+ {1\02} 
   \lk \rule{0pt}{4mm} \lk {\bf T}\, , \, P \rk , P \rk 
   + {\bf V} \quad , \quad
\ee 
the functional $P=P_1+P_2+P_3 + \ldots \,$ can now be determined
term by term. Due to ${\bf T} \sim m$, ${\bf V} \sim m^{-1}$, 
and $P_j \sim  m^{-2j}\;$ ($j=1,2, \ldots$) we have 
\bea{15.6}
  m^{-1} \; : \quad & 0 = \lk {\bf T} , P_1 \rk 1 + {\bf V} 
          \hspace{6.7cm} 
   & \;\;\folg \; P_1  \\[1mm]
  m^{-3} \; :\quad & 0 = \lk {\bf T} , P_2 \rk 1 + \2  \lk 
    \rule{0pt}{4mm} \lk {\bf T}\, , \, P_1 \rk , 
    P_1 \rk  \hspace{3.7cm}
   & \;\;\folg \; P_2  \nonu \\ 
 m^{-5} \; :\quad & 0 = \lk {\bf T} , P_3 \rk 1 + \2 \lk 
    \rule{0pt}{4mm} \lk {\bf T} ,  P_2 \rk , P_1 \rk 
    + \2 \lk  \rule{0pt}{4mm} \lk {\bf T} ,  P_1 \rk , 
    P_2 \rk \ & \;\;\folg \; P_3 \hspace*{15mm} \nonu  
\eea 
and so on. In \ekk{23} KKN state the result for $P_1$, $P_2$
and $P_3\,$. But here, we shall be satisfied with $P_1\,$ and
rather study the delicate normal ordering. Also, we switch to
the Fourier version \eq{13.23}. $P_1$ cannot contain a third power
of $I\,$, because ${\bf T}_1\sim I\,\d$ reproduces the $I$ power
and otherwise there is only ${\bf V}\sim I^2$ in \eq{15.6}.
An additive constant in $P_1$ would drop out in \eq{15.6}.
Hence, we set
\be{15.7}
  P_1 \; = \;\a \int \! d^2q \; \schl I^a (\vc q)\, 
  \schl I^a (-\vc q)   \quad . \quad
\ee 
\eq{15.6} has no contribution from ${\bf T}_3\sim I^c\,f^{cab}
\d^a\d^b\,$, because via $\lk \d\, \d , I^2\rk = \d \lk \d, 
I^2\rk + \lk \d, I^2\rk \d = 2 + 4I\d$ the second term $4I\d$
vanishes when applied to 1, and ``$\!$2'' actually is $2 \d^{ab}\,$
and gives $f^{caa}=0$. There remains
\bea{15.8}
  0 &=& \a\,  m \int\! d^2k \; \schl I^a (\vc k ) \, 
      \lk \d_{ \schl I^a (\vcsm k )} \,\hbox{ \large\bf ,} 
      \int \! d^2q \; \schl I^a (\vc q)\, \schl I^a (-\vc q) 
      \rk 1 \nonu \\
  & &  - \;\a\,\pi m N \!\int\! d^2k \; k^2 \; 
     \lk \d_{\schl I^a (\vcsm k)} \d_{\schl I^a (-\vcsm k)} 
     \,\hbox{ \large\bf ,} \int \! d^2q \; 
     \schl I^a (\vc q)\, \schl I^a (-\vc q) \rk 1 \nonu \\
  & &  + \;{1\0 4\pi m N} \int \! d^2k \,\; 
       \schl I^a (\vc k ) \, \schl I^a (-\vc k ) \nonu \\
  &=& 2 \,\a\,  m \int\! d^2k \; \schl I^a (\vc k )\, 
       \schl I^a (-\vc k)  - \;\a\,\pi m N \!\int\! d^2k \; 
      k^2 \;  2\,\d_{\schl I^a (\vcsm k)} \schl I^a (\vc k)
        \nonu \\
  & &  + \;{1\0 4\pi m N} \int \! d^2k \,\; 
       \schl I^a (\vc k ) \,\schl I^a (-\vc k ) \quad\quad
\eea 
with an obvious catastrophe in the overlast term. It derived from 
the second line through $\lk \d\, \d , I^2\rk 1 = \d \lk \d, 
I^2\rk 1 =\,$``$\!2$''. $\,\d_{\schl I^a (\vcsm k)} 
\schl I^a (\vc k) = n \,\d(\vc 0 )$ \ -- \  ?\gef \ -- \  
this is unthinkable. By including {\bf V} we runned into a typical
field theoretical infinity. If the groundstate energy was to
be kept at $E=0$ then {\bf V} had to be normal ordered from the
outset. Concentrating to the $I^2$ terms of \eq{15.8}, one 
is led to $\a = - 1/(8\pi m^2 N)$ and
\be{15.9}
   P_1 = -\, {1\0 8\pi m^2 N} \int \! d^2q \; 
   \schl I^a (\vc q)\,\schl I^a (-\vc q) \quad . \quad
\ee 
To maintain this conclusion, one misses one term in the last 
line of \eq{15.8}, namely $-{1\0 8 m }\!\int\! d^2k\, k^2\, 
2\,\d_{\schl I^a (\vcsm k)} \schl I^a (\vc k)\,$. By including
this missing term, one turns from {\bf V} to 
$\,:\!{\bf V}\!:\,$. In the next subsection we make an attempt 
to understand this a bit better.

No problems arise in transporting the result \eq{15.9} from 
its Fourier underworld back to the $J$ version. By means of 
\eq{13.16} this looks as
\bea{15.10}
   P_1 &=& -\, {(2\pi)^2\0 8\pi m^2 N} \int\! d^2r \, 
    I^a (\vc r)\, I^a (\vc r)   \nonu \\ 
   &=& -\, {\pi \0 2 m^2 N } \int\! \( \ov{\6} J^a \) 
       \, \ov{\6} J^a    = -\, {\pi\0 m^2N}  
       \,\Sp \( \int\! ( \ov{\6} J ) \ov{\6} J \)
       \quad . \quad
\eea 
In the last step concatenation was used and the definition
$J^a \glr 2 \,\Sp ( T^a J )\,$.
There is no perfect agreement of our last expression
in \eq{15.10} with \ekk{23} (first term), because KKN 
have normal ordering colons around it. 

To summarize, to its leading order, the vacuum wave functional
is obtained as
\bea{15.11} \!
   \psi_0 &=& e^{P}\, 1 = e^{\, P_1 + O(m^{-4})}
   \approx e^{P_1} \, = \,
     \underline{ \rule[-2mm]{0pt}{5mm} e^{\dis -{\pi \0 2m^2N} 
     \int (\ov{\6} J^a)\,\ov{\6} J^a}}\,  \\
   &=& e^{-{1\02m} {\bf V}} = e^{ -{1\02me^2} \int\! 
     {\rm Sp} (BB)} \;\;\; . \;\; \nonu
\eea 
In the second line (less important) \ekk{24} is reached, 
and just \eq{13.10} was inserted for {\bf V}. But the underlined
version in the first line will be basic to the derivation 
of confinement in \S~17.
\pagebreak[2]

\subsection{ Analogy with a 1D oscillator } 
 
\nopagebreak[3]
From the viewpoint of a poor man, familar with non--relativistic 
quantum mechanics, the above was a somewhat strange procedure.
Assume, someone works with the norm $\lw1|2\rw = \pi^{-1/2} \int
\! dx\, e^{-x^2}\,\ph_1^* \,\ph_2$ and would like to solve
\be{15.12}
  H \ph = E \ph \quad \hbox{with} \quad
  H = mx\6_x - {m\02} \6_x^2 + {k^2 \0 2m} x^2 
  \; \glr\, T + V  \quad .\quad
\ee 
Obviously, this was inspired by \eq{13.23} (with $\schl I 
= k\wu{2\pi N}x$ in mind). $\ph \equiv 1$ is correctly
normalized, but only for $V=0$ it is a solution to the eigenvalue
equation.

For solving exactly, he turns back to the usual norm by 
$\ph=\pi^{1/4} e^{x^2/2} \ta\,$, hence to the Hamiltonian 
$\ov{H} = e^{-x^2/2} H e^{x^2/2}$. Then, through 
$e^{-x^2/2} \6_x e^{x^2/2} = \6_x + x\,$, he obtains
\bea{15.13}
  \ov{H} \ta &=& E \ta  \quad , \quad 
  \ov{H} = -{m\02} \6_x^2 + {m\02} 
   \,\Big[ 1 + {k^2\0m^2} \Big]\, x^2 - {m\02} 
  \quad . \quad  n=0,1,2, \ldots \, :  \qquad\quad \nonu \\
  E_n &=& {1\02} \( -m + \wu{ m^2 + k^2}\) + n \, \wu {m^2 + k^2}
  \quad , \quad \ta_0 = e^{-{1\02}\wu {1+k^2/m^2} x^2} 
  \quad , \qquad
\eea 
because he recognized an ``oscillator with $\hbar=m$ and 
$\o=\wu{1+k^2/m^2}\,$''. If $k^2=0$ one would define the
annihilators $\ov{b}=(x+\6_x)/\wu{2}$ and creators 
$\ov{b}^\dagger=(x-\6_x)/\wu{2}$. Herewith, and using
$b = e^{x^2/2}\,\ov{b}\, e^{-x^2/2}$ etc., we now turn
back to the original space, obtaining
\be{15.14} \!
  \ph_0 = e^{{1\02}\( 1 - \wu {1+k^2/m^2}\) x^2} 
  = e^{-{k^2\04m^2} x^2 + O(k^4) }  \;\; , \;\; 
    b= {1\0\wu 2} \6_x \; , \;  
    b^\dagger = {1\0\wu 2} (2x-\6_x ) \;\; . \;\;  
\ee 
Note that, by analoguous simplifications, $P_1$ from
\eq{15.9} turns into $- k^2x^2/(4m^2)\;$ and agrees with
the above exponent of $\ph_0$. The two kinetic terms
of \eq{15.12} are $mb^\dagger b$, so normal ordering has no
effect. But in $V$ it has, because due to $x=(b+b^\dagger)/\wu 2$
it is
\be{15.15}
   x^2 = {1\02} \( b^2 + b^{\dagger 2} + b b^\dagger 
   + b^\dagger b\) \quad , \quad 
   \dep \, x^2 \dop\; = {1\02} \( b^2 + b^{\dagger 2} 
   + 2 b^\dagger b\) = \ldots = x^2 - {1\02} \quad . \quad
\ee 

Let us now treat the oscillator with KKN's $P$ method, however
using $\,\;\dep\,V\dop\;\,$ in \eq{15.12} from the outset.
Then, the ansatz $P_1=\a x^2$ (no matter whether normal ordered
or not) leads to the \eq{15.8} analogue
\be{15.16}
  0 \; = \; m \,\( 2 x^2 - 1 \) \, \( \a + {k^2\0 4 m^2} 
  \) \quad  \quad
\ee 
and to the lowest--order groundstate function $\ph_0$ 
of \eq{15.14}, indeed. Applying  $\,\;\dep\,H\dop\;\,$ to
this function, the eigenvalue $E=0$ might be obtained to lowest 
order in $k^2$. This is indeed the case~:
\be{15.17 } \!
  \dep\, H\dop\;\;\, e^{-{k^2\04m^2}x^2} =
  \lb - {k^2\02m}x^2 - {m\02} \Big[ {k^4 x^2\0 4m^4} 
      - {k^2\02m^2}\Big] + {k^2\02m} \( x^2-{1\02}\) \rb 
  \,e^{-{k^2\04m^2}x^2}  \, = \, 0 \;\; .
\ee 
\vskip -1.55cm \hspace*{5.46cm} \unitlength 1cm 
\begin{picture}(1,1)
\put(0,0){\line(-1,3){.4}} \put(.14,0){\line(-1,3){.4}}
\end{picture} \\
We do not claim that herewith the \S~15.1 is fully
understood.
\pagebreak[2] 

\subsection{\boldmath$\sqrt{m^2+k^2\raise 12pt\hbox{$_{}
            $}\,}^{\!\!\lower .16mm\hbox{\ft $\neg$}}$ } 
 
\nopagebreak[3]
By turning to the excited states, we shall run into certain 
difficulties. So, let this subsection 15.3 be tentative and
end up with open questions, see the remarks at its end.

In search of a starting point, there are two possible 
brute--force simplifications of the Hamiltonian \eq{13.23}~: 
\lower 1pt\hbox{\frame{\rule[-.4mm]{0pt}{3.8mm}\,1.\,}} 
$\,{\bf V}=0$ and 
\,\lower 1pt\hbox{\frame{\rule[-.4mm]{0pt}{3.8mm}\,2.\,}} 
\, NT$=0$, where NT 
stands for ''nonlinear term'', i.e. for the $f^{cab}$ term 
in \eq{13.23}. 

$ $
\lower 1pt\hbox{\frame{\rule[-.4mm]{0pt}{3.8mm}\,1.\,}} 
$\,{\bf V}=0\,$:
hence ${\bf H}={\bf T}\,$, and we have ${\bf T}1=0\,1$,
and ${\bf T} \schl I^a(\vc k) = m \schl I^a(\vc k)$, while
in higher excited states the NT is involved, perhaps
forming glue ball states with increasing content. We
are interested in the first excited state and the fate of
its infinite degeneracy under inclusion of ${\bf V}\,$.  
Including ${\bf V}$ to its first order leads to the
modification $m \to m + k^2/(2\pi)$ of the eigenvalue
(we jumb over the derivation, because this will be seen shortly 
to be contained in the following at small--$k$). But to
second order in ${\bf V}$ the NT comes into play, and we 
run in essence into the same difficulties as
\lower 1pt\hbox{\frame{\rule[-.4mm]{0pt}{3.8mm}\,2.\,}} 
So, 
\lower 1pt\hbox{\frame{\rule[-.4mm]{0pt}{3.8mm}\,1.\,}} 
is a poor starting point.

The first step towards the inclusion of {\bf V} or NT, or both,
is exact. In general, for solving the stationary Schr\"odinger 
equation ${\bf H} \psi = E\psi$, one can write it as
\be{15.18}
  e^{-P} {\bf H} e^P \, \(e^{-P}\psi\) =
  \schl H \, \(e^{-P} \psi\) = E \, \(e^{-P}\psi\)
\ee 
with an arbitrary function(al) $P\,$. In particular, we are
allowed to fix $P$ by $\schl H\, 1=0\,$, as in \S~15.1.
Assume that $P$ has been determined exactly this way, then
the $V$ term can be eliminated from the Hamiltonian \ --- \ 
by subtracting the number zero \eq{15.5}~:
\bea{15.19}
 \schl H &=& {\bf T} + \lk {\bf T} , P \rk 
   + {1\02} \lk \rule{0pt}{4mm} \lk {\bf T}\, , \, 
     P \rk , P \rk   + {\bf V} \nonu \\
  & & -\;\lk {\bf T} , P \rk 1 
   - {1\02} \lk \rule{0pt}{4mm} \lk {\bf T}\, , \, 
     P \rk , P \rk  - {\bf V} \nonu \\[1mm]
  &=& {\bf T} + \,\Big\{ \lk {\bf T} , P \rk - \lk {\bf T} , 
   P \rk 1 \Big\} \quad . \quad
\eea 
The last term just cancels terms of $\lk {\bf T} , P \rk$,
which do not differentiate. Hence $\schl H\, 1=0\,$ is fulfilled, 
indeed. The potential is now hidden in $P$ through \eq{15.5}.
Using the exact $P\,$, all powers of $I$ will appear in 
front of the first differentiation. Symbolically, the 
Schr\"odinger equation reads
\be{15.20}
  \schl H \chi = 
 \lk \( I + I^2 + I^3 + \, \ldots\, \)\,\d + \d\,\d  + 
   I f \d\,\d \rk \chi \, = \, E \,\chi \quad . \quad
\ee 
Already the inclusion $I^2$ in the round bracket (but no more 
$I^3$), requires $\chi$ to contain all powers of $I\,$.

$ $
\lower 1pt\hbox{\frame{\rule[-.4mm]{0pt}{3.8mm}\,2.\,}}\,
Omitting the NT, the Hamiltonian becomes a quadratic form
in $I$ and $\d$. Without NT, each contribution $P_1$, $P_2$, 
$P_3$ etc. reduces to its term quadratic in the field.
Instead of summing them up, it is quite easier, of course,
to solve \eq{15.5} directly with the ansatz
\be{15.21}
  P_{\rm qu} \; = \; \int \! d^2q \; \schl I^a (\vc q)\, 
  \,\a ( q^2 )\,  \schl I^a (-\vc q)   \quad . \quad
\ee 
for the function $\a\,$. Using $P_{\rm qu}$ in \eq{15.19} 
we have
\bea{15.22}
  \lk {\bf T} , P_{\rm qu} \rk &=& m \int\! d^2k \; 
           \schl I^a (\vc k ) 
  \, \lk \d_{ \schl I^a (\vcsm k )} \,\hbox{ \large\bf ,} 
  \, P_{\rm qu} \rk  - \,\pi m N \!\int\! d^2k \; k^2 \; 
     \lk \d_{\schl I^a (\vcsm k)} \d_{\schl I^a (-\vcsm k)} 
     \,\hbox{ \large\bf ,} \, P_{\rm qu}  \rk  \nonu \\
  &=& 2\, m \, P_{\rm qu} \, - 4\pi m N \int\! d^2k \; k^2 \;
      \, \a (k^2)\,  \schl I^a (\vc k ) 
      \,\d_{ \schl I^a (\vcsm k )} \quad , \quad
\eea 
where the familar infinity is suppressed. Note that the latter 
would drop out in the wavy bracket of \eq{15.19} anyway. Thus, 
$\lk {\bf T} , P_{\rm qu} \rk 1 = 2 m P_{\rm qu}$. The doubly 
commutator gives 
\be{15.23}
   {1\02} \lk \rule{0pt}{4mm} \lk {\bf T}\, , \, 
    P_{\rm qu} \rk , P_{\rm qu} \rk = - 4\pi m N \int\! d^2k \; k^2 \;
      \, \a^2 (k^2)\,  \schl I^a (\vc k )\, 
      \schl I^a (-\vc k ) \quad , \quad
\ee 
and \eq{15.5} takes the form
\be{15.24}
  0 = \int\! d^2k \;  \schl I^a (\vc k )\, \schl I^a 
      (-\vc k ) \;\,\lk \, {1\0 4\pi m N} + 2 m\,\a (k^2)
      - 4\pi m N k^2 \,\a^2 (k^2) \,\rk \quad . \quad
\ee 
As the square bracket must vanish, one obtains
\be{15.25}
  \a (k^2) = {m - \wu {m^2 + k^2 } \0 4\pi m N k^2}  
  \; =\;  - \,{1\0 4\pi m N} \; {1\0  m + \wu {m^2 + k^2 } } 
  \quad . \quad
\ee 
The second solution, having the opposite sign in front of the root,
can be presumably ruled out towards normalizability. Inserting 
\eq{15.25} into \eq{15.21}, the latter equation can be rewritten
to be \ekk{31}.

Having found the function $\a(k^2)\,$, not only $P_{\rm qu}$ is
specified, but also the commutator $\lk {\bf T} , P_{\rm qu} \rk$,
\eq{15.22}. This in turn specifies the Hamiltonian $\schl H\,$ 
due \eq{15.19}. There we are~:
\bea{15.26}
  \schl H &=& m \int\! d^2k \; 
  \schl I^a (\vc k )\, \d_{ \schl I^a (\vcsm k )}
  + {\bf T}_2 \; 
  - 4\pi mN \int\! d^2k \; k^2 \, \a (k^2) \,
  \schl I^a (\vc k )\, \d_{ \schl I^a (\vcsm k )} \nonu \\
  &=& \int\! d^2k \; \wu {m^2 + k^2}  \,
  \schl I^a (\vc k )\, \d_{ \schl I^a (\vcsm k )} 
  \;  - \; \pi m N \!\int\! d^2k \; k^2 \;\d_{\schl I^a (\vcsm k)} 
   \d_{\schl I^a (-\vcsm k)}  \quad . \quad 
\eea 

For the first excited state (in field--square approximation)
we now might look for ``the first Hermite polynomial'', i.e.
the above $\schl H\,$ might be applied to a lonely $I$~:
\be{15.27}
  \schl H \;\; \schl I^a (\vc k )\,\; = \,\;\wu {m^2 + k^2} 
  \;\;\schl I^a (\vc k ) \quad , \quad
\ee 
which exhibits the effect of {\bf V} to the formerly degenerated 
eigenvalue $m$. Note that index $a$ and wave vector $\vc k$ have
converted to be quantum numbers of the wave functional. The 
corresponding eigenfunction of the original Hamiltonian $H$ is 
$I\,e^{P_{\rm qu}}\,$. Even $\schl I^a (\vc k_1 ) \schl I^b (\vc k_2 )
\,e^{P_{\rm qu}}\,$ ($\vc k_2\neq\pm\vc k_1$) solves the 
stationary Schr\"odinger equation, with eigenvalue 
$\wu {m^2 + k_1^2} + \wu {m^2 + k_2^2}\,$, and
so on. Without NT, one just arrives at a wave--functional version 
of a free massive $\phi^2$ theory. But, unfortunately, this holds 
true only without NT. This nonlinear term, Ara Sedrakyan says, is 
the most important one. 

Well. The root spectrum is very appealing. Also, it would be 
very welcome as the potential effect in lifting the degeneracy. 
But there is the NT. Could the effect of the NT be anyhow 
subdominant~? (so noticed as an expectation at the end of 
\cite{nach}). The next step, including the NT, amounts to summing 
up the field--cubic terms in $P_1 + P_2 + P_3 + \ldots\,\;$. KKN 
state the result in \cite{nach} as \ekk{33}, \ekk{34} (and we 
have verified). The expression shows that $P$ reduces to $P_1$ 
at $k\to0$, indeed. So, the sorting according to field powers 
is reasonable in the vacuum wave function and towards the Wilson 
loop analysis, see \S\S~16, 17. But for obtaining the spectrum, 
we had no success (through \eq{15.20} and some additional work) 
in recognising any subdominance of the NT. Further analysis is 
required. See also \S 18.3.

Perhaps, with respect to the excitation spectrum, we stand
merely at the beginning. From naive physical imagination 
we would like to obtain {\sl white} glue balls (as the lowest 
states), having some mass $m_\bullet$, and being in plane wave 
states with momentum $\vc k$ and energy $\wu {m_\bullet^2 
+ k^2}\,$. For some first steps see \S~6 of \cite{kkn}.


\sec{Wilson loop} 

In non-abelian theory, where even the field tensor transforms, 
gauge invariant (hence physical) quantities are of particular 
high value. Moreover, the construction $W \( \cl C \) 
= \Sp \( \,\cl P \, e^{-\oint_{\cl C} \! d\vcsm r\, \vcsm A} \,\)$ 
is even a continuum of gauge invariants, since the closed curve 
$\cl C$ may be chosen at will. Everyone knows. Really? $W$ is not 
a speciality of lattice theory. We follow the text book\footnote{\ 
        Start chapter 15 of \cite{ps} from the beginning.} 
of Peskin and Schroeder \cite{ps}, pedestrianize a bit and come to 
the restricted dependence of $W$ on the WZW current in subsection 
16.5. So, in this \S~16, $W$ remains a functional of fields (and 
of $\cl C$, too, of course). For the ground state expectation 
value $\lw W \rw$ see \S~17. 
\pagebreak[2]

\subsection{ Potential on a curve } 
      
\nopagebreak[3]

Mechanics. A force $\vc K$ with $\nabla\times \vc K \neq 0$ has
no potential. Since having a curl to the right, but none to the left,
the equation ``$\nabla V = - \vc K$'' is wrong \ --- \ but ``not
very wrong''. Given a curve $\cl C$ through the field of force,
one may calculate the work done as $\int_{\cl C} ds \; 
\vc r_{\!\prime s}\, \vc K\,$. On 
\\
\parbox[t]{10.3cm}{
$\cl C$ things become one-dimensional, 
and in 1D there is always a potential. Of course, it depends
on the choice of the curve~: $V(\vc r, \cl C, \vc r_0)\;$. 
Let $s$ be the length on $\cl C$. Then $\vc r_{\!\prime s}$
is a variable unit vector, and $ \vc r_{\!\prime s}\cdot \vc K$ 
the component of $\vc K$ along the curve~:}
\parbox[t]{5cm}{\vskip -.3cm\hspace*{.7cm}
\unitlength 1cm
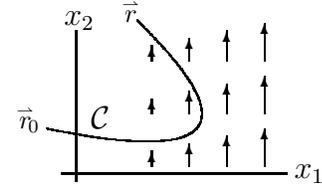
\begin{picture}(2,2) 
  \put(.8,-.4){\line(1,0){3}}  \put(1,-.5){\line(0,1){2}}
  \put(2.0,-.3){\vector(0,1){.2}} \put(2.5,-.3){\vector(0,1){.3}}
  \put(3.0,-.3){\vector(0,1){.4}} \put(3.5,-.3){\vector(0,1){.5}}
  \put(2.0,.4){\vector(0,1){.2}} \put(2.5,.4){\vector(0,1){.3}}
  \put(3.0,.4){\vector(0,1){.4}} \put(3.5,.4){\vector(0,1){.5}}
  \put(2.0,1.1){\vector(0,1){.2}} \put(2.5,1.1){\vector(0,1){.3}}
  \put(3.0,1.1){\vector(0,1){.4}} \put(3.5,1.1){\vector(0,1){.5}}
  \qbezier(.6,.2)(4,-.5)(1.8,1.64)
  \put(1.2,.2){$\cl C$}  \put(.2,.2){\ft $\vc r_{\!\!0}$}
  \put(1.6,1.62){\ft $\vc r$} 
  \put(3.9,-.46){$x_1$}  \put(.84,1.58){$x_2$}
\end{picture}}
\be{16.1}
  \6_s \; V(\vc r(s), \cl C, \vc r_0)
 \; = \;\vc r_{\!\prime s}\cdot \nabla V 
 \; = \; - \vc r_{\!\prime s} \cdot 
 \vc K\big(\vc r(s)\big)  \quad . \hspace*{2cm}
\ee 
Hence, ``$\nabla V = - \vc K$'' needs only the interpretation
of providing all information, to be extracted by multiplication
with an arbitrary (!) $\vc r_{\prime s}\,$, to the curve--dependent
object $V$. Integration over $s$ gives
\be{16.2}
 V(\vc r (s), \cl C, \vc r_0)
 = V(\vc r_0, {\bf 0}, \vc r_0)
 - \int_0^s \! ds'\;  \vc r_{\!\prime s} (s' )\,  
   \vc K\big(\vc r(s')\big) \quad . \quad
\ee 

For the next, assume the force to be coupled (anyhow, by computer, say)
to the potential by $\vc K = \vc k V\,$. Then ``$\nabla V 
= - \vc k V$'' allows for the interpretation
\bea{16.3}
  \6_s \; V(\vc r(s), \cl C, \vc r_0)
   &=&  - \lk \vc r_{\!\prime s} \cdot \vc k\big(\vc r(s)\big) 
  \rk V(\vc r(s), \cl C, \vc r_0)
  \hspace*{2cm} \nonu \\
          \hbox{with solution} \ \ \ \
  V(\vc r (s), \cl C, \vc r_0)
  &=& e^{- \int_0^s \! ds'\;  \vcsm r_{\!\prime s} (s' )\,  
   \vcsm k\big(\vcsm r(s')\big)} \,\;
   V(\vc r_0, {\bf 0}, \vc r_0) \quad , \quad
\eea 
or, relaxing the notation~:
\be{16.4}
 V(\vc r, \cl C, \vc r_0)
 = {\rm exp} \( - \int_{\cl C} \! d\vc r\,  
   \vc k\, \)\; V(\vc r_0, {\bf 0}, \vc r_0) \quad . \quad
\ee 
 
By the next example (next subsection) we shall be led into field 
theory. For convenience, but other then in \cite{ps}, we keep track 
with 2D Euclidean coordinates $(x_1,x_2)\glr\vc r$, metrics $++$, 
two (times $n$) real fiels $A_1^a$, $A_2^a\,$, antihermitean 
matrices $A_j \gll -i A_j^a T^a\,$, $(A_1,A_2)\glr \vc A\,$ and 
with $(D_1,D_2)\glr \vc D\,$. Herewith, the gauge transformation 
and the covariant derivative may be recapitulated as
\bea{16.5} 
  \vc A{\!}\ou &=& U \vc A U^{-1} - \(\nabla U \) U^{-1}
  \quad , \quad U=e^{-i\L^a T^a} \nonu \\
  \vc D &=& \nabla + \vc A \quad , \quad
  \vc D{\!}\ou \,\gll\, \nabla + \vc A{\!}\ou = U \vc D U^{-1} 
  \quad . \quad
\eea 
\vskip -3mm
\pagebreak[2]

\subsection{ Wilson line } 
      
\nopagebreak[3]
Let this subsection head give a name to the on--$\cl C$ solution 
$V$ of the ``wrong'' differential equation
\be{16.6}
  \hbox{\bf`` } \vc D\, V \,=\, 0 \hbox{ \bf''} 
  \quad \hbox{, \ or}\quad
  \hbox{\bf`` } \nabla\, V \,=\, - \vc A\; V \hbox{ \bf''} 
\ee 
to a given initial condition $V(\vc r_0, {\bf 0}, \vc r_0)\,$.
Here, other than in ``$\nabla V=-\vc k V$'', $\vc A$ as well as $V$
are $\,N\times N$ matrices. Of course, if applied to an $N$--component
spinor, \eq{16.6} is nothing but the Dirac equation of motion
of a massless quark field. The intermediate step
\be{16.7}
   V(\vc r(s), \cl C , \vc r_0) = 
   V(\vc r_0, {\bf 0} , \vc r_0) - \int_0^s \! ds'\;
   \vc r_{\prime s} (s' ) \vc A (\vc r (s' ))
  \,\; V(\vc r(s'), \cl C , \vc r_0) \quad
\ee 
shows that the $V$ matrix keeps standing at the right end, even
while iterating \eq{16.7}. Recognizing the time development
in cases of a time dependent Hamiltonian, the solution
to \eq{16.7} can be dealt with as a path ordered (symbol $\cl P$) 
exponential~:
\be{16.8}
  V(\vc r, \cl C , \vc r_0) = \,\cl P
  \, e^{-\int_{\cl C} \! d\vcsm r\; \vcsm A}
  \; V(\vc r_0, {\bf 0} , \vc r_0) \quad . \quad
\ee 
\vskip -3mm
\pagebreak[2]

\subsection{ The Wilson loop} 
     
\nopagebreak[3]
From \eq{16.8} the curve--$\cl C$ specific number $W$,
called Wilson loop or Polyakov loop\footnote{\ 
    M. Thies \cite{lt} feels well with this term. Ara 
    Sedrakyan knows of a corresponding manuscript of 
    Polyakov. Obviously, the gauge invariant loop had 
    been discovered by Polyakov and Wilson independently, 
    such as it happened with the exact solution to the 
    Kondo problem by Wiegmann and Andrei.}, 
is obtained by the following three simple manipulations~:
\be{16.9}
  \matrix{ \hbox{\ft 1. \ specify the initial condition as } 
            V(\vc r_0, \cl C , \vc r_0) = 1 \hfill \cr
  \hbox{\ft 2. \ close the curve } \cl C\;: \;\; 
          \vc r = \vc r_0 \hfill  \cr
  \hbox{\ft 3. \ perform the trace of the so far 
          $N\times N$ fold result \ : } \hfill \cr}
  \hspace*{3cm}
\ee 
\be{16.10}
  W \( \cl C \) = \Sp \( \;\cl P
  \, e^{-\oint_{\cl C} \! d\vcsm r\; \vcsm A}
  \; \) \quad . \quad
\ee 

This functional $W(\cl C)$ is the one, which turns into itselves
by an arbitrary gauge transformation. Presumably, this invariance
can be shown anyhow directly with the expression \eq{16.10}.
But the pleasant and short way goes back to the differential
equation that determines $W\,$ uniquely. While retaining the
unit initial condition, someone starts from \eq{16.6} with
a regauged field  $\vc A{\!}\ou$, calls his solution $V\ou$
and omits quotation marks for brevity~:
\bea{16.11}
 \nabla \,V\ou 
 &=& - \vc A{\!}\ou\, V\ou
      \hbox{ \ \ \ \ \ {\bf to} \ \ \ } V\ou_0 =1  \nonu \\
 U \vc D U^{-1} \,V\ou 
 &=& 0 \quad ; \qquad U^{-1}V\ou \glr \mho \quad \nonu \\
 \vc D \,\mho 
 &=& 0  \hbox{ \ \ \ \ \ {\bf to} \ \ \ } \mho_0 
       = U^{-1}(\vc r_0 ) \quad . \quad
\eea  
Fortunately, in \eq{16.8} the initial condition was kept arbitrary. 
So, the solution to \eq{16.11} can be booked down immediately
or, equivalently, $V\ou= U(\vc r)\,\mho(\vc r )$~:
\be{16.12}
  V\ou (\vc r, \cl C , \vc r_0) 
  = U(\vc r )\;\,\cl P
  \, e^{-\int_{\cl C} \! d\vcsm r\; \vcsm A}
  \;\; U^{-1}(\vc r_0) \quad . \quad
\ee 
Now close the curve $\cl C$, take the trace and enjoy obtaining 
the number $W(\cl C)$ of \eq{16.10} again, {\sl quod erat 
demonstrandum}. Yes, $W$ is a functional of $\cl C$ {\sl and} 
of the fields $A_1$, $A_2$. With respect to the $A$--fields, 
$W$ separates between regaugings (no change) and the physical 
orbit. So far what ``everyone knows''. 
\pagebreak[2]

\subsection{ Reduction of \boldmath$W(\cl C)$ to 
            physical degrees of freedom } 

\nopagebreak[3]
For the 2+1~D Yang--Mills system we know more, namely the physical
degrees $\rho$ of freedom. Using $A = - (\6 M)M^{-1}\,$ and  
$M=U\rho$, we may \ e x p l o i t \ the gauge invariance just 
derived by simpy turning to $U\equiv 1$ in the last of the
following four lines. Remember that $z\gll x_1-ix_2$, 
$\ov{z}\gll x_1+ix_2$, ${1\02}(A_1+iA_2)\glr A$, 
${1\02}(A_1-iA_2)\glr\ov{A}$ $\;\folg\; A^\dagger = - \ov{A}\,$ and
$\6={1\02}(\6_1+i\6_2)\,$. Hence, the Wilson loop exponent
may be rewritten as
\bea{16.13}
  - \oint d\vc r\; \vc A 
    &=&  \oint \( - dz\, A - d\ov{z}\, \ov{A} \)  
        = \oint \( - dz\, A + d\ov{z}\, A^\dagger \) \nonu \\ 
    &=& \oint \( \, dz\, (\6 M) M^{-1} 
      - d\ov{z}\, M^{\dagger -1} \ov{\6} M^\dagger \, \) \nonu \\
    &=& \oint \(\, dz\, (\6 U) U^\dagger 
           + dz\, U (\6 \rho)\rho^{-1} U^\dagger 
           - d\ov{z}\, U \rho^{-1} (\ov{\6} \rho) U^\dagger 
           - d\ov{z}\, U \ov{\6} U^\dagger \, \) \nonu \\
    &=& \oint \(\,  dz\, (\6 \rho) \rho^{-1} 
            - d\ov{z}\, \rho^{-1} \ov{\6} \rho \, \) 
    \quad . \quad
\eea 
Of course, the last step, namely setting $U\equiv 1$, is only 
allowed after the exponent is already placed into the traced 
and ordered exponential function~:
\be{16.14}
  W \( \cl C \) = \Sp \( \;\cl P
  \, e^{\oint \(\,  dz\, (\6 \rho) \rho^{-1} 
           - d\ov{z}\, \rho^{-1} \ov{\6} \rho \, \)} 
  \; \) \quad . \quad
\ee 
\vskip -3mm
\pagebreak[2]

\subsection{ \boldmath$W(\cl C)$ is a functional of only $J$} 
     
\nopagebreak[3]
The exponent $-\oint_{\cl C} \! d\vc r\, \vc A\,$ of \eq{16.10}
was an antihermitean matrix, as is the exponent of \eq{16.14}, 
of course. But this property has gone in KKN's expression 
\ek{2.32} ($\;\equiv$ \eq{16.21} below) for $W(\cl C)\,$. How 
this? To tackle this problem, we might go back again to the 
differential equation, but this time to that of the physical 
version \eq{16.14}.

It may well be guessed how \eq{16.14} might derive from
differential equations, namely from the following ``wrong'' 
pair~:
\be{16.15}
   \(\, \6 - (\6 \rho) \rho^{-1}\,\)\, V_{\rm ph} = 0
   \quad , \quad
   \(\, \ov{\6} + \rho^{-1} (\ov{\6} \rho) \,\)\, V_{\rm ph} = 0
   \hbox{ \ \ \ \ \ {\bf to} \ \ \ } V_{{\rm ph}\;0} =1 
   \quad . \quad 
\ee 
But now go slowly, dear pedestrian, to enjoy the details.
At first, the line \eq{16.15} is a bit rewritten for harmony~:
\be{16.16}
    \6 \,\rho^{-1}\, V_{\rm ph} = 0  \quad , \quad
    \ov{\6}\, \rho \, V_{\rm ph} = 0
   \hbox{ \ \ \ \ \ {\bf to} \ \ \ } V_{{\rm ph}\;0} =1 
   \quad . \quad 
\ee 
From this we are led to the idea of introducing $\rho V_{\rm ph}$
as a new Wilson--line matrix. Also, this brings the matrix
$H=\rho^2$ into play~:
\be{16.17}
   \rho V_{\rm ph} \glr \Upsilon \quad : \quad
    \6 \,H^{-1}\, \Upsilon = 0  \;\;\; , \;\;\;
    \ov{\6}\, \Upsilon = 0
   \hbox{ \ \ \ \ \  {\bf to}  \ \ \ } \Upsilon_0 
         =\rho (\vc r_0) \quad . \quad 
\ee 
Writing this as
\be{16.18}
   \6 \, \Upsilon = (\6 H) H^{-1}\, \Upsilon  
   \quad , \quad \ov{\6}\, \Upsilon = 0
   \hbox{ \ \ \ \  {\bf to}  \ \ } \Upsilon_0 
         =\rho (\vc r_0) \quad , \quad 
\ee 
\eq{16.8} can be applied once more to get the solution
\be{16.19}
  \Upsilon (\vc r, \cl C , \vc r_0)  \,= \;\cl P
  \, e^{\,\int_{\cl C} \! dz \; (\6 H) H^{-1} } 
  \, \rho(\vc r_0) \quad , \quad 
\ee 
having no $d\ov{z}$ term any more. To form $W(\cl C)$
we need $V_{\rm ph} = \rho^{-1}(\vc r )\,\Upsilon(\vc r)\,$, 
\be{16.20}
  V_{\rm ph}(\vc r, \cl C , \vc r_0)  
  \, = \, \rho^{-1}(\vc r) \;\,\cl P
  \, e^{\,\int_{\cl C} \! dz \; (\6 H) H^{-1} } 
  \; \rho(\vc r_0) \quad . \quad
\ee 
Close the curve $\cl C$, take the trace \ --- \ 
and, surprisingly, KKN are right with their mysterious
statement \ek{2.32}~:
\be{16.21}
  W \( \cl C \) \; = \;\Sp \( \;\cl P
  \, e^{\oint  dz\; (\6 H) H^{-1} } \; \) \; 
  = \;\Sp \( \;\cl P\, e^{{\pi\0N} \oint  dz\; J } 
  \; \) \quad . \quad
\ee 
The Wilson loop, reduced to physical variables, is a functional
of only (apart from $\cl C$) the WZW current $J\,$ or,
equivalently, of $J^a=2 \,\Sp \( T^a J\) = {N\0\pi}\, 2\,
\Sp\( T^a (\6 H) H^{-1}\)\,$.

The wisdom \eq{16.21} has now to be combined with an other one,
namely that, according to \cite{ps}, {\sl in fact, all 
gauge--invariant functions of $A^\mu$ can be thought of
as combinations of Wilson loops for various choices of the 
path $\cl P\,$}. For a few more explanations see \cite{ps}.
The combination of this argument with \eq{16.21} means, that
any physics (including normalizable $\psi$'s) might have 
functional dependence of only $J$.


\sec{Confinement} 

Sometimes, the ``well known'' things make the trouble, while 
the advanced matter can be done merely straightforwardly. 
Here it is the first subsection \S~17.1, which made the
problems \ --- \ answered more or less good, hopefully. 
Then, through the other subsecitons, $\lw W \rw$ is calculated
explicitly for a large loop $\cl C$. \ --- \ Why $\,\lw W \rw$~?
\pagebreak[2]

\subsection{ Two--quark state }

\nopagebreak[3]
Let us first state what we need. If the pure--gluonic
full--interacting--groundstate expectation value 
$\lw 0 |\, W(\cl C) \, | 0\rw
= \lw \, W \, \rw_{\psi_0}\,$, ~or $\lw W \rw$ for brevity,
can be analysed in the limit of large area $\MMA$ surrounded 
by the closed (non--intersecting) curve $\cl C$, and if this
leads to the ``area law''
\be{17.1}
  \lw 0 |\, W(\cl C) \, | 0\rw \; \glr \;
  \lw \, W \, \rw \quad  \unitlength 1cm \begin{picture}(4.8,.2) 
  \put(-.1,.1){\vector(1,0){4.6}}
  \put(.03,-.3){\ft both $\MiA$--dimensions $\,\to\;\infty$}
  \end{picture} \;\;
  e^{-\s\,\MMA} \quad , \quad
\ee 
then there is ``confinement'', $\s$ is the ``string tension'' and 
the energy $E_0$ in the gluon system, which surrounds and connects 
a fixed quark--antiquark pair at large separation $R\,$, increases 
as $E_0 = \s\,R\,$. Let the area $\MMA$ lie in a plane. But it is 
irrelevant, whether this plane is fully spatial or extends in time 
direction, because the latter is Euclidean anyway. Show all this.

To keep a quark fixed at (or near) the position $\vc R$ (in 
3--dimensional space, say) we need hypothetical ``other'' forces.
The quark being bounded at $\vc R$, its field operator 
$\ov{q}_\a(\tau, \vc r)$ contains some real function 
$\psi_{\vcsm R}(\vc r)$, which is strongly localized around $\vc R\,$ 
and takes the place of the plane wave, times $b_\a^\dagger$, the quark 
creator, times $\gamma^0\,$. Conveniently, we shall rather state
its center $\vc R$ in place of the variable $\vc r$, in the sequel.
The time is rotated into the Euclidean by $t\to -i\tau\,$. Let also
an antiquark be fixed at the origin with field operator $q_\b
(\tau,\vc r)\,$, containing some function $\psi_{\vcsm 0}\,$ times
$d_\b^\dagger$, the antiquark creator. We follow Bander \cite{ban} 
(a delicate article of 1981, blind copies from are found in two text 
books) to write down a gauge invariant ``meson state'' as
\be{17.2}
  |\hbox{ meson at $\tau$ } \rangle \; = \; \ov{q}_\a 
   (\tau \vc R) \; V_{\a\b} ( \tau \vc R , \cl C , \tau \vc 0 ) 
   \; q_\b (\tau \vc 0 )\;  | 0 \rangle \; \glr \; \Gamma (\tau)
   \, | 0 \rangle \quad . \quad
\ee 
Here, $| 0 \rangle$ is a bare quark vacuum, and $V$ is the 
non--abelian object \eq{16.8} with matrix indices made explicit 
(and suitably generalized to include time). For the present,
the gauge fields in $V$ are considered classical. 
\\ \unitlength 1cm 
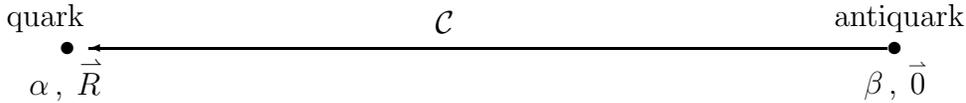
\begin{picture}(13,1)
  \put(1.3,.2){quark}  \put(12.3,.2){antiquark}  
  \put(1.6,-.72){$\a\,$, $\vc R$}  \put(12.7,-.7){$\b\,$, $\vc 0$}  
  \put(2,-.2){$\bullet$}  \put(13,-.2){$\bullet$}  
  \put(13,-.1){\vector(-1,0){10.6}}  \put(7,.1){$\cl C$}  
\end{picture} 

\vskip 7mm
There are two circumstances in favour of the expression \eq{17.2}. 
First of all, it is gauge invariant, indeed, since according to 
\eq{16.12} the $U$ matrices recombine with those splitting off 
from the quark fields at the right space--time points. Secondly, 
imagine a quark--antiquark pair produced at the origin and 
the quark then be moved to $\vc R$ along the path $\cl C$. Thereby, as 
we know from nonrelativistic quantum mechanics in given classical
fields (see Aharonov Bohm effect), it develops just the phase
shown in \eq{17.2}. Admittedly, since depending on $\cl C$, the 
term ``meson state'' needs its parantheses. A sum (sum???) over paths 
is seen in \cite{ban} together with some unspecified weight function
of $\cl C$. But please, Dr. Bander, one cannot state this sum
on one page and simply omit it on the next. Presumably, this detail
(this arbitrarity) becomes irrelevant in the limit of large
separation. We shall keep track with a straight line connecting
the quark pair.

So far, there were no conclusions, hence all the above is 
$\glr\,$right. Furthermore, we may con\-sider what we want. 
Consider the overlap
\bea{17.3}
   & & \Omega_{\rm classical} \; \gll \;  
       \langle \hbox{ meson at $T$ } | \hbox{ meson at $0$ } 
       \rangle \, = \, \langle 0 | \,\Gamma^\dagger (T) 
       \,\Gamma(0)\, | 0  \rangle \hspace*{3cm} \nonu \\[1mm]
   &=&  \langle \, 0 \, | \,  q^\dagger_\b (T \vc 0 )\; 
        V_{\a\b}^* ( T \vc R , \cl C , T \vc 0 )\; 
        \ov{q}^\dagger_\a  (T \vc R) \; \;
     \ov{q}_\g  (0 \vc R) \; V_{\g\l} ( 0 \vc R , \cl C , 0 \vc 0 ) 
     \; q_\l ( 0 \vc 0 )\;  | \, 0 \, \rangle \qquad \nonu \\[1mm]
   &=&  \langle \, 0 \, | \,  \ov{q}_\b (T \vc 0 )\; 
        V_{\b\a} ( T \vc 0 , \cl C , T \vc R )\; 
        q_\a  (T \vc R) \; \;
     \ov{q}_\g  (0 \vc R) \; V_{\g\l} ( 0 \vc R , \cl C , 
      0 \vc 0 ) \; q_\l ( 0 \vc 0 )\;  | \, 0 \, \rangle 
      \qquad  \nonu \\[1mm]
   &=&  \langle \, 0 \, | \,  \ov{q}_\b (T \vc 0 )\; 
         q_\l ( 0 \vc 0 )\, | \, 0 \, \rangle
        \; \langle \, 0 \, | \, q_\a  (T \vc R) \;  
        \ov{q}_\g (0 \vc R) \, | \, 0 \, \rangle
       \; V_{\b\a} ( T \vc 0 , \cl C , T \vc R )\; 
       \; V_{\g\l} ( 0 \vc R , \cl C ,  0 \vc 0 ) \nonu \\[1mm]
   &=& \langle \, 0 \, | \,  \ov{q}_\s (0 \vc 0 )\; 
         q_\l ( 0 \vc 0 )\, | \, 0 \, \rangle \, 
         e^{-mT}\, V_{\s\b}(0\vc 0 , \cl C , T\vc 0 ) \nonu \\[1mm]
   & & \langle \, 0 \, | \, q_\a  (T \vc R) \;  
        \ov{q}_\rho (T \vc R) \, | \, 0 \, \rangle
                e^{-mT}\, V_{\rho\g}(T\vc R , \cl C , 0\vc R ) 
       \;\; V_{\b\a} ( T \vc 0 , \cl C , T \vc R )\; 
       \; V_{\g\l} ( 0 \vc R , \cl C ,  0 \vc 0 )  \qquad \nonu \\[1mm]
   &\sim&  e^{-2mT}\; V_{\l\b}(0\vc 0 , \cl C , T \vc 0)
       \; V_{\a\g}(T\vc R , \cl C , 0 \vc R)
       \; V_{\b\a} ( T \vc 0 , \cl C , T \vc R )
       \; V_{\g\l} ( 0 \vc R , \cl C ,  0 \vc 0 )  \qquad  
\eea 
Third line. The c.c. of $V$ is the same with reversed 
matrix indices and reversed curve (exercise). Also,
$(q_\a^\dagger \g^0)^\dagger = \g^0 q_\a$ had been used. Do not 
smile. Rather remember that in $q_a^\dagger$ the term $\sim 
\psi_{\vcsm R} \,b_\a^\dagger$ was relevant, and this has now
turned into $\sim \psi_{\vcsm R} \,b_\a\,$. Hence,
$ q_\a  (T \vc R)$ in the third line annihilates a quark at $\vc R\,$.
Moving $\g^0$ to the left, $\ov{q}_\b$ is formed. It annihilates
an antiquark at origin.

Fourth line. Here, Wicks theorem has only one term. The second term 
vanishes, since there is no overlapp between the localizing functions 
$\psi$ at different positions. Both the two bare quark propagators 
contain an annihilator to the left and a creator to the right.

Fifth line. $q_\b(0\vc 0) \sim \psi_{\vcsm 0} d_\b^\dagger$ had
created the antiquark. In the presence of classical fields it
develops as $V_{\b\s}(T\vc 0, \cl C , 0 \vc 0) q_\s (0\vc 0 )$,
or daggered as $\ov{q}_\b(T\vc 0) = \ov{q}_\s (0\vc 0 ) \,
V_{\s\b}(0\vc 0, \cl C , T \vc 0)\,e^{-mT}\,$, where the exponential 
stems from $e^{iH_0 t} d_\b e^{-iH_0t} = e^{-imt} d_\b \to 
e^{-mT} d_\b\,$. It is, by the way, irrelevant towards the area law. 

Last line. Here, only the Kroneckers $\d_{\s\l}$ and \\
$\d_{\a\rho}$ are retained from the propagators. The four \\
Wilson lines combine to the square shown, which \\
is a special Wilson loop in the $x$--$\tau$ plane, say. \\[-3cm] 
\hspace*{11.4cm}
\unitlength 1cm 
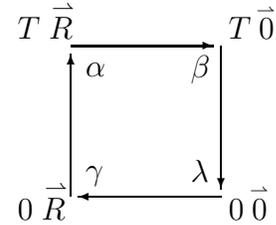
\begin{picture}(3,2.9)
  \put(0,0){\vector(0,1){1.9}}  \put(2,0){\vector(-1,0){1.9}}
  \put(0,2){\vector(1,0){1.9}}  \put(2,2){\vector(0,-1){1.9}}
  \put(0.2,1.6){$\a$}  \put(1.6,1.58){$\b$}
  \put(1.6,.2){$\l$}  \put(0.2,.22){$\g$}
   \put(-.7,2.1){$T\,\vc R$}   \put(2.1,2.1){$T\,\vc 0$}
   \put(2.1,-.3){$0\,\vc 0$}   \put(-.7,-.3){$0\,\vc R$}
\end{picture} \vskip 3mm

So far the gauge fields have remained classical. But now let 
them become operators and \eq{17.3} be put in the full gluonic
ground state expectation value. The two ground states, 
$|0\rangle_{\rm gluon}$ and the bare quark vacuum $|0\rangle\,$,
may be combined to a product state $|0_\bullet\rangle\,$.
The reason for the sandwiching is seen as follows~:
\bea{17.4}
  \O &\gll&  \langle 0 |_{\rm gluon}\,\; \O_{\rm classical}\;\, 
            | 0 \rangle_{\rm gluon} 
     \; = \; \langle 0_\bullet |\,\Gamma^\dagger (T) 
             \,\Gamma(0)\, | 0_\bullet \rangle  \nonu \\[1mm]
   &=&  \sum_n \langle 0_\bullet | \,\Gamma^\dagger (T)\, 
      | n_\bullet \rangle\, \langle n_\bullet | \,
        \Gamma(0)\, | 0_\bullet \rangle   
     \, = \, \sum_n e^{-E_{\bullet\; n}\, T} | \langle n_\bullet |
          \,\Gamma (0)\, | 0_\bullet \rangle\, |^2  \; \nonu \\
   &\sim& e^{-E_{\bullet\,0}\, T} \, = \, e^{-2mT - E_0 T}
       \quad \big(\, T \to \infty \,\big)  \quad , \quad      
\eea 
where $E_0$ is the gluonic part of the meson (i.e. the energy
in the {\sl string} of field lines, if this picture makes sense 
at all). Note that, as  $\G$ creates and $\G^\dagger$ 
annihilates the whole meson, we have $\,e^{iHt}\G^\dagger(0) 
e^{-iHt} = e^{-iEt} \G^\dagger(0) \to e^{-E\tau}\G^\dagger(0)\,$. 

If the gluon energy increases as $E_0=\s R$ with large spatial 
separation $R$, then the exponent $-E_0T$ turns into $-\s RT 
= -\s \MiA\,$, as announced. Forming the gluon sandwich with the 
last line of \eq{17.3}, $\O$ may be also written as
\be{17.5}
  \O \, \sim \, e^{-2mT} \, \langle \, W(\cl C) \,\rangle 
  \quad . \quad
\ee 
By comparison of \eq{17.5} with \eq{17.4}, the statement \eq{17.1}
is derived. ~Really~?  
\pagebreak[3]

\subsection{ A special 2D Euclidean \boldmath$J$ theory with action \$}
     
\nopagebreak[3]
We return to the 2+1~D Yang--Mills system in Weyl gauge.
As in \S~16, the closed loop $\cl C$ lies in the 2D $xy$--plane.
We trust in \eq{17.1} and the statements around it.
In wave quantum mechanics, the ground state expectation
value \eq{17.1} must be written as a functional integral
over physical variables, i.e. with the measure $d\mu(\cl C )$.
Oh, two $\cl C$'s with different meaning. We change the
notation of the measure from $d\mu(\cl C )$ to $d\mu_{\rm phys}\;$:
\be{17.6}
   \lw\, W\,\rw =  \lw 0 |\, W(\cl C) \, | 0\rw 
   \, = \,\int \! d\mu_{\rm phys}  \; \psi_0 \,W(\cl C)\, \psi_0 
   \, = \,\int \! d\mu_{\rm phys}  \; e^{2P}\; W(\cl C) 
   \quad . \quad 
\ee 
Here $\psi_0[J]=e^{P[J]}\, 1$ is the exact ground state functional, 
as introduced in \eq{15.2} together with the functional $P$ 
(exact as well). $\psi_0$ is chosen real. 

In the limit of large area in $\cl C$ (large in both directions),
we may argue that the functional $P$ is probed at short wavevectors 
only. This is just the limiting case that could be worked out
in \S~15.1 with the result \eq{15.11}~:
\be{17.7}
 2P \;\to\; 2P_1 = - {{\bf V}\0m} =
  - {\pi \0 m^2N} \int (\ov{\6} J^a)\,\ov{\6} J^a
   =  - {2\pi \0 m^2 N} \int \Sp \( (\ov{\6}J)\, \ov{\6}J \) 
  \,\glr\,  - \$  \quad . \quad
\ee 
Remember that $J=J^aT^a\,$. Reading \$ as one more auxiliary 2D 
action, but this time with the currents $J^a$ as fields and 
containing no interaction, the Wilson loop expectation value 
$\langle W \rangle$ becomes a functional integral of a free 
theory, namely~:
\be{17.8}
  \lw\, W(\cl C)\,\rw  
  = \int \! d\mu_{\rm phys}\; e^{ - {2\pi \0 m^2 N} 
     \int \Sp \( (\ov{\6}J)\,\ov{\6}J \) }
   \; \Sp \( \;\cl P\, e^{{\pi\0N} \oint  dz\; J } 
  \; \) \quad . \quad
\ee 
Its evaluation might be anyhow possible. Note that the WZW 
action $S$ (KKN should distinguish \$ from $S$) is quite another 
object. It contains no derivatives of $J$ and is, via $e^{2NS}$, 
part of the measure $d\mu_{\rm phys}\,$ in the above.
\pagebreak[2]

\subsection{ The \boldmath$J$ propagator } 
     
\nopagebreak[3]
We shall need the propagator of the \$ theory for matrix 
currents $J_{\a\b}(\vc r)$, path ordered along the curve $\cl C$~:
\pagebreak[3]
\bea{17.9}
 \lw\, \cl P \, J_{\a\b}(\vc r) \, J_{\l\tau}(\vc r') \,\rw
 &=& T^a_{\a\b} T^b_{\l\tau} \lk 
     \theta_{\ueb{\vcsm r > \vcsm r'}{{\rm on\,\;} \cl C} } 
         \lw J^a(\vc r) J^b(\vc r') \rw +
     \theta_{\ueb{\vcsm r < \vcsm r'}{{\rm on\,\;} \cl C} } 
         \lw J^b(\vc r') J^a(\vc r) \rw  \rk \nonu \\[-2.4mm]
 & & \hspace*{2.6cm} \unt{2.4}{.5}{2}{\d^{ab}\D(\vc  r - \vc r')} 
        \nonu \\[-3.4mm]
 &=& T^a_{\a\b} T^a_{\l\tau} \;\D ( \vc r - \vc r') \nonu \\
 &=& {1\02} \( \d_{\a\tau} \d_{\b\l} - {1\0N}\d_{\a\b}\d_{\l\tau}\) 
     \,\D ( \vc r - \vc r') \quad , \quad
\eea 
because $\D (\vc r)=\D (-\vc r)$ will turn out shortly. The
average in the first line is defined by $\langle \ldots \rangle
\,\gll \int d\mu_{\rm phys} \, e^{-\$} \, \ldots\;$. For the
function $\D$ we first Fourier transform the action by inserting
$J^a = \int\! {d^2k\0(2\pi)^2} e^{i\vcsm k \vcsm r}\schl J^a (\vc k)$ 
and introduce the propagator $\schl\D ^{ab}(\vc k)$ by the last line~:
\bea{17.10}
  \$ &=& {\pi \0 m^2N} \int (\ov{\6} J^a)\,\ov{\6} J^a
   \, = \, {\pi\04m^2N} \int\! {d^2k\0(2\pi)^2}\; \schl J^a (\vc k) 
        \, (k_1-ik_2)^2 \,\schl J^a (-\vc k)  \nonu \\
  &\ueb{!}{=}& - {1\02} \int\! {d^2k\0 (2\pi)^2} \; 
     \schl J^a (\vc k) \; \Bigg( {1\0 \schl\D (\vc k) } 
     \Bigg)^{\!\!ab} \schl J^b (-\vc k)  \quad . \quad
\eea 
Clearly, the Fourier transformed propagator is given by
\be{17.11}
   \schl \D^{ab}(\vc k) = \d^{ab}\, \schl \D (\vc k )
   \qquad \hbox{with} \qquad \schl \D (\vc k ) = {2m^2N\0\pi}\, 
   {-1 \0 (k_1-ik_2 )^2 } \quad . \quad
\ee 
The transformation back to real space needs no additional work, 
since it runs along the five lines of \eq{13.21}. Just $k_1$, 
$k_2$ are to be interchanged with $x$, $y$, and an additional
factor of $1/(2\pi)^2$ must be included here. Hence, the 
result is
\be{17.12}
  \D^{ab} (\vc r) = \d^{ab}\, \D(\vc r)  \qquad \hbox{with} 
  \qquad \D (\vc r) = {m^2N\0 2\pi^2}\, {x+iy \0 x-iy} 
  = \D(-\vc r) \quad . \quad
\ee 
It specifies \eq{17.9}.
\pagebreak[2]

\subsection{ Varying the area~: 
            ~\boldmath$\6_{\MA} \lw W \rw$}
     
\nopagebreak[3]
For obtaining the desired differential equation for $\lw W \rw$
a paper of Kasakov and Kostov \cite{kasako} was useful. Possibly,
Gross \cite{gross} knows of another shorter way to derive the 
area law. But \cite{gross} is so far not understood with respect to
both, prerequisites and stability. We favour explicit calculation.

Increasing the area $\MiA$ means elongating its border line
$\cl C$, e.g. by an infinitesimal bump at position $\vc r$ on
$\cl C\,$. Hence, the derivative $\6_{\MiA(\vcsm r)}$ carries
$\vc r$ as an index. But the result will, of course, not depend on
this position.  Moreover, the bump can be given any convenient
form. The circle version to the right in the figure turned out 
most convenient.

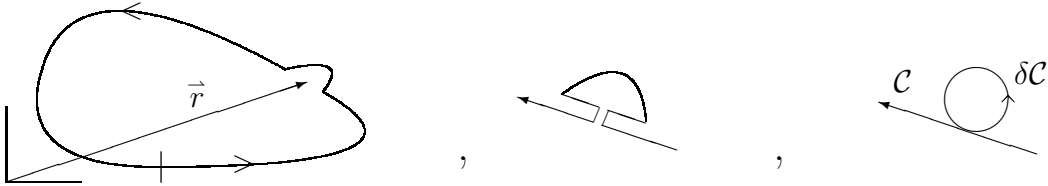
\unitlength 1cm 
\begin{picture}(10,2.6)
   \put(0,0){\line(1,0){1}}    \put(0,0){\line(0,1){1}}
   \put(0,0){\vector(3,1){4}}  \put(2.4,1){$\vc r$}
   \put(3,.14){$>$}  \put(1.5,2.17){$<$}  \put(2,.1){$|$}
   \qbezier(2,.2)(6,.2)(4.2,1.2)  \qbezier(4.2,1.2)(4.6,1.7)(3.7,1.5)
   \qbezier(3.7,1.5)(1,3)(.5,1.6) \qbezier(.5,1.6)(0,.2)(2,.2)
\put(6,.3){{\Large ,}} \put(10.2,.3){{\Large ,}}
    \put(7.8,.8){\vector(-3,1){1}}  \put(7.92,.76){\line(3,-1){1}}
      \put(7.78,.84){{\tiny $/$}}  \put(7.9,.8){{\tiny $/$}}
    \put(7.88,1){\line(-3,1){.5}}  \put(8,.96){\line(3,-1){.5}}
        \qbezier(7.38,1.17)(8.4,1.9)(8.5,.81) 
  \put(13.7,.37){\vector(-3,1){2.1}} \put(12.9,1.1){\circle{.8}}
  \put(13.13,.66){{\LARGE $\hat{\phantom{a}}$}}
  \put(11.8,1.2){$\cl C$}  \put(13.4,1.3){$\d \cl C$}
\end{picture} 

\vskip 2mm
Let us write $\cl C + \d \cl C\,$ for the enlarged curve. Of course,
in the differentation to be performed, 
\be{17.13}
  \6_{\MiA} \lw W \rw = {1\0 \d \MiA}\; \lw \;
   W ( \cl C + \d \cl C ) - W( \cl C ) \; \rw \quad , \quad
   \; W( \cl C )= \,\Sp \( \cl P\, e^{{\pi\0N} \oint\! dz\; J } \) 
   \quad , \quad
\ee 
the path ordering runs through to the circle, if any. Due to
the ordering $\cl P$ in front of the exponential, we may relax 
inside and write $\oint_{ \cl C + \d \cl C }\, =  \,\oint_{ \cl C } 
+ \oint_{\d \cl C }\,$. Keeping $dz {\pi\0 N}J$ in mind, we may 
even write
\be{17.14}
  e^{\oint_{ \cl C + \d \cl C }} - e^{\oint_{ \cl C }}
  = \lk\, e^{\oint_{\d \cl C }} - 1 \,\rk e^{\oint_{ \cl C }}
  = \lk \; \oint_{\d \cl C } \; + \; {1\02} 
    \oint_{\d \cl C } \oint_{\d \cl C } \; \rk 
    e^{\oint_{ \cl C }} \quad . \quad
\ee 
Note that $\MiA \sim \d r^2\,$, with $\d r$ the radius of the circle.
But it is $\oint_{\d \cl C } \sim \d r\,$, only. So, the second term
in \eq{17.14} must be retained. Corrrespondingly, there are two 
contributions~: $ \6_{\MiA} \lw W \rw \,\glr\, 
\6_{\MiA} \lw W \rw\bigg|_1 + 
\6_{\MiA} \lw W \rw\bigg|_2\,$. More explicitly they read
\bea{17.15}
  \6_{\MiA} \lw W \rw \bigg|_1 &=& {1\0\d\MiA}\, {\pi \0 N}
  \oint_{\d\cl C}\! dz\; \,\Sp \( \lw \cl P \; J(\vc r)\,
  e^{{\pi\0N} \oint_{\cl C}\! dz' \; J(\vcsm r')} \rw \)
        \\   \label{17.16} 
  \6_{\MiA} \lw W \rw \bigg|_2 &=& {1\0\d\MiA}\, {\pi^2 \0 2N^2}
  \oint_{\d\cl C}\! dz\; \oint_{\d\cl C}\! dz'\; \,\Sp \( \lw \cl P \, 
  J(\vc r)\, J(\vc r')\,  e^{{\pi\0N} \oint_{\cl C}\! dz'' 
  \; J(\vcsm r'')}  \rw \) \;\;\; . \qquad
\eea 
As an outlook, the second contribution \eq{17.16} will in fact
be argued to vanish for the regularised (!) theory at hand
(but not otherwise). We shall understand this in the course
of evaluating the first contribution.

Imagine the exponential of \eq{17.15} be expanded. The
functional average $\lw\phantom{AA}\rw$ will ``pair'' each
product of $J$'s, i.e. decompose it into a sum of propagators
$\lw\, J\,J\,\rw\,$ (Wick's theorem in functional integral 
language). With this fact in mind, we may even leave the
exponential intact. To pair the extra $J(\vc r)$ in
all possible ways with one in the exponential we put
the operator
\be{17.17}
 \int\! d^2r_0 \; J_{\l\tau}(\vc r_0) \, 
  \d_{J_{\l\tau}(\vcsm r_0)} \qquad
\ee 
in front of it. Note that it selects each inner $J$ once.
Hence
\bea{17.18}
  \6_{\MiA} \lw W \rw \bigg|_1 &=& {1\0\d\MiA}\, {\pi \0 N}
  \oint_{\d\cl C}\! dz\, \int\! d^2r_0 \; \bigg\langle \,\cl P\,
   J_{\a\b} (\vc r) \, J_{\l\tau}(\vc r_0) \, \bigg\rangle \cdot 
      \nonu \\ 
  & & \hspace*{2.8cm} \cdot \lw \,\d_{J_{\l\tau}(\vcsm r_0)} 
    \( \, \cl P \, e^{{\pi\0N} \oint_{\cl C}\! dz' \; J(\vcsm r')}
    \)_{\b \a} \, \rw \quad . \qquad
\eea 
The first average is just the matrix propagator as detailed in
\eq{17.9}. Performing the functional differentiation in the
second average, is left as an exercise to the reader (expand the
exponential up to the third order term, at least. Better way?). 
But let us consider the first step into it, namely the first order 
term of the expansion~:
\be{17.19} 
 \d_{J_{\l\tau}(\vcsm r_0)} \,
 {\pi\0N}\int_{\vcsm 0}^{\,\rm end}\!dz'\;J_{\b\a}(\vc r')
   = {\pi\0N} \, \d_{\l\b}\, \d_{\tau\a}\,
     \int_{\vcsm 0}^{\,\rm end}\!dz'
     \; \d (\vc r_0  - \vc r' ) \quad . \quad
\ee 
This integral, removing ``half of'' the 2--dimensional
delta function, occurs once in each of the other terms, too.
The end point along the closed curve $\cl C$ is, of course, 
identical with the point $\vc 0\,$. The exercise results in
\bea{17.20}
  \d_{J_{\l\tau}(\vcsm r_0)} \, \( \, \cl P \, 
  e^{{\pi\0N} \oint_{\cl C}\! dz' \; J(\vcsm r')} \)_{\b \a}
  &=&  {\pi\0N} \, \int_0^{\,\rm end}\!dz''
   \; \d (\vc r_0  - \vc r'' ) \; \nonu \\
  & &  \hspace*{-12mm}
  \( \, \cl P \,  e^{{\pi\0N} \int_{\vcsm r_0}^{\,\rm end} dz' 
          \; J(\vcsm r')} \)_{\b \l}
  \( \, \cl P \, e^{{\pi\0N} \int_{\vcsm 0}^{\vcsm r_0} dz' 
          \; J(\vcsm r')} \)_{\tau \a}
\eea 
In the intermediate result, to be booked down next, we get rid of 
the above delta function through integrating over $d^2r_0$, see 
\eq{17.18}. Then, after inserting \eq{17.9} and using the 
Kroneckers there, we arrive at
\bea{17.21}
  \6_{\MiA} \lw W \rw \bigg|_1 
  &=& {1\0\d\MiA}\, {1\02} \, \({\pi \0 N}\)^2
  \oint_{\d\cl C}\! dz\, \int_0^{\,\rm end}\!dz'' \;
  \D (\vc r - \vc r'') \, \cdot \nonu \\ 
  & & \hspace*{-26mm}  \cdot\,\lw \, 
    \(\,\cl P\, e^{\,{\rm from}\;\vcsm r'' } \)_{\b\b} 
    \(\,\cl P\, e^{\,{\rm to}\;\vcsm r'' } \)_{\a\a} \, 
    - \, {1\0N}\,  \(\,\cl P\, e^{\,{\rm from}\;\vcsm r'' } 
    \)_{\a\l} \(\,\cl P\, e^{\,{\rm to}\;\vcsm r'' } \)_{\l\a} 
   \,\rw  \quad . \quad
\eea 
The two matrices at the right end may be combined to $\Sp \( 
\cl P  e^{\ldots} \) = W(\cl C)\,$, immediately.

The doubly integral in \eq{17.21},
\be{17.22}
 \cl N \,\gll\, \oint_{\d\cl C}\! dz\, \int_0^{\,\rm end}\!dz'' \;
  \D (\vc r - \vc r'') \bigg|_{
  {\rm both,\;} \vcsm r {\;\rm and\;} \vcsm r'' ,\; {\rm on\;} \cl C} 
  \quad , \quad  
\ee 
will turn out $\sim \d r^2$ in the next subsection. This means, 
that only some infinitesimal short range on $\cl C$ makes
$\vc r''$ different from $\vc r\,$. Hence, in the second line
of \eq{17.21}, we are allowed to replace $\vc r''$ by $\vc r\,$,
the position where we differentiate with respect to $\MiA\,$.
Now remember that the start position $\vc 0$ on the closed curve
could have been chosen at will. Taking it near to $\vc r\,$ we
get $\(\phantom{\int aa} \)_{\a\a} \to 1_{\a\a} = N\,$ and
$\(\phantom{\int aa} \)_{\b\b} \to W(\cl C)\,$, in the second 
line of \eq{17.21}\footnote{\ 
    Admittedly, we do not like such dangerous arguments. 
    Is there a better way?}. 
So, up to the evaluation of $\cl N$, we have that
\be{17.23}
  \6_{\MiA} \lw W \rw \bigg|_1 = {1\0\d\MiA}\, 
  {1\02} \, \({\pi \0 N}\)^2  \;\cl N\; {n\0N} \;\lw W \rw 
  \quad . \quad
\ee 
In fact, evaluating the prefactor $\cl N$ means calculating 
the string tension.
\pagebreak[2]

\subsection{ The integral for the string tension } 

\nopagebreak[3]     
To start with, insert the Greens function \eq{17.12}
into \eq{17.22},
\be{17.24}
  \cl N \glr  {m^2N\0 2\pi^2}\,  \cl N_0 \quad , \quad  
   \cl N_0 = \oint_{\d\cl C}\! dz \int \! dz'\; 
   {(x-x') + i (y-y') \0 (x-x') - i (y-y') } 
   \quad , \quad
\ee 
and let the relevant piece of the curve $\cl C $ be a vertical,
parallel to the $y$--axis with distance $\d r\,$. With the
center of the circle $\d \cl C$ at origin, the four cartesian
variables may be parametrized as
\be{17.25}
  \matrix{ x=\d r \,\cos (\ph) \qquad \hfill & 
           y= \d r \,\sin (\ph )  \qquad \hfill & 
    dz = dx - i dy = -i \d r\, e^{-i\ph}\, d\ph \qquad \hfill  \cr
          x'=  - \d r  \qquad \hfill & 
          y' = \d r\; t  \qquad \hfill &
    dz' = dx' - i dy' = -i \d r \; dt  \qquad \hfill \cr }
\ee 
to give
\be{17.26}
  \cl N_0 = (-i \d r )^2 \int_{-\pi}^{\pi} \! d\ph \; 
  e^{-i\ph} \int_{-\infty}^\infty \! dt \;\, 
  { e^{i\ph} + 1 - it \0  e^{-i\ph} + 1 + it }
\ee 
There are several possibilities of further evaluation.
We set $1+it \glr \wu {1+t^2} e^{i\a}$ and shift
$\ph \to \ph - \a$~:
\bea{17.27}
  \cl N_0 &=& - (\d r)^2 \int\! dt\; e^{-i\a} 
  \int_{(2\pi)}\! d\ph\; { 1 + \wu {1+t^2} e^{-i\ph} 
       \0 e^{-i\ph} + \wu {1+t^2}} \nonu \\
  &=&  - (\d r)^2 \int\! dt\; e^{-i\a}
       \int_{(2\pi)}\! d\ph\; \( \wu {1+t^2} - { t^2 
       \0 e^{-i\ph} + \wu {1+t^2} } \) \nonu \\
  &=&  - {\d \MiA \0 \pi} \int\! dt\; e^{-i\a}
       \( 2\pi \wu {1+t^2} - t^2 \, {2\pi \0 \wu {1+t^2}} 
       \) \qquad \;\; (*) \nonu \\
  \cl N_0
  &=& - 2\, \d \MiA \int\! dt \; {1-it \0 1+t^2 } 
       \; = \; - 2\,\pi\, \d \MiA \quad , \quad
\eea 
where the line $(*)$ was reached by contour integration
with $e^{i\ph}\equiv z$ and a pole at $\,- 1/\wu{1+t^2}\,$
inside the unit circle.

$\cl N$ is proportional to the area $\d \MiA\,$, indeed. 
But it specifies the first contribution $\6_{\MiA} \lw W 
\rw \bigg|_1$ only. Let us now think about the fate of
$\cl N$ under regularization of the theory. Short distances
might be smoothed. So we add a length $1/\L$ to both,
the numerator and the denominator of the propagator
in \eq{17.24}. The calculation runs through the above steps
and results in
\be{17.28}
  \cl N_0^{\rm regularised} =  - 2\, \d \MiA \int\! dt \; 
      {1+\l -it \0 (1+\l)^2+t^2 } 
       \; = \; - 2\,\pi\, \d \MiA \quad \hbox{with} \quad 
  \l \gll {1\0 \L\, \d r} \quad . \quad
\ee 
Thus, regularization (of the above kind, at least) has no
effect on $\cl N$, although the factor $\l$ tends to infinity 
under $\d r \to 0$. Note that by omitting the propagator in 
\eq{17.24} entirely ($\d r \to 0$ before integrating), we would
have $\oint_{\d\cl C}\! dz \int \! dz'\; =$ ``$0\cdot\infty$''.

For the second term, which is $\6_{\MiA} \lw W \rw \bigg|_2$
and given by \eq{17.16}, a similar analysis leads one to
consider
\be{17.29}
  \oint_{\d\cl C}\! dz \oint_{\d\cl C}\! dz' \;  
   {(x-x') + i (y-y') \0 (x-x') - i (y-y') } 
  \; = \;\; \ldots \;\; = 4\pi\, \d \MiA \qquad  
\ee 
in place of $\cl N_0$ (more precisely, in place of $2n\cl N_0/N^2\,$). 
But the fate of this expression under regularization is quite different. 
It simply vanishes for $\l \to \infty\,$. For this, after all, the rough 
argument $\oint_{\d\cl C}\! dz \oint_{\d\cl C}\! dz'\; =$ ''$0\cdot 0$''
might be sufficient. We might state, finally, that \cite{kasako} 
(e.g. in treating intersecting curves $\cl C\,$) goes far ahead over 
what we have used here.
\pagebreak[2]

\subsection{ Area law } 

\nopagebreak[3]     
There we are. Combining \eq{17.23}, \eq{17.24} and \eq{17.27}
the differential equation for the gluon--averaged Wilson loop
turns about as
\be{17.30}
   \6 _{\MiA} \lw W(\cl C)\rw = - {\pi\,n \0 2 \, N^2} 
   \, m^2 \, \lw W(\cl C)\rw \quad . \quad
\ee 
The initial condition is $\lw W(\cl C =0)\rw=1\,$.
Will we be able to solve this?  
\be{17.31}
    \lw W(\cl C)\rw \; = \; e^{-{\pi\, n \0 2\, N^2}\, m^2 \;\MMA } 
    \quad . \quad 
\ee 
The string tension $\s$ of 2+1~D QCD is thus given by
\be{17.32}
  \s = {\pi\, n \0 2\, N^2} \, m^2 \; = \; e^4\, 
  { N^2-1 \0 8\pi}  \quad . \quad
\ee 
This is, we think, an exact result, because the approximation
\eq{17.7} used for the vacuum wave function $\psi_0$ becomes 
accurate for long wavelengths. Afterwards there were no 
further approximations towards a large loop. \eq{17.32} is
\ekk{26}. In \cite{nach}, the comparison with lattice data
exhibits excellent agreement.


\def\bulli{{\tiny $\;\; _\bullet \; _\bullet \; _\bullet\;\;$}}
\def\num#1{\noindent{\bf #1.}\ }

\sec{Magnetic mass} 

There is 4D reality, where all knowledge on the 3D YM system
gets application. Moreover, for the high--temperature gluon gas
it solves the outstanding Linde problem. So, it is quite natural,
to adress a final section to this detail of thermal field theory.
However, if continuing with the foregoing pedestrianity, a text 
book would arise. We therefore decide for the other extreme, 
refer to our own material in \S~18.1, add a few general remarks
in \S~18.2 and end up with a hairraising conjecture in \S~18.3.
\pagebreak[2]

\subsection{ The Linde sea } 

\nopagebreak[3]
\cite{tau}, {\small\sl Magnetic screening in the 
hot gluon system}, Introduction~: 

\leftskip 4mm {\small  
Twenty years ago it was observed by Linde \cite{linde} that the 
perturbative treatment of the high--temperature Yang--Mills 
system runs into a serious problem. If a magnetic mass $m$, 
the system might be able to generate thermally, falls short of 
$g^2T$ in magnitude, the series would diverge, and a phase of 
deconfined gluons could not exist. But even if $m \sim g^2T$,
the perturbation series becomes an (unknown) numerical series. 
Due to this phenomenon \cite{gpy}, no one was able so far to 
calculate the pressure at order $g^6$ or the gluon self-energy 
at $g^4$ \ -- \ a shame for analytical theoretical physics. Today, 
however, there is a way out. \ \bulli  

\begin{figure}[bth]  
 \unitlength 1cm  
 \hspace*{2cm} \begin{picture}(11,1)
 \put(4.5,.3){\oval(4,1)[l]}  \put(4.7,.3){\oval(8,1)[r]}   
 \put(2,.3){\line(1,0){.5}}  \put(8.7,.3){\line(1,0){.5}}  
 \put(4.2,.28){$_\bullet$}     \put(4.5,.28){$_\bullet$}
 \put(4.8,.28){$_\bullet$}
\put(3,.8){\line(2,-3){.66}} \put(3,-.2){\line(2,3){.66}}
\put(6.66,.8){\oval(.6,.6)[t]} \put(6,-.2){\line(2,3){.66}}
\put(6.42,-.2){\oval(1.3,1.3)[tr]} 
\put(1.5,.19){$Q$}  \put(2,.3){\vector(1,0){.3}} 
\put(9.34,.19){$Q$} \put(8.8,.3){\vector(1,0){.3}}
\end{picture}
\caption{\ft An arbitrary 2--leg n--loop skeleton diagram
   with one line added$\,$: the half circle on top, say, or,
   equivalently, the one below. The outer momentum $Q$ 
   is static ($Q_0=0$) and supersoft ($q \sim m$).}
\end{figure}
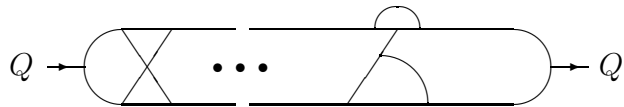

The ``Linde sea'' of diagrams is easily understood from figure 1. 
If one more line is added to an arbitrary skeleton diagram, e.g.
in the manner shown in the figure, then, in the sense of power 
counting, it has two more 3--vertices ($\sim p^2 g^2$), three 
more propagators ($\sim (p^2 + m^2)^{-3}$) and one more loop 
integration ($T\int\! d^3p$, if reduced to the term with zero 
Matsubara frequency). Thus, the (n+1)--loop and n--loop 
differ by a factor $\sim g^2 T \int\! d^3p \, p^2 (p^2+m^2)^{-3} 
\sim g^2T/m\,$. For $m \sim g^2 T$ this factor has order 1 
in magnitude. Once the zero--frequency modes become relevant, 
{\it all} skeletons contribute with the same order of magnitude. 
Any finite--n--loop calculation of the magnetic mass is thus 
inconsistent. 

Bosonic fields live on a cylinder with circumference $\b=1/T$. 
Each loop integration $\sum_P \equiv T\sum_n \int^3_p$, $\int^3_p 
\equiv (2\pi)^{-3} \int\! d^3p$, has its zero--frequency part 
$T\int^3_p\,$. A field depen\-ding on $P$ looses dependence on its 
time coordinate in this part. Irrespective of the physical 
quantity under study, the subset of contributions with $P_0=0$ in 
{\sl all} loops might be the full set of an Euclidean physics at 
$T=0$ in three dimensions. However, this theory needs regulators 
to be derived from the underlying 4D setup.  \ \bulli }

\pagebreak[2]   \leftskip 0pt 

\subsection{ A few remarks } 

\nopagebreak[3]  
\num 1
All about the 3D Yang--Mills system is in its functional
Hamiltonian ${\bf H}\,$, \eq{13.23}, and the associated 
scalar product $\lw 1 | 2 \rw\,$, \eq{14.1}, involving 
$e^{2NS}\,$. Work on this structure remains to be done. 
Before all, however, the KKN setup is waiting for 
quantities to be calculated, because they are relevant
in 4D reality. 

\num 2 
4D. The hot gas of deconfined interacting gluons. Given
a physical quantity of interest, the pressure $p$, say,
or the frequency $\o$ of running waves with wavevector
$\vc q$ (longitudinal or transverse). This quantity has
a prefactor (carrying the dimension, $T^4$ in $p$)
times a dimensionless function $f$ of the dimensionless
coupling $g\,$. $g$ is small. By the term {\sl asymptotic
expansion} we mean a splitting of $f$ into terms of decreasing
order of magnitude \ --- \ but \ n o t \ necessarily
in powers of $g\,$. \ E a c h \ such term can in principle
be measured. It is thus a physical quantity separately, 
hence gauge--fixing independent.

\num 3
A l l \ diagrams that contribute to a term of the asymptotic
expansion (as specified above) must be found and taken into
account. It is just this type of consistency we had to learn
(and to follow up in the sequel) from the Braaten--Pisarski
analysis in 1990. A gauge--fixing independent (gfi) subset
may be found, of course, to consist of several gfi subsubsets.
In particular, this may happen by decomposing the Matsubara sum 
into zero-- and other modes.

\num 4
Once a Matsubara zero--mode is part of your subsubset, then, for 
consistency, \ \hbox{a l l} \ diagrams with the same number of 
external legs, and with the zero--mode selected in each loop,
have to be included: the Linde sea (for the quantity at hand). 
Note that, for the power counting of \S~18.1, the number of 
external legs was irrelevanct. 

\num 5
If, in general covariant gauges, the gluon propagator is reduced 
to its transverse part plus the one with gauge--fixing parameter 
in front of, then the Linde sea diagrams are precisely those of 
Euclidean YM$_3$ \cite{gpy}. This is a superrenormalizable theory. 
It needs regulators to keep the mass finite. Since being a 
substructure of 4D, these regulators are to be found in the 4D 
embedding, namely by watching the other parts of the gluon 
propagator, see \cite{tau,jens}. Working this way, the Linde sea 
turns into a gfi subset, because now it has become a physics by 
itself (though in a ``wrong'' dimension).

\num 6 
Such a Linde sea, now representing a physical quantity (or being 
part of its asymptotic expansion), is a suitably question
to the KKN theory. Rather avoid asking KKN ``for the exact gluon
propagator in covariant gauges''. The latter is gf--dependent,
even in the self--energies (except at pole). So, the answer 
could be: ask physically, you have not yet done your job. 

\num 7
Pressure $p$ of the gluon plasma. The zeroth approximation to 
$p$ is  that of 4D free Bosons, hence $\sim T^4\,$, while the 
infinity of equal--order diagrams occurs at $g^6T^4\,$. The 
dimensional reduction analysis of Braaten and Nieto \cite{branie}
ends up at the supersoft scale by adressing some reduced free 
energy to lattice work. No. Today, this question is to KKN.

\num 8
Static magnetic screening. The zeroth approximation to the 
dispersion lines of the gluon plasma is the Braaten--Pisarski 
setup, hence $\sim g^2T^2$ in the self--energy. This gives the 
Debye mass, but zero for the magnetic one. Here, the Linde sea 
occurs at order $g^4T^2\,$. There is no other subsubset, as 
shown in \cite{tau}. Therefore, using $e^2=g^2T\,$, the 
magnetic screening mass agrees with the {\sl propagator mass}.
Arguments, why the latter could be just $m\,$, are collected
in the next, last subsection.

\num 9
Real photon production. This is one of the LAPTH domains. An 
exciting detail is found in \cite{au}, namely the power counting 
in equation (1) there. Suitably generalized, it might mean that 
the zeroth approximation of the production rate of real photons 
is a Linde sea problem immediately.

\vskip -3mm
\rightline{Annecy--Le--Vieux, 26. June 2000}
\pagebreak[2]

\subsection{ A speculative way out } 

\nopagebreak[3]     
Apparently, we runned into some conflict. On one hand, at the 
end of \S~15.3, the true YM$_{2+1}$ spectrum was expected to 
start with white glue balls. Hence, colored one--gluon--states 
do not exist. On the other hand, 4D TFT needs information about 
a diagrammatic subset summed up exactly. So, in particular, TFT 
claims for an exact {\bf one}--gluon spectrum \ --- \ which does 
not exist. A disaster~?

The 2+1~D theory can be expanded diagrammatically as well, 
thereby exhibiting the perfect agreement with the Linde sea.
Perturbation theory, even if summed up, must not agree with truth 
(e.g. $e^{-1/e^2} \neq 0+0+0+ \ldots\;$). Glue balls are bound 
states. As such, they are outside the realm of diagrammatics. 
The glue ball mass $m_\bullet$ cannot be read off from a 
two--gluon propagator, or in other words, one can n o t \ learn 
about the magnetic mass from the true YM$_{2+1}$ spectrum (how 
to rotate a glue ball into the Euclidean?).   

\vskip 1mm
YM$_{2+1}$ has two faces. We might them separate~: \\
\be{18.1}
\unitlength 1cm  \begin{picture}(15,1.8)
\put(0,2){\parbox[t]{6.9cm}{
  \ct{\bf The perturbative range} \\[2mm]   \ft 
  Colored objects exist, propagate and have mass $m\,$.
  Rotation into the Euclidean is possible, 
  as is identification with corresponding subsets
  of 4D TFT.  
  The whole Linde sea is on this side. Its summation
  reads \\[2mm]
  $ \dis m^2 = \hbox{${1\02}$} \Sp \( \MMA \sum \lk
  \parbox{3.1cm}{\scriptsize all 2PI self--energy \\[-1mm]
                skeleton diagrams with \\[-1mm]
                $(m^2+q^2)^{-1}$ in each line} \rk \,\) $  }}
\multiput(7.3,-2.28)(.01,0){4}{\line(0,1){4.06}}
\put(7.7,2){\parbox[t]{6.4cm}{
  \ct{\bf The YM$_{2+1}$ physical reality} \\[2mm] \ft 
  The energetically lowest states are white glue balls. Their 
  construction as eigenstates of {\bf H} remains a challenge 
  for future time. Vibrations and interaction of balls might 
  complicate the spectrum only at higher energies. Colored 
  objects do not exist. Their energy has turned to infinity. }}
\end{picture} \qquad
\ee 
\\[8mm]

By the above separation (ignoring the details in \eq{18.1} 
for a moment), we are led to the next question, namely, how 
to separate the two regions, or, how to make shure working inside 
one half. For being on the right side, one just has to work with 
the full Hamiltonian {\bf H}. For being on the left side, ``avoid 
bound state formation'', and ``keep colored states at finite 
energy''. But how~? ~For the present, with shaking knees, we
escape into a   
\be{18.2}
  \parbox{11cm}{
  {\bf conjecture~:} \ to reach the (full) perturbative 
  range  \\
  just omit the NT (which is the nonlinear $f^{cab}$ 
  term in {\bf H}).}
\ee 

\vskip -2mm
We have two arguments (both a bit weak) in support of this
conjecture. Nair \cite{dubna} calls $J^a$ {\sl the 
gauge-invariant definition of the gluon}. Consider the 
propagator made up of two such fields (as if they 
could propagate). In Schroedinger picture language it might
read 
\bea{18.3}
  i\,G^{ab} &=& \int \! dt \; e^{ik_0t} \int \! d^2r 
  \; e^{-i \vcsm k \vcsm r} \, \( \,\bigg[ \int \! d\mu (\cl C)
  \( e^P J^a(\vc r)\)^* \, e^{-i{\bf H}t} \, e^P J^b(\vc 0)
  \bigg] \,\theta (t) \; + \right. \nonu \\ 
  & & \hspace*{-2cm} \left . 
     + \,\bigg[ \phantom{\int}  \bigg]^* \theta (-t) \) 
     \;\, , \;\, \bigg[ \phantom{\int} \bigg]  =
     \!\int\!\!\! {d^2q \0 (2\pi)^2} \; e^{-i \vcsm q \vcsm r}
     \!\int\!\!\! {d^2p \0 (2\pi)^2}\!\int\!\! d\mu (\cl C) 
     \, e^{2P} \schl J^{a\,*}(\vc q)\, e^{-i \schl H t} 
     \schl J^b (\vc p)      \;\;\; , \quad
\eea 
where $\,{\bf H}\, e^P=e^P \schl H\,$ led to the second line.
Without NT, we have $\,\schl H\,\schl J^b(\vc p) = 
\wu {m^2+p^2}$ $\schl J^b (\vc p)\,$ from \eq{15.27} and thus 
$G^{ab} \sim \d^{ab}/(k_0^2 - m^2 - k^2)\,$.
With NT, however, $G^{ab}$ has probably no poles 
(perhaps, this can be cleared up by calculation). In passing, 
rotation $k_0\to iq_3$ makes $-k_0^2+\vc k^2$ to become 
$\vc q^2$ as it occurs in \eq{18.1}. 

The second argument arises from a calculation, which 
includes the NT in a first (still inconsistent) way in the 
eigenvalue equation. For brevity, let us report it as if it 
were some (lengthy) exercise. Determine the weight $C$
in the cubic term $P_{\rm cub}=\int_q\int_p\int_o \, 
C^{abc} (\vc q,\vc p,\vc o ) $ $\schl I^a(\vc q) 
\schl I^b(\vc p) \schl I^c(\vc o)\,$ of $P$ by comparing 
the $I^3$ terms in \eq{15.5}. Obtain $C\sim\,$\ekk{34}. 
Use $P = P_{\rm qu} + P_{\rm cub}$ to get $\schl H = 
e^{-P}{\bf H} e^P\,$ from \eq{15.19}. It contains terms 
$I\d$, $I^2\d$, $I^3\d$, $\d\d$ and $I\d\d\,$\footnote{\ 
   Drop the $I^3\d$ term. Then, $\schl H$ is correct to 
   first order in the NT (as if $f^{cab}$ would be small). 
   But $T_3 R I^2 \sim ({\rm NT})^2\,$ is included in 
   $\schl H \chi=E\chi\,$. Possibly, \eq{18.4} corresponds 
   to (is part of) a divergent 1-loop self--energy of a 
   $\phi^3$ theory, whose renormalization, however, 
   is forbidden.}. 
Solve $\schl H \chi = E\chi$ by restricting $\chi$, and 
hence $\schl H \chi$ too, to terms $\sim I$ and $\sim I^2$. 
This is the announced inconsistency (remember \eq{15.20}). 
To be specific, $\chi= \schl I^d(\vc \kappa) + \int_q 
\int_p R^{dab} (\vc \kappa, \vc q, \vc p) \schl I^a(\vc q) 
\schl I^b(\vc p)$. Obtain $R$ from comparing the $I^2$ terms. 
Then, equalizing terms $\sim I\,$, an equation for the 
eigenvalue $E$ derives. Simplify towards large $q\,$, angular 
integrate and put a cutoff $\L$ in by hand. One obtains
\bea{18.4}
  E &=& \wu {m^2+\kappa^2}\, + \,\hbox{${7\02}$} 
  \, m \int_{m_0}^\L \! dq \; {1\0 2q - E} \\  \label{18.5}
  \hbox{\small with solution} & & E \;= \; \hbox{${7\04}$}
  \, m \lk \ln\({\L\0m}\) \; + \; O \Big(\, \ln 
  ( \ln (\L/m)\,\Big) \rk   \quad , \quad
\eea 
showing the irrelevance of the lower cutoff $m_0=O(m, 
\kappa)\,$. By including more and more oscillators (of higher 
and higher wave vector: $\L\to\infty\,$), the energy of this 
colored state $\chi^d(\vc \kappa)$ tends to infinity. So, 
this (first step of a) calculation points in the right 
direction, as it reveals a dramatic effect of the NT and 
the removal of colored states, so announced in the right 
half of \eq{18.1}. 

There is still all the mystery in the conjecture \eq{18.2}, 
of course. Perhaps, the above provokes anyone to give a 
better answer. \rule[-4mm]{0pt}{0pt} 
 
\pagebreak[3]

To end up, let us return to the left half of \eq{18.1}, in
particular to the magnetic screening mass $m_{\rm scr}\,$. 
It is defined as the position of the pole (at imaginary $q$ 
\cite{tau,frosch}) of the transverse piece of the thermal 
gluon propagator in the static limit $Q_0=0$. Pts. 2 to 6 of 
\S~18.2 tell us that $m_{\rm scr}$ is a physical quantity~: 
\be{18.6}
  m_{\rm scr}^2 = \P_t (0, q^2 = - m_{\rm scr}^2) 
 \quad , \quad
 \P_t \gll \hbox{${1\02}$} \Tr \( \MMA \P \) \;\; , \;\;
 \MMA \gll {\vc q \circ \vc q \0 q^2} - 1 \;\; . \quad
\ee 
As a 4 by 4 matrix $\MMA$ has no zeroth components.
Note the rotational invariance of $\P_t\,$ with respect
to $\vc q = (q_1,q_2,q_2)\,$. Diagrammatics and some 
labour concerning possible additional terms and regulators 
\cite{jens} tell us that the function $\P_t$ perfectly 
agrees with the exact gluon self--energy $\P$ of Euclidean 
YM$_3~$, i.e. $\P_t (0, q^2) = \P(q^2)\,$. From YM$_{2+1}$, 
treated diagrammatically in covariant gauges, one would 
read off the spectrum $k_0$ from $0=k_0^2 - \vc k^2 - 
\P(-k_0^2+\vc k^2)$. Here, $\vc k=(k_1,k_2)$ is two--component. 
If one is told the spectrum as $k_0=\wu {m^2+k^2}$ from 
whatsoever non--perturbative treatment [\,only here we now 
make use of the conjecture \eq{18.2}\,], then $0=k_0^2 
- \vc k^2 - \P(-k_0^2+\vc k^2)$ turns into
\be{18.7}
  m^2 = \P(-m^2) \qquad \folg \quad   m_{\rm scr}^2 
  = m^2 \bigg|_{e^2=g^2T} \, = \; {g^4 N^2 T^2 \0 
  4 \pi^2}  \quad . \quad
\ee 
For the mass identification, compare \eq{18.7} (left) with 
\eq{18.6} (left). The formula in the left half of \eq{18.1} 
is the consistency condition for an IR regulator mass 
\cite{rs}, still valid when one is forced (by Linde) to 
become exact. It illustrates that the Linde sea diagramms 
are summed up, indeed.


\subsection*{Acknowledgements}

\vskip -3mm
{\small\sl
In the text, several people are mentioned, helping me with 
difficulties. A big thank to all of them. The people are ~F. 
Brandt, N. Dragon, M. Flohr, O. Lechtenfeld, S. Ketov, 
J. Reinbach, Y. Schr\"oder, J. Schulze and M. Thies. 
I am also indebted to D. Karabali and V.P. Nair for helpful 
discussions during the TFT'98 in August 1998 at Regensburg.  

Special thanks are due to Paul Sorba, LAPTH at Annecy--Le--Vieux 
in France, for his hospitality from April to June 2000. Here, thanks 
to Patrick Aurenche, the English translation was initiated and 
a series of lectures given. I enjoyed detailed discussions 
with him and also with Thomas Binoth, Merab Eliashvili, Ara 
Sedrakyan and Haitham Zaraket. } 


\end{document}